\documentclass[a4paper,11pt]{article}
\pdfoutput=1 

\usepackage{jheppub} 

\usepackage[T1]{fontenc} 

\usepackage{amsmath}    
\usepackage{amssymb}    %
\usepackage{graphicx}   
\usepackage{verbatim}   
\usepackage{color}      
\usepackage{subfigure}  
\usepackage{hyperref}
\usepackage{multirow}
\usepackage{comment}
\usepackage{pdflscape}
\usepackage{rotating}
\usepackage{float}	
\usepackage{color}
\definecolor{rosso}{RGB}{210,0,0}

\newcommand\sss{\scriptscriptstyle}

\newcommand{\gev}{\,\textrm{GeV}}

\newcommand{\TO}{\rightarrow}

\newcommand{\tth}{t\bar{t}H}

\newcommand{\ord}{\mathcal{O}}

\def\beq{\begin{equation}}
\def\beqn{\begin{eqnarray}}
\def\eeq{\end{equation}}
\def\eeqn{\end{eqnarray}}
\def\beal{\begin{align}}
\def\endal{\end{align}}

\newcommand\aNLO{{\sc\small MadGraph5\_aMC@NLO}}

\newcommand\Madspin{{\sc\small Madspin}}
\newcommand\Pythiae{{\sc\small Pythia8}}
\newcommand\FastJet{{\sc\small FastJet}}
\newcommand{\pt}{p_{\sss T}}

\newcommand{\ttbar}{t\bar{t}}
\newcommand{\ttVV}{\ttbar VV}
\newcommand{\ttWW}{\ttbar W^{+}W^{-}}
\newcommand{\ttZZ}{\ttbar ZZ}
\newcommand{\ttaa}{\ttbar \gamma\gamma}
\newcommand{\ttWa}{\ttbar W^{\pm}\gamma}
\newcommand{\ttWZ}{\ttbar W^{\pm}Z}
\newcommand{\ttV}{\ttbar V}
\newcommand{\ttZa}{\ttbar Z\gamma}
\newcommand{\ttZ}{\ttbar Z}
\newcommand{\ttW}{\ttbar W^{\pm}}
\newcommand{\tta}{\ttbar \gamma}
\newcommand{\tttt}{\ttbar \ttbar}
\newcommand{\mua}{\mu_{a}}
\newcommand{\mug}{\mu_{g}}

\allowdisplaybreaks

\title{Associated production of a top-quark pair with vector bosons at NLO in QCD: impact on $t \bar{t} H$ searches at the LHC}
 \author{Fabio Maltoni,}
 \author{Davide Pagani,}
 \author{Ioannis Tsinikos}
 \affiliation{Centre for Cosmology, Particle Physics and Phenomenology (CP3), Universit\'e Catholique de Louvain, B-1348 Louvain-la-Neuve, Belgium}

\emailAdd{fabio.maltoni@uclouvain.be}
\emailAdd{davide.pagani@uclouvain.be}
\emailAdd{ioannis.tsinikos@uclouvain.be}

\abstract{We study the production of a top-quark pair in association with one and two vector bosons, 
$\ttV$ and $\ttVV$ with $V=\gamma, Z, W^\pm$, at the LHC. We provide predictions at next-to-leading order in QCD for total cross sections and top-quark charge asymmetries as well as for differential distributions. A thorough discussion of the residual theoretical uncertainties related to missing higher orders and to parton distribution functions is presented. As an application, we calculate the total cross sections for this class of processes (together with $t \bar t H$ and $t \bar t t \bar t$ production) at hadron colliders for energies up to 100 TeV. In addition,  by matching the NLO calculation to a parton shower, we determine the contribution of $\ttV$ and $\ttVV$ to final state signatures (two-photon and two-same-sign-, three- and four-lepton) relevant for $t \bar t H$ analyses at the Run II of the LHC.}

\preprint{CP3-15-20}

\begin{document} 
\maketitle
\flushbottom

\section{Introduction}
\label{sec:intro}

With the second run of the LHC at 13 TeV of centre of mass energy, the Standard Model (SM) is being probed at the highest energy scale ever reached in collider experiments. At these energies, heavy particles and high-multiplicity final states are abundantly produced, offering the opportunity to scrutinise the  dynamics and the strength of the interactions among the heaviest particles discovered so far: the $W$ and $Z$ bosons, the top quark and the recently observed scalar boson \cite{Aad:2012tfa, Chatrchyan:2012ufa}.  The possibility of measuring the couplings of the top quark with the $W$ and $Z$ bosons and the triple (quadruple) gauge-boson couplings will  further test the consistency of the SM and in case quantify possible deviations.  In addition, the couplings of the Higgs with the $W$ and $Z$ bosons and  the top quark, which are also crucial to fully characterise the scalar sector of the SM, could possibly open a window on   Beyond-the-Standard-Model (BSM) interactions. 

Besides  the study of their interactions,  final states involving the heaviest states of the SM are an important part of the LHC program, because they naturally lead to high-multiplicity final states (with or without missing transverse momentum). This kind of signatures are typical in BSM scenarios featuring new heavy states that  decay via long chains involving, e.g., dark matter candidates.  Thus, either as signal or as background processes, predictions for this class of SM processes need to be known at the best possible accuracy and precision to maximise the sensitivity to deviations from the SM. In other words, the size of higher-order corrections and the total theoretical uncertainties have to be under control. In the case of future (hadron) colliders, which will typically reach higher energies and luminosities, the phenomenological relevance of this kind of processes and the impact of higher-order corrections on the corresponding  theoretical predictions are expected to become even more relevant \cite{Torrielli:2014rqa}.  

In this work we  focus on a specific class of high-multiplicity production process in the SM, i.e., the associated production of a top-quark pair with either one ($\ttV$) or two gauge vector bosons ($\ttVV$). The former includes the processes $\ttW(\ttbar W^+ + \ttbar W^-)$, $\ttZ$ and $\tta$, while the latter counts six different final states, i.e., $\ttWW$, $\ttZZ$, $\ttaa$, $\ttWa$, $\ttWZ$ and $\ttZa$. In addition, we consider also the associated production of two top-quark pairs ($\tttt$), since it will be relevant for the phenomenological analyses that are presented in this work.

The aim of our work is twofold. Firstly, we perform a detailed study of the predictions at fixed NLO QCD accuracy for all the $\ttV$ and $\ttVV$ processes, together with  $\ttbar H$ and $\tttt$ production, within the same calculation framework and using the same input parameters. This approach allows to investigate, for the first time, whether either common features or  substantial differences exist  among the theoretical predictions for different final states. More specifically, we investigate the impact of  NLO QCD corrections on total cross sections and differential distributions. We systematically study the residual theoretical uncertainties due to missing higher orders by considering the dependence of key observables on different definitions of central renormalisation and factorisation scales and on their variations. NLO QCD corrections are known for $\ttbar H$ in \cite{Beenakker:2001rj, Beenakker:2002nc, Dawson:2002tg, Dawson:2003zu}, for $\tta$ in \cite{Melnikov:2011ta,Hirschi:2011pa}, for $\ttZ$ in \cite{Hirschi:2011pa,Lazopoulos:2008de,Garzelli:2011is,Kardos:2011na,Garzelli:2012bn}, for $\ttW$ in  \cite{Hirschi:2011pa,Garzelli:2012bn,Campbell:2012dh, Maltoni:2014zpa} and for $\tttt$ in \cite{Bevilacqua:2012em,Alwall:2014hca}. NLO electroweak and QCD  corrections have also already been calculated for $\ttbar H$ in \cite{Frixione:2014qaa, Yu:2014cka, Frixione:2015zaa} and for $\ttW$ and $\ttZ$ in \cite{Frixione:2015zaa}. Moreover, in the case of $\ttbar H$, NLO QCD corrections have been matched to parton showers \cite{Frederix:2011zi, Garzelli:2011vp} and calculated for off-shell top (anti)quarks with leptonic decays in \cite{Denner:2015yca}. In the case of $\ttbar \gamma$, NLO QCD corrections have been matched to parton showers in \cite{Kardos:2014zba}. For the $\ttVV$ processes a detailed study of NLO QCD corrections has been performed only for $\ttaa$ \cite{Kardos:2014pba,vanDeurzen:2015cga}. So far, only representative results at the level of total cross sections have been presented for the remaining $\ttVV$ processes \cite{Alwall:2014hca,Torrielli:2014rqa}. When possible, i.e. for $\ttV$, $\ttbar H$ and $\ttaa$, our results have been checked against those available in the literature in previous works~\cite{Hirschi:2011pa, Frederix:2011zi,Alwall:2014hca,Kardos:2014pba,Frixione:2015zaa,Garzelli:2011vp,Garzelli:2012bn,Kardos:2014zba,Campbell:2012dh}, and we have found perfect agreement with them. This cross-check can also be interpreted as a further verification of the correctness of both the results in the literature and of the automation of the calculation of NLO QCD corrections in {\aNLO}. 

Secondly, we perform a complete analysis, at NLO QCD accuracy including the matching to parton shower and decays, in a realistic experimental setup, for both signal and background processes involved in the searches for $\ttbar H$ at the LHC. Specifically, we consider the cases where the Higgs boson decays either into two photons ($H\TO\gamma\gamma$), 
 or into leptons 
(via $H\TO WW^{*}$, $H\TO ZZ^{*}$, $H\TO \tau^+  \tau^-$), which have already been analysed by the CMS and ATLAS collaborations at the LHC with 7 and 8 TeV~\cite{Khachatryan:2014qaa,ATLAS:2014mla,Aad:2015iha}. In the first case, the process $\ttaa$ is the main irreducible background. In the second case, the processes $\ttWW$, $\ttZZ$,  $\ttWZ$ are part of the background, although their rates are very small, as we will see. However,  $\ttWW$ production, e.g, has already been  taken into account at LO in the analyses of the CMS collaboration at 7 and 8 TeV, see for instance~\cite{Khachatryan:2014qaa}. A contribution of similar size can originate also from $\tttt$ production \cite{Craig:2013eta}, which consequently has to be included for a correct estimation of the background.\footnote{Triple top-quark production, $t\bar t t W$ and $t\bar t tj$, a process mediated by a weak current, is characterised by a cross section that is one order of magnitude smaller than $\tttt$ at the LHC and it is usually neglected in the analyses.} Furthermore, depending on the exact final state signature, the $\ttV$ processes can give the dominant contribution, which is typically one order of magnitude larger than in $\ttVV$ and $\tttt$ production.
 
In this work, the calculation of the NLO QCD corrections and the corresponding event generation has been performed in the {\aNLO}  framework \cite{Alwall:2014hca}. This code allows the automatic calculation of tree-level amplitudes, subtraction terms and their integration over phase space \cite{Frederix:2009yq} as well as of loop-amplitudes \cite{Ossola:2007ax,Hirschi:2011pa,Cascioli:2011va} once the relevant Feynman rules and UV/$R_2$ counterterms for a given theory are provided~\cite{Degrande:2011ua,Alloul:2013bka,deAquino:2011ub}.  Event generation is obtained by matching short-distance events to the shower employing the MC@NLO method~\cite{Frixione:2002ik}, which is implemented for  {\sc Pythia6}~\cite{Sjostrand:2006za}, {\sc Pythia8}~\cite{Sjostrand:2007gs}, {\sc HERWIG6}~\cite{Corcella:2000bw} and {\sc HERWIG++}\cite{Bahr:2008pv}. The reader can find in the text all the inputs and set of instructions that are necessary to obtain the results presented here.  

The  paper is organised as follows. In section \ref{sec:prod} we present a detailed study of the predictions at NLO QCD accuracy for the total cross sections of $\ttV$, $\ttVV$ and $\tttt$ production. We study their dependences on the variation of the factorisation and renormalisation scales. Furthermore, we investigate the differences among the use of a fixed scale and two possible definitions of dynamical scales. Inclusive and differential   $K$-factors are also shown.  As already mentioned above, these processes are backgrounds to the $\ttbar H $ production with the Higgs boson decaying into leptons, which is also considered in this work. To this purpose, we show also the same kind of results for $\ttbar H$ production. In addition, in the case of $\ttV$ and $\ttbar H$,  we provide  predictions at NLO in QCD for the corresponding  top-charge asymmetries and in order to investigate the behaviour of the perturbative expansion for some key observables, we also compute $\ttbar Vj$ and $\ttbar Hj$ cross sections at NLO in QCD. Such results appear here for the first time. 
In section \ref{sec:prod} we also study the dependence of the total cross sections and of global $K$-factors for $\ttbar VV$ and $\ttbar V$ processes as well as for $\ttbar H$ and $\tttt$ production on the total energy of the proton--proton system, providing predictions in the range from 8 to 100 TeV. 

In section \ref{sec:analysis} we present results at NLO accuracy for the background and signal relevant for $\ttbar H$ production. In subsection \ref{sec:photon}  we consider the signature where the Higgs decays into photons.  In our analysis we implement a selection and a definition of the  signal region that are very similar to those of the corresponding CMS study~\cite{Khachatryan:2014qaa}.  For the signal and background processes $\ttaa$, we compare LO, NLO results and LO predictions rescaled by a global flat $K$-factor for production only, as obtained in section \ref{sec:prod}. We discuss the range of validity and the limitations of the last approximation, which is typically employed in the experimental analyses.
In subsection \ref{sec:leptons} we present an analysis at NLO in QCD accuracy for the searches of $\ttbar H$ production with the Higgs boson subsequently decaying into leptons (via vector bosons), on the same lines of subsection \ref{sec:photon}. In this case, we consider different signal regions and exclusive final states, which can receive contributions from $\tttt$ production and from all the $\ttV$ and $\ttVV$ processes involving at least a heavy vector boson. Also here, we compare LO, NLO results and LO predictions rescaled by a global flat $K$-factor for production only. In section \ref{sec:conclusion} we draw our conclusions and present an outlook.

\section{Fixed-order corrections at the production level}
\label{sec:prod}

In this section we describe the effects of fixed-order NLO QCD corrections at the production level for  
$\ttV$ processes and $\tth$ production (subsection \ref{sec:ttvh}), for  $\ttVV$ processes (subsection \ref{sec:ttvv}) and then for $\tttt$ production (subsection \ref{sec:tttt}). All the results are shown for 13 TeV collisions at the LHC. In subsection \ref{sec:energy} we provide total cross sections   and global $K$-factors for proton--proton collision energies from 8 to 100 TeV.   With the exception of $\ttaa$, detailed studies at NLO  for $\ttVV$ processes are presented here for the first time. The other processes have already been investigated in previous works, whose references have been listed in introduction. Here, we (re-)perform all such calculations within the same framework, {\aNLO}, using a consistent set of input parameters and paying special attention to features that are either universally shared or differ among the various processes. Moreover, we investigate aspects that have been only partially studied in previous works, such as the dependence on (the definition of) the factorisation and renormalisation scales, both at integrated and differential level. To this aim we define the variables that will be used as renormalisation and factorisation scales.

Besides a fixed scale, we will  in general explore the effect of dynamical scales that depend on the transverse masses $(m_{T,i})$ of the final-state particles. Specifically, we will employ the arithmetic mean of the $m_{T,i}$ of the final-state particles ($\mua$) and the geometric mean ($\mug$), which are defined as
\beqn
\mua=\frac{H_{T}}{N}:=\frac{1}{N}\sum_{
   i=1,N(+1)  }m_{T,i}\, ,\label{mua}\\
\mug:=\left(\prod_{\substack{i=1,N}} m_{T,i}\right)^{1/N}\, .\label{mug}
 \eeqn
In these two definitions $N$ is the number of final-state particles at LO and with $N(+1)$ in eq. \eqref{mua}  we understand that, for the real-emission events contributing at NLO, we take into account the transverse mass of the emitted parton.\footnote{This cannot be done for $\mug$; soft real emission would lead to $\mug\sim 0$. Conversely,  $\mua$ can also be defined excluding the partons from real emission and, in the region where $m_{T,i}$'s are of the same order, is numerically equivalent to $\mug$. We remind that by default in {\aNLO} the renormalisation and factorisation scales are set equal to $H_T/2$.} There are two key aspects in the definition of a dynamical scale: the normalisation and the functional form. We have chosen a ``natural'' average normalisation in both cases leading to a value close to $m_t$ when the transverse momenta in the Born configuration can be neglected. This is somewhat conventional in our approach as the information on what could be considered a good choice (barring the limited evidence that a NLO calculation can give for that in first place) can be only gathered a posteriori by explicitly evaluating the scale dependence of the results. For this reason, in our studies of the total cross section predictions, we vary scales over a quite extended range, $\mu_c /8 < \mu < 8 \mu_c$. 
More elaborate choices of even-by-event scales, such as a CKKW-like one~\cite{Catani:2001cc} where factorisation and renormalisation scales are ``local'' and evaluated by assigning a parton-shower like history to the final state configuration, could be also considered. Being ours the first comprehensive study for this class of processes and our aim that of gaining a basic understanding of the dynamical features of these processes, we focus on the simpler definitions above and leave possible refinements to specific applications. 

All the NLO and LO results have been produced with the {\sc\small MSTW2008} (68\% c.l.) PDFs \cite{Martin:2009iq} respectively at NLO or LO accuracy, in the five-flavour-scheme (5FS) and with the associated values of $\alpha_s$. $\ttWW$ production, however,  has been calculated  in the four-flavour-scheme (4FS) with 4FS PDFs, since the 5FS introduces intermediate  top-quark resonances that need to be subtracted and thus unnecessary technical complications.

The mass of the top quark has been set to $m_t=173 \gev$ and the mass  of the Higgs boson to $m_H=125  \gev$, the CKM matrix is considered as diagonal. NLO computations are performed by leaving the top quark and the vector bosons stable. In simulations at NLO+PS accuracy, they are decayed by employing {\sc MadSpin}~\cite{Artoisenet:2012st,Frixione:2007zp} or by \Pythiae.  If not stated otherwise photons are required to have a transverse momentum larger than 20 GeV ($p_T(\gamma)> 20 \gev$) and Frixione isolation \cite{Frixione:1998jh} is imposed for jets and additional photons, with the technical cut $R_0=0.4$. The fine structure constant $\alpha$ is set equal to its corresponding value in the $G_{\mu}$-scheme for all the processes.\footnote{This scheme choice for  $\alpha$  is particularly suitable for processes involving $W$ bosons \cite{Denner:1991kt}.  Anyway, in our calculation, no renormalisation is involved in the  electroweak sector, so results with different values of $\alpha$ can be obtained by simply rescaling the numbers listed in this paper.}

\subsection{$\ttV$ processes and $\tth$ production}
\label{sec:ttvh}

\begin{figure}[t]
\centering
\includegraphics[width=0.8\textwidth]{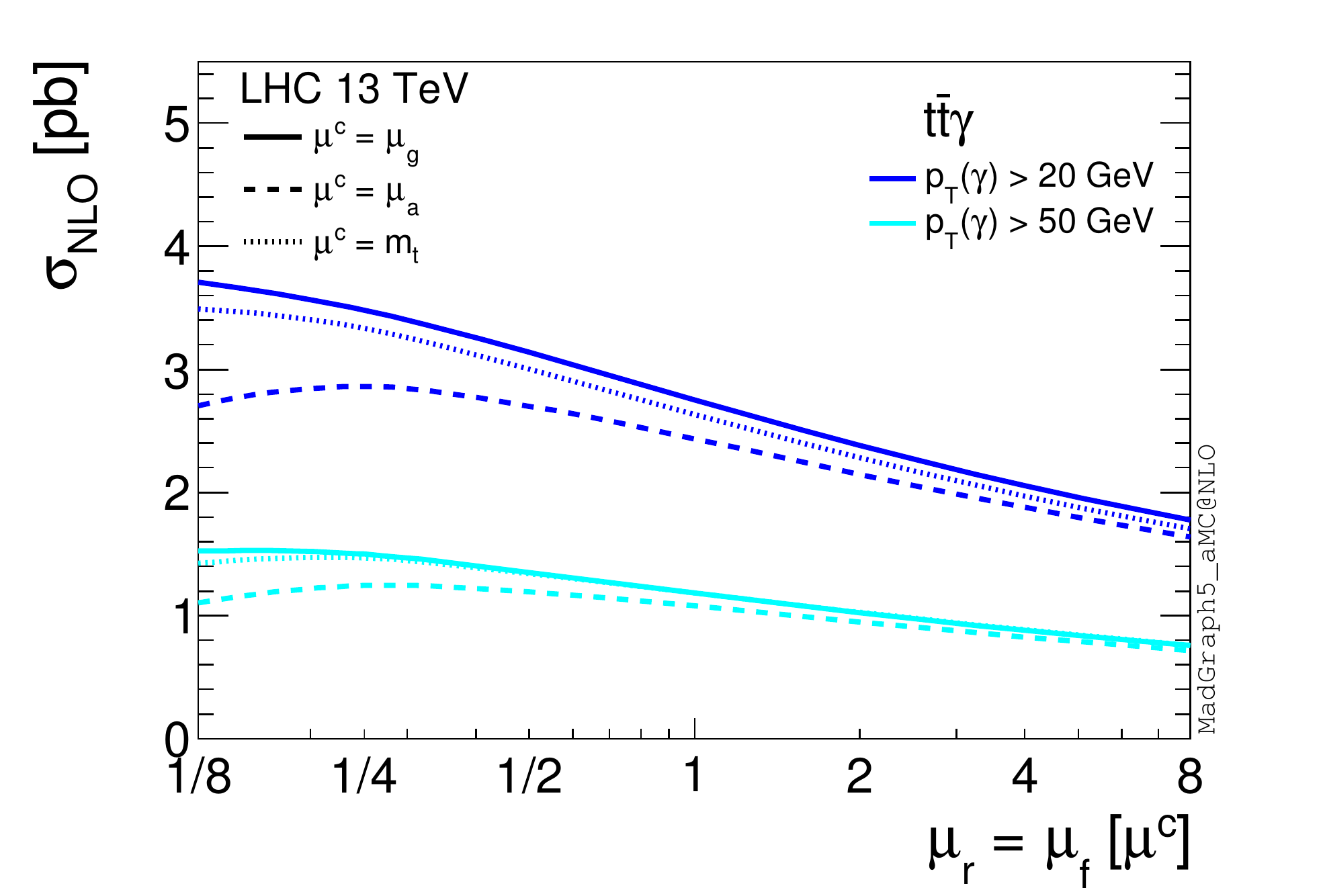}
\includegraphics[width=0.8\textwidth]{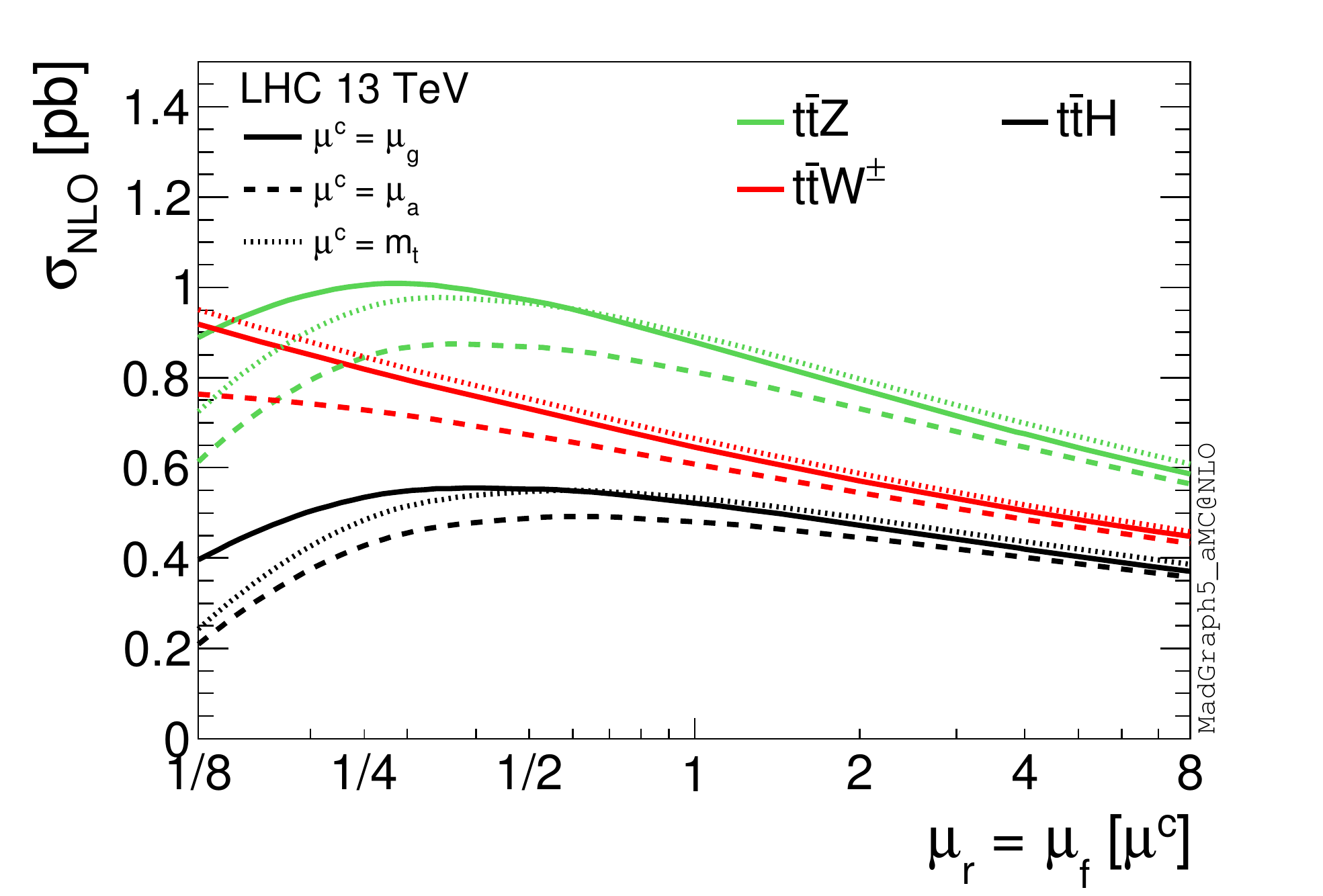}
\caption{Cross sections at 13 TeV. Comparison of the NLO scale dependence in the interval $\mu^c/8<\mu<8\mu^c$ for the three different choices of the central value $\mu^c$: $\mu_g$, $\mu_a$, $m_t$. The upper plot refers to $\tta$ production, the lower plot to $\ttW$, $\ttZ$ and $\ttbar H$ production.}
\label{fig:scales_ttVH}
\end{figure}

As first step, we show for $\tth$ production and all the $\ttV$ processes the dependence of the NLO total cross sections,  at 13 TeV, on the variation of the renormalisation and factorisation scales $\mu_r$ and $\mu_f$. This dependence is shown in fig.~\ref{fig:scales_ttVH} by keeping $\mu=\mu_r=\mu_f$ and varying it by a factor eight around the central value $\mu=\mug$ (solid lines), $\mu=\mua$ (dashed lines) and $\mu=m_t$ (dotted lines). The scales $\mua$ and $\mug$ are respectively defined in eqs. \eqref{mua} and \eqref{mug}. 
 
As typically $\mua$ is larger than $\mug$ and $m_t$, the bulk of the cross sections originates from phase-space regions where $\alpha_s(\mua)<\alpha_s(\mug) ,\, \alpha_s(m_t)$. Consequently, such choice
gives systematically smaller cross sections. On the other hand, the dynamical scale choice $\mug$ leads
to results very close in shape and normalisation to a fixed scale of order $m_t$. 

Driven by the necessity of making a choice, in the following of this section and in the analyses of section \ref{sec:analysis} we will use $\mug$ as reference scale. Also, we will independently vary $\mu_f$ and $\mu_r$ by a factor of two around the central value $\mug$, $\mug/2<\mu_f , \mu_r<2\mug$, in order to estimate the uncertainty of missing higher orders. This generally includes, e.g., almost the same range of values spanned by varying $\mu=\mu_r=\mu_f$ by a factor of four around the central value $\mu=\mua$, $\mua/4<\mu<4\mua$ (cf. fig.~\ref{fig:scales_ttVH}) and thus it can be seen as a conservative choice. In any case, while certainly justified a priori as well as a posteriori, we stress that the $\mu=\mug$ choice is an operational one, i.e. we do not consider it as our ``best guess'' but just use it as reference for making meaningful comparisons with other possible scale definitions and among different processes.   

\begin{figure}[t]
\centering
\includegraphics[width=0.8\textwidth]{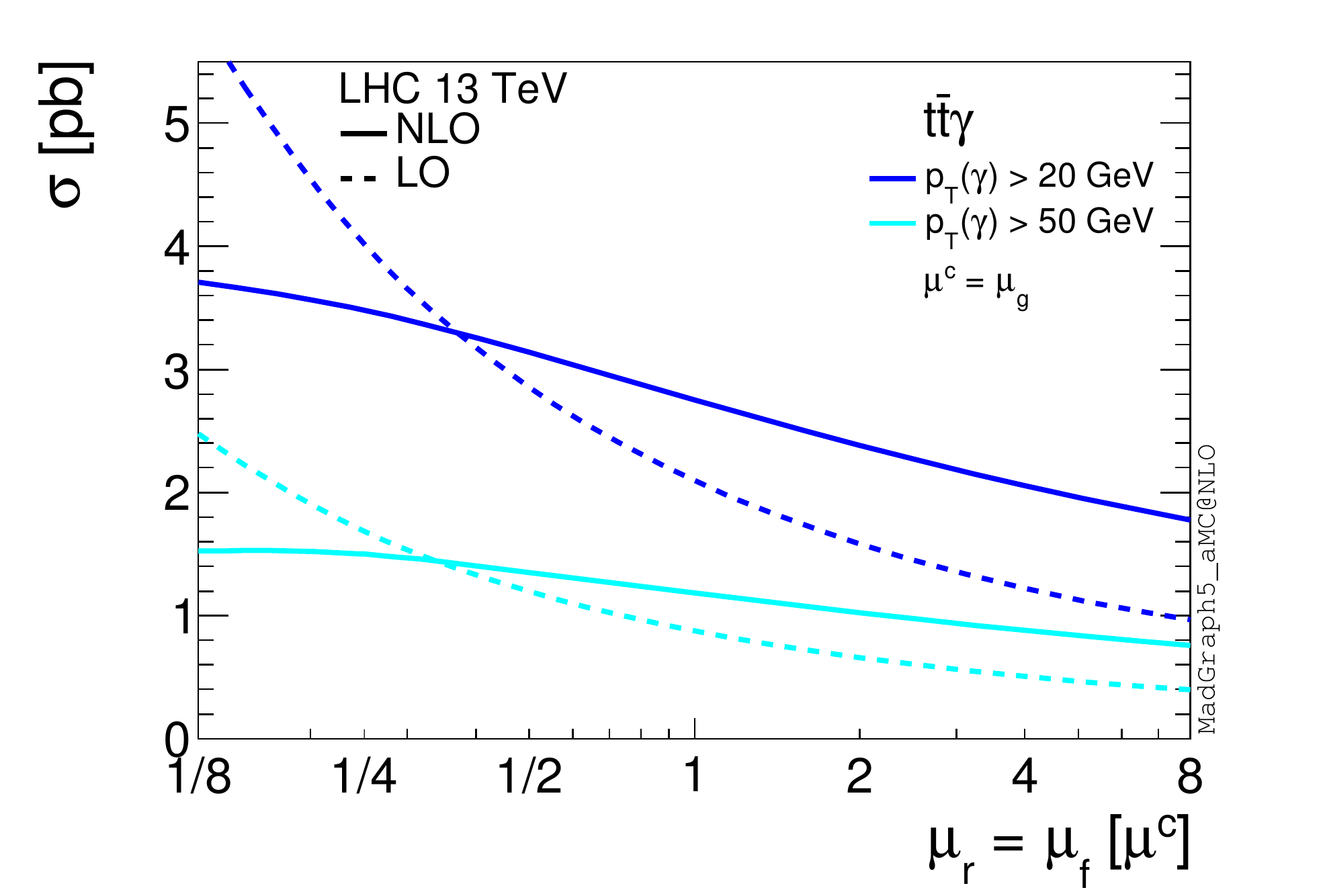}
\includegraphics[width=0.8\textwidth]{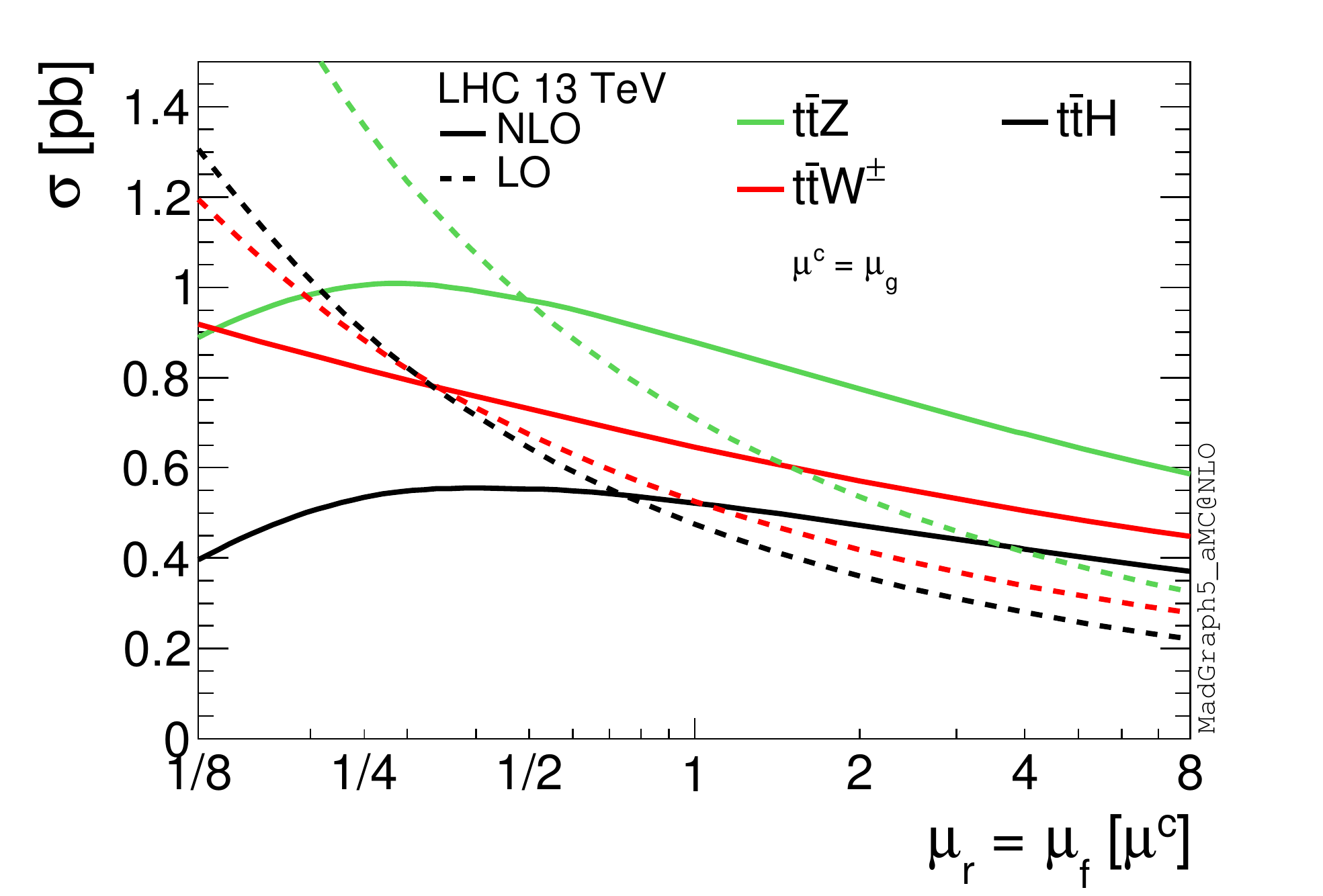}
\caption{LO and NLO cross sections at 13 TeV. Scale dependence in the interval $\mu^c/8<\mu<8\mu^c$ with $\mu^c=\mu_g$. The upper plot refers to $\tta$ production, the lower plot to $\ttW$, $\ttZ$ and $\ttbar H$ production.}
\label{fig:diagttVH}
\end{figure}

Using the procedure described before, in table \ref{table:13tevttv} we list, for all the processes, LO and NLO cross sections together with PDF and scale uncertainties, and $K$-factors for the central values.
The dependence of the LO and NLO cross sections on $\mu=\mu_r=\mu_f$ is also shown in  fig.~\ref{fig:diagttVH}  in the range $\mug/8<\mu<8\mug$. As expected,  for all the processes, the scale dependence is strongly reduced from LO to NLO predictions both in the standard interval $\mug/2<\mu<2\mug$ as well as in the full range $\mug/8<\mu<8\mug$. For $\tta$ process (upper plots in figs.~\ref{fig:scales_ttVH} and~\ref{fig:diagttVH}), we find that in general the dependence of the cross-section scale variation is not strongly affected by the minimum $p_T$ of the photon, giving similar results for $p_T(\gamma)> 20 \gev$ and $p_T(\gamma)> 50 \gev$. As already stated in section \ref{sec:intro}, with $\ttW$ we refer to the sum of the $t\bar{t}W^+$ and $t\bar{t}W^-$ contributions.

\noindent

\begin{table}[t]
\small
\renewcommand{\arraystretch}{1.5}
\begin{center}
\begin{tabular}{  c | c c c c c }
\hline\hline
13 TeV $ \sigma$[fb] & $t \bar t H$ &  $t \bar t Z$ & $\ttW$ & $t \bar t \gamma$ \\
\hline
NLO & $522.2^{+6.0 \%}_{-9.4 \%}~^{+2.1 \%}_{-2.6 \%}$ & $873.6^{+10.3 \%}_{-11.7 \%}~^{+2.0 \%}_{-2.5 \%}$ & $644.8^{+13.0 \%}_{-11.6 \%}~^{+1.7 \%}_{-1.3 \%}$ & $2746^{+14.2 \%}_{-13.5 \%}~^{+1.6 \%}_{-1.9 \%}$ \\
\hline
LO & $476.6^{+35.5 \%}_{-24.2 \%}~^{+2.0 \%}_{-2.1 \%}$ & $710.3^{+36.1 \%}_{-24.5 \%}~^{+2.0 \%}_{-2.1 \%}$ & $526.9^{+28.1 \%}_{-20.4 \%}~^{+1.7 \%}_{-1.8 \%}$ & $2100^{+36.2 \%}_{-24.5 \%}~^{+1.8 \%}_{-1.9 \%}$ \\
\hline
$K$-factor & 1.10 & 1.23 & 1.22 & 1.31 \\

\hline
\end{tabular}
\caption{NLO  and LO cross sections for $\ttV$ processes and $\ttbar H$ production at 13 TeV for $\mu=\mu_g$. As already stated in the text, with $\ttW$ we refer to the sum of the $t\bar{t}W^+$ and $t\bar{t}W^-$ contributions. The first uncertainty is given by the scale variation within $\mug/2<\mu_f,\mu_r<2\mug$, the second one by PDFs. The relative statistical integration error is equal or smaller than one permille.}
\label{table:13tevttv}
\end{center}
\end{table}

We now show the impact of NLO QCD corrections on important distributions and we discuss their dependence on the scale variation as well as on the definition of the scales.
For all the processes we analysed the distribution of the invariant mass of the top-quark pair and the  $\pt$ and the rapidity of the (anti)top quark, of the top-quark pair and of the vector or scalar boson. 
Given the large amount of distributions, we show only representative results. All the distributions considered and additional ones can be produced via the public code \aNLO. 

For each figure, we display together the same type of distributions for the four different processes: $\tta$, $\ttbar H$, $\ttW$ and $\ttZ$. Most of the plots for each individual process will be displayed in the format described in the following.

In each plot, the main panel shows  the specific distribution  at LO (blue) and NLO QCD (red) accuracy, with $\mu=\mu_f=\mu_r$   equal to the reference scale $\mug$.
In the first inset we display  scale and PDF uncertainties normalised to the blue curve, i.e., the LO with $\mu=\mug$. The mouse-grey band indicates the scale variation at LO in the standard range  $\mug/2<\mu_f, \mu_r<2\mug$, while the dark-grey band shows the PDF uncertainty. The black dashed line is the central value of the grey band, thus it is by definition equal to one.
The solid black line is the NLO QCD differential $K$-factor at the scale $\mu=\mu_g$, the red band around it indicates the scale variation in the standard range  $\mug/2<\mu_f, \mu_r<2\mug$. The additional blue borders show the PDF uncertainty.
We stress that in the plots, as well as in the tables, scale uncertainties are always obtained by the independent variation of the factorisation and renormalisation scales, via the reweighting technique introduced in \cite{Frederix:2011ss}.  
 The second and third insets show the same content of the first inset, but with different scales.
In the second panel both LO and NLO have been evaluated with $\mu=\mua$, in the third panel with $\mu=m_t$.

The fourth and the fifth panels show a direct comparison of NLO QCD predictions using the scale $\mug$ and, respectively, $\mu_a$ and $m_t$. All curves are normalised to the  red curve in the main panel, i.e., the NLO with $\mu=\mug$. The mouse-grey band now indicates the scale variation dependence of NLO QCD with $\mu=\mug$. Again the dashed black line, the central value, is by definition equal to one and the dark-grey borders represent the PDF uncertainties. The black solid line in the fourth panel is the ratio of the NLO QCD predictions at the scale $\mua$ and $\mug$. The red band shows the scale dependence of NLO QCD predictions at the scale $\mua$, again normalised to the central value of NLO QCD at the scale $\mug$, denoted as $R(\mua)$. Blue bands indicate the PDF uncertainties.
The fifth panel, $R(m_t)$, is completely analogous to the fourth panel, but it compares NLO QCD predictions with $\mug$ and $m_t$ as central scales.\\

\begin{figure}[t]
\centering
\includegraphics[width=0.475\textwidth]{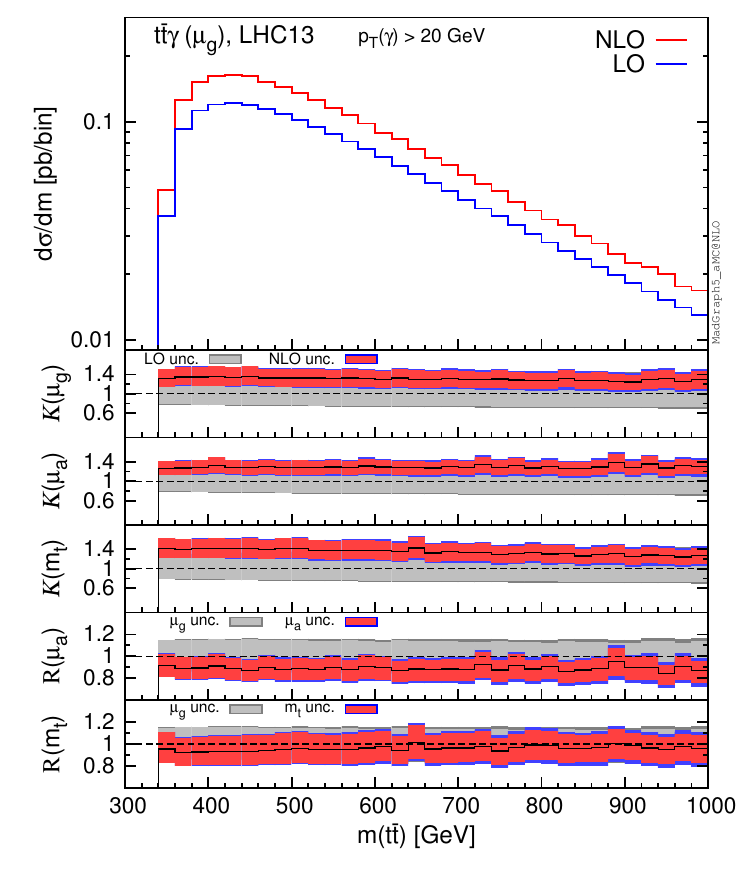}
\includegraphics[width=0.475\textwidth]{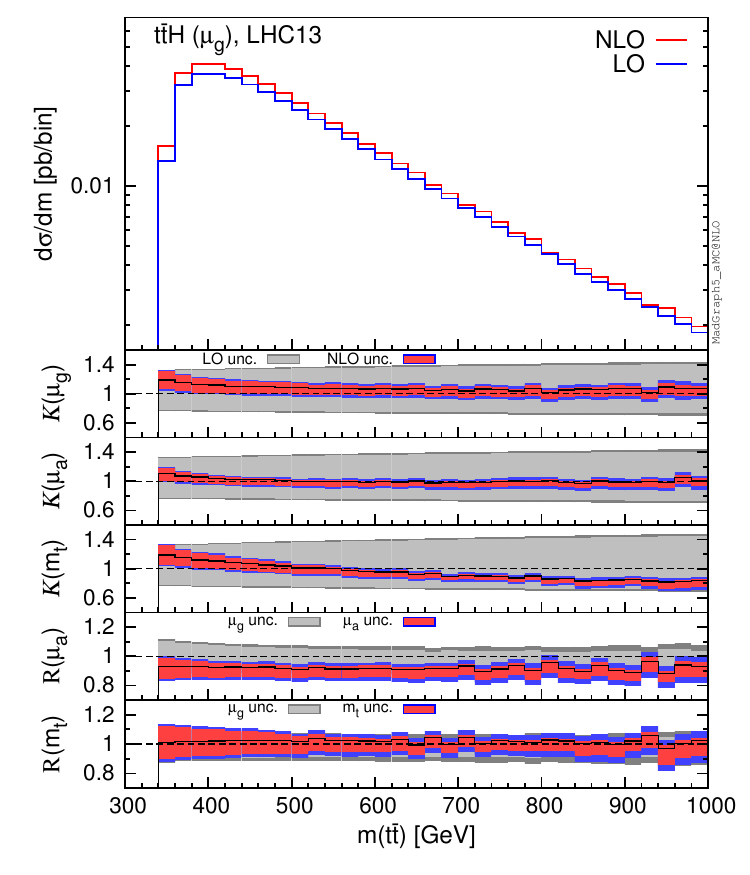}
\includegraphics[width=0.475\textwidth]{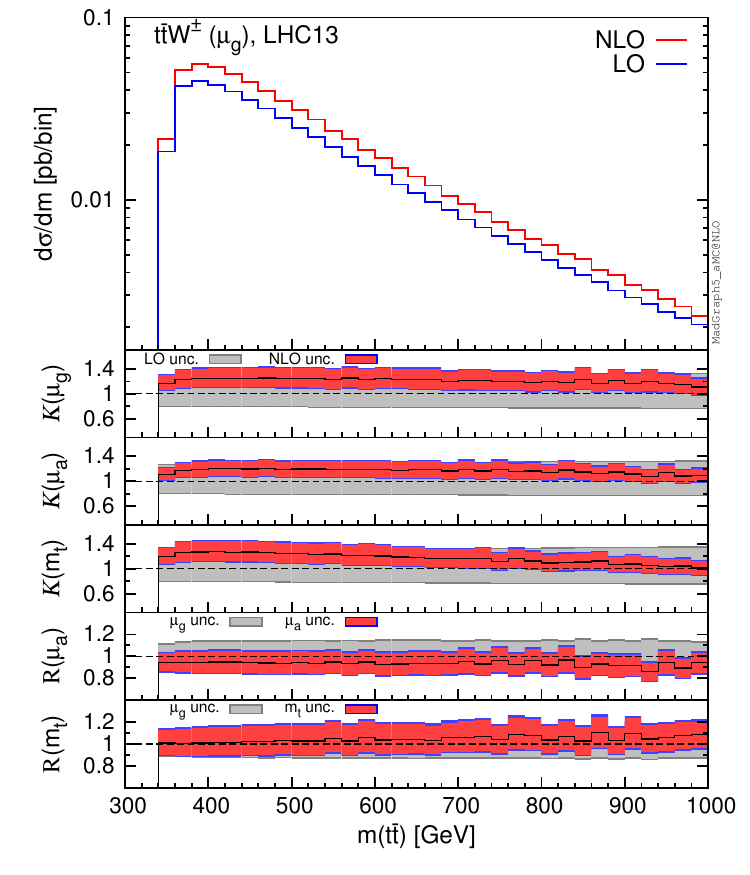}
\includegraphics[width=0.475\textwidth]{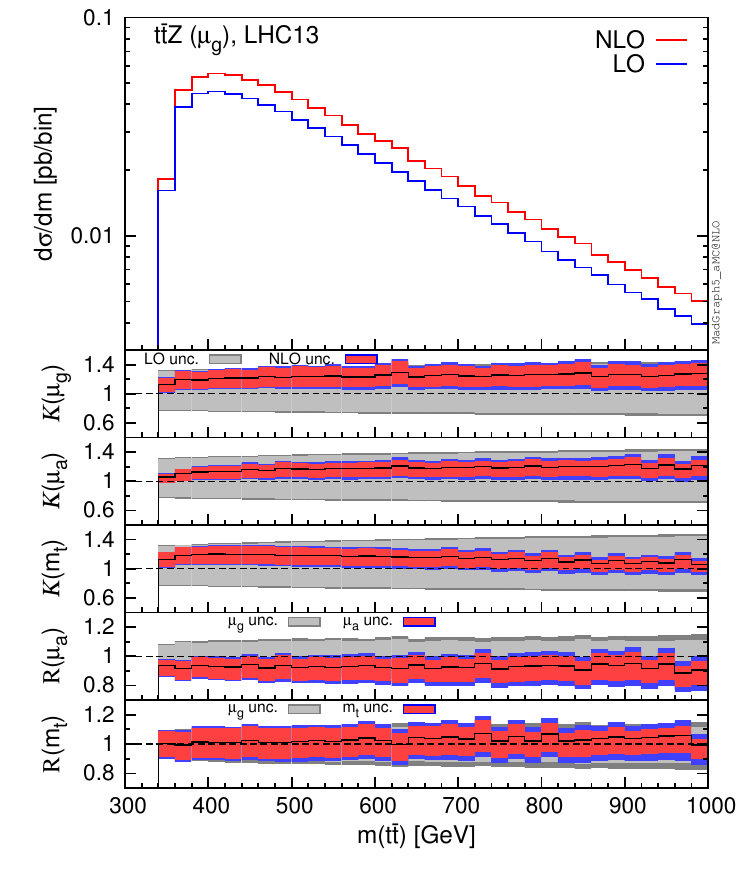}
\caption{Differential distributions for the invariant mass of top-quark pair, $m(\ttbar)$. The format of the plots is described in detail in the text.}
\label{fig:ttV_inv}
\end{figure}

\begin{figure}[t]
\centering
\includegraphics[width=0.475\textwidth]{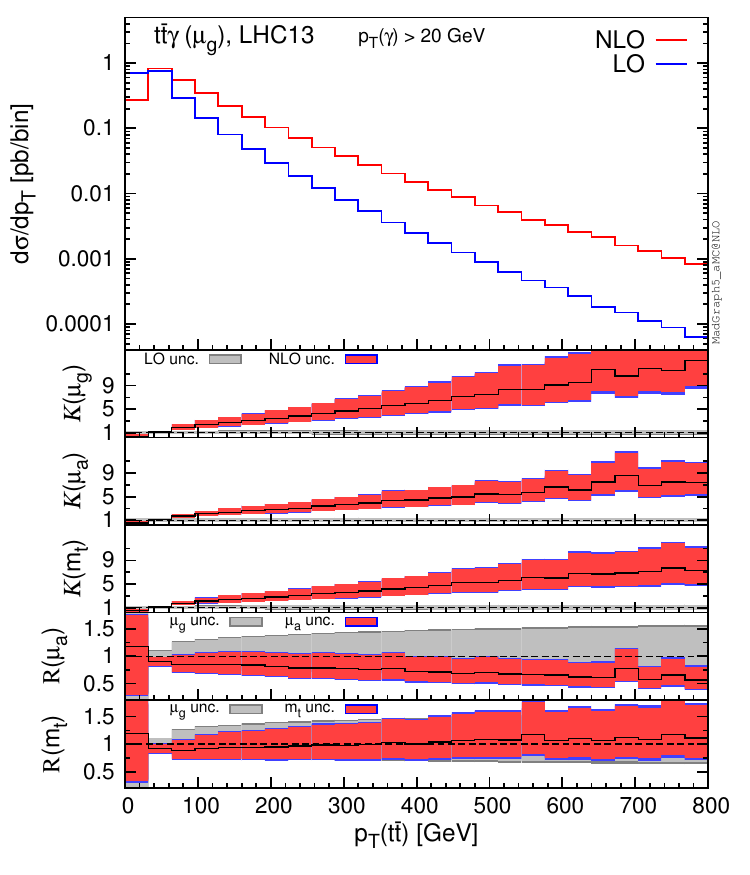}
\includegraphics[width=0.475\textwidth]{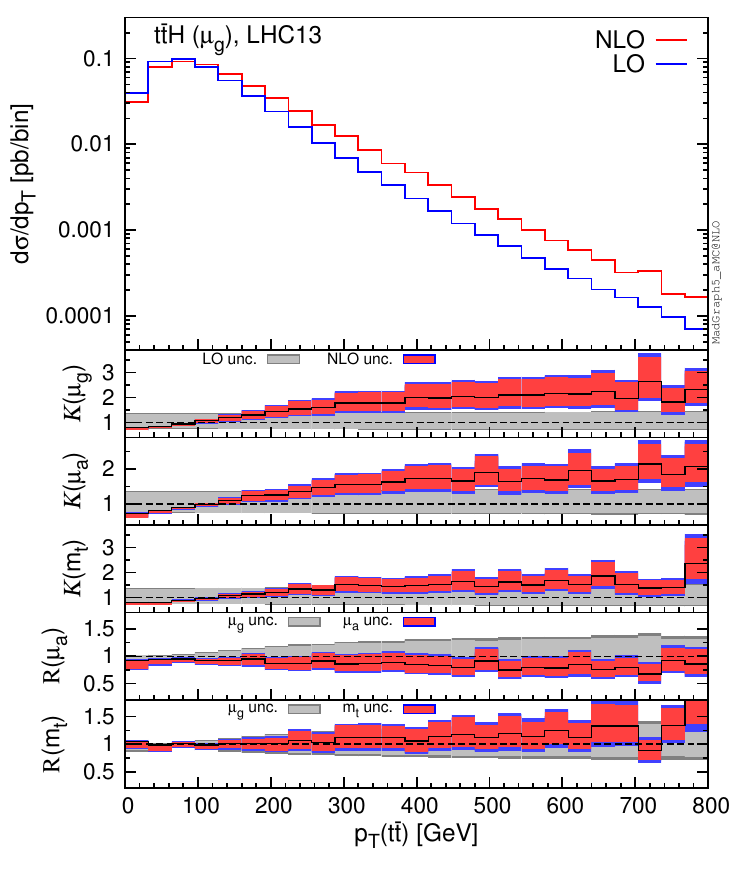}
\includegraphics[width=0.475\textwidth]{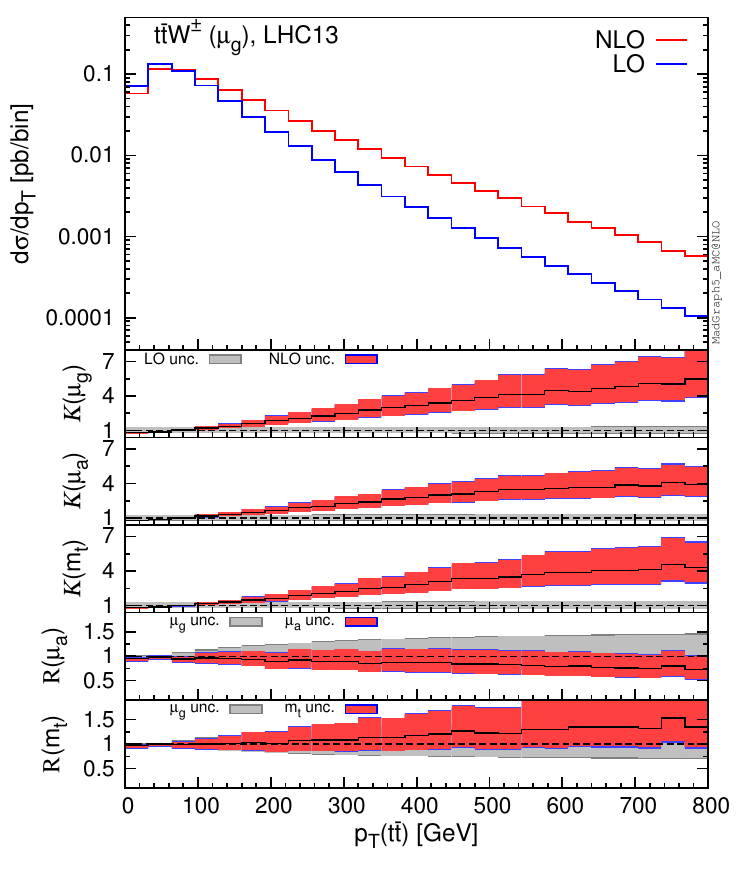}
\includegraphics[width=0.475\textwidth]{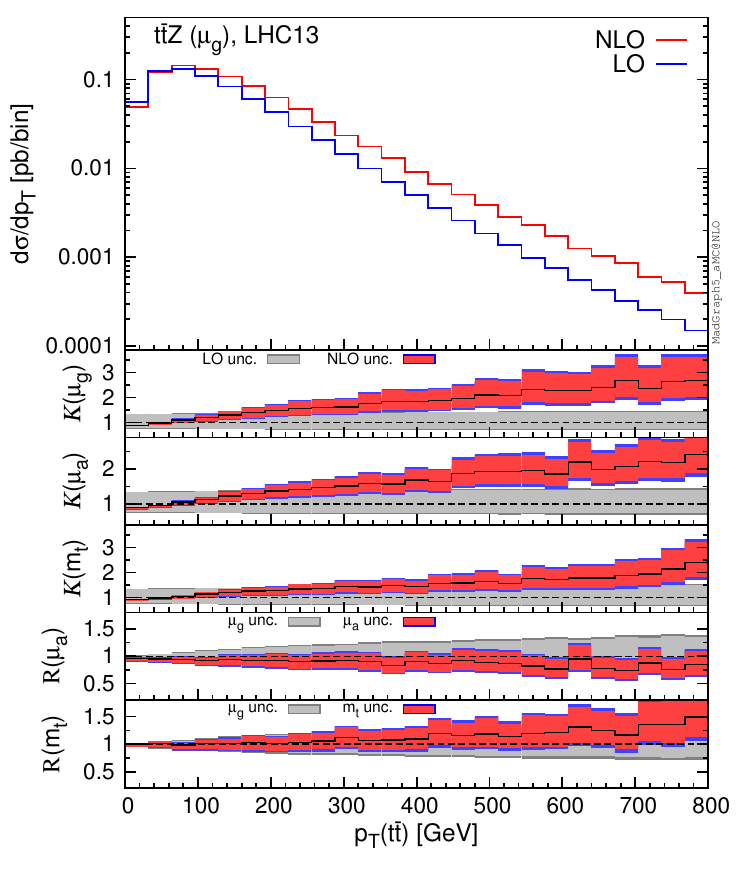}
\caption{Differential distributions for the $\pt$ of top-quark pair, $\pt(\ttbar)$. The format of the plots is described in detail in the text.}
\label{fig:ttV_ptttx}
\end{figure}

We start with fig.~\ref{fig:ttV_inv}, which shows the distributions for the invariant mass of the top-quark pair ($m(\ttbar)$) for the four production processes. From this distribution it is possible to note some features that are in general true for most of the distributions. As can be seen in the fourth insets, the use of $\mu=\mua$ leads to NLO values compatible with, but systematically smaller than, those obtained with $\mu=\mu_g$. Conversely, the using $\mu=m_t$ leads  to scale uncertainties bands that overlap with those obtained with $\mu=\mu_g$.  By comparing the first three insets for the different processes, it can be noted that the reduction of the scale dependence from LO to NLO results is stronger in $\ttbar H$ production than for the $\ttV$ processes. As we said, all these features are not peculiar for the $m(\ttbar)$ distribution, and are consistent with the total cross section analysis presented before, see fig.~\ref{fig:scales_ttVH} and table \ref{table:13tevttv}. From fig.~\ref{fig:ttV_inv} one can see that the two dynamical scales $\mug$ and $\mua$ yield flatter $K$-factors than those from the fixed scale $m_t$, supporting a posteriori such a reference scale. While this feature is  general,  there are important exceptions. This is particular evident for the distributions of the $\pt$ of the top-quark pair ($\pt(\ttbar)$) in fig.~\ref{fig:ttV_ptttx}, where the differential $K$-factors strongly depend on the value of $\pt(\ttbar)$ for both dynamical and fixed scales. The relative size of QCD corrections grows with the values of $\pt(\ttbar)$ and this effect is especially large in $\ttW$ and $\tta$ production. In the following we investigate the origin of these large $K$-factors.

\begin{figure}[t]
\centering
\includegraphics[trim=00 500 0 50,width=0.8\textwidth]{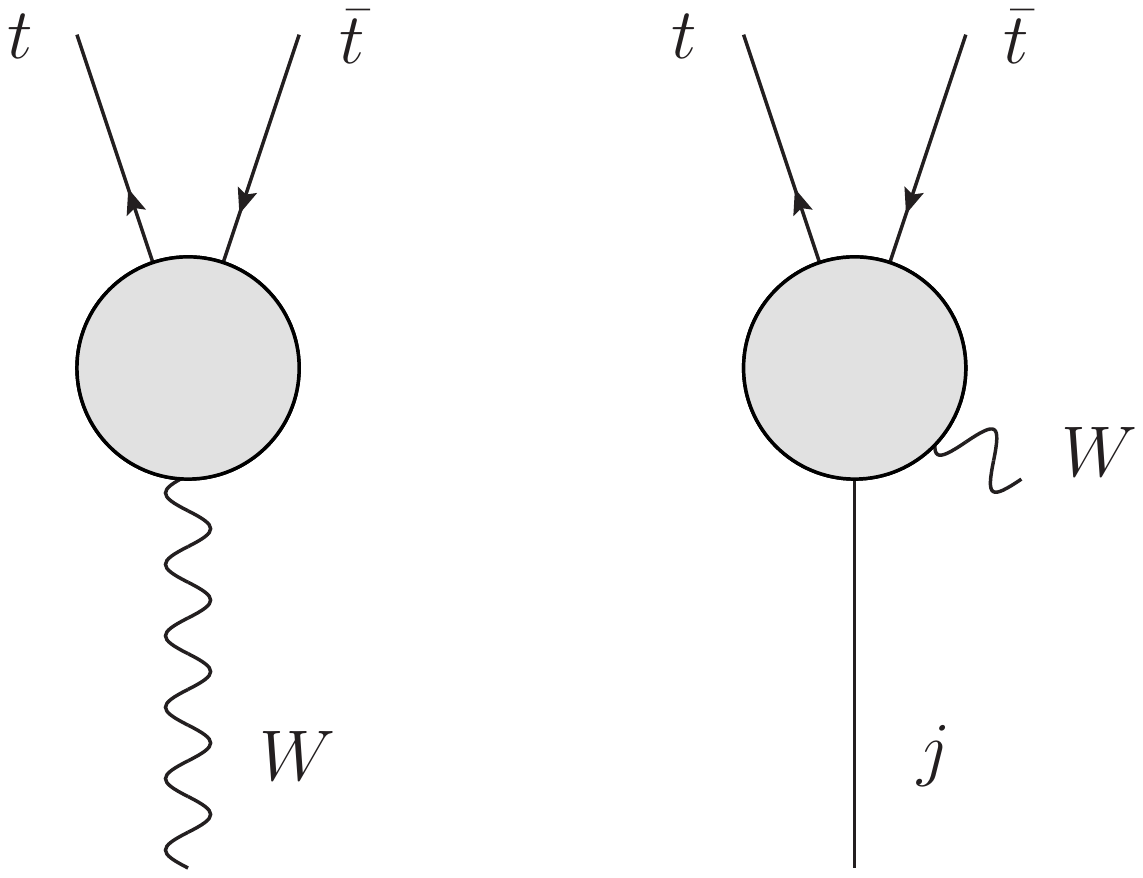}
\caption{Representative kinematical configurations for $\ttbar W$ final state. At LO (left) a high-$\pt$ $\ttbar$ pair recoils against the $W$ boson. At NLO (right), the dominant configuration is the one where the jet takes most of the recoil and the $W$ boson is soft.}
\label{fig:ttW_sketch}
\end{figure}

Top-quark pairs with a large $\pt$ originate at LO from the recoil against a hard vector or scalar boson. Conversely, at NLO, the largest contribution to this kinetic configuration emerges from the recoil of the top-quark pair against a hard jet and a soft scalar or vector boson (see the sketches in fig.~\ref{fig:ttW_sketch}). In particular, the cross section for a top-quark pair with a large $\pt$ receives large corrections from (anti)quark--gluon initial state, which appears for the first time in the NLO QCD corrections. This effect is further enhanced in $\ttW$ production for two different reasons. First, at LO $\ttW$ production does not originate, unlike the other production processes, form the gluon--gluon initial state, which has the largest partonic luminosity. Thus, the relative corrections induced by (anti)quark--gluon initial states have a larger impact. Second, the emission of a $W$ collinear to the final-state (anti)quark in $qg\rightarrow \ttW q'$ can be approximated as the $qg\rightarrow \ttbar q$ process times a $q\rightarrow q' W^\pm$ splitting. For the $W$ momentum, the splitting involves a soft and collinear singularity which is regulated by the $W$ mass. Thus, once the $W$ momentum is integrated, the  $qg\rightarrow \ttW q'$ process yields contributions to the $\pt(\ttbar)$ distributions that are proportional to $\alpha_s\log^2\left[\pt(\ttbar)/m_W\right]$.\footnote{In $\ttZ$ the same argument holds for the  $q\rightarrow q Z$ splitting in $qg\rightarrow \ttZ q$. However, the larger mass of the $Z$ boson and especially the presence of the gluon--gluon initial state at LO suppress this effect.}  The same effect has been already observed for the $\pt$ distribution of one vector boson in NLO QCD and EW corrections to $W^\pm W^\mp,W^\pm Z$ and $ZZ$ bosons hadroproduction  \cite{Frixione:1992pj,Frixione:1993yp,Baglio:2013toa}.

\begin{figure}[t]
\centering
\includegraphics[width=0.8\textwidth]{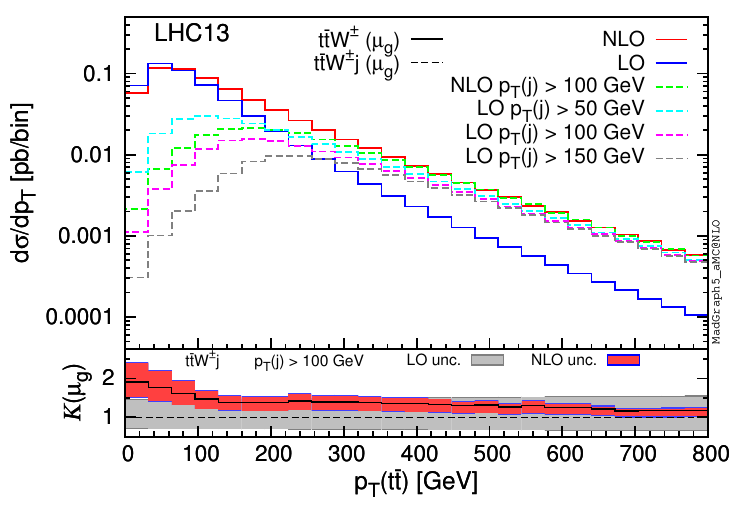}
\caption{Comparison between differential distribution of the $t\bar t$ transverse momentum in $\ttW$
as obtained from calculations performed at different orders in QCD.  The blue and red solid histograms are obtained from the $\ttW$ calculation at LO and NLO respectively. The dashed histograms are obtained  from the $\ttW j$ calculation at LO (light blue, purple, and mouse-grey) and at NLO (green), for different minimum cuts (50, 100, 150 GeV) on the jet $\pt$. The lower inset shows the differential $K$-factor as well as the residual uncertainties as given by  the $\ttW j$ calculation.}
\label{fig:ttVj}
\end{figure}
\noindent

The argument above clarifies the origin of the enhancement at high $\pt$ of the $\ttbar$ pair, yet it raises the question of the reliability of the NLO predictions for $\ttbar V$ in this region of the phase space. In particular the giant $K$-factors and the large scale dependence call for better predictions. At first,  one could argue that only a complete NNLO calculation for $\ttbar V$ would settle this issue. However, since the dominant kinematic configurations (see the sketch on the right in fig.~\ref{fig:ttW_sketch}) feature a hard jet, it is possible to start from the  $\ttbar Vj$ final state and reduce the problem to the computation of NLO corrections to $\ttbar Vj$. Such predictions can be automatically obtained within \aNLO.  We have therefore computed results for different minimum $\pt$ for the extra jet both at NLO and LO accuracy. In fig.~\ref{fig:ttVj} we summarise  the most important features of the
$\ttW(j)$ cross section  as a function of the $\pt(\ttbar)$ as obtained from different calculations and orders. Similar results, even though less extreme, hold for $\ttZ$ and $\ttbar H$ final states and therefore
we do not show them for sake of brevity. 
In fig.~\ref{fig:ttVj}, the solid blue and red curves correspond to the predictions of $\pt(\ttbar)$ as obtained from $\ttW$ calculation at LO and NLO, respectively. The dashed light blue, purple and mouse-grey curves are obtained by calculating $\ttW j$ at LO (yet with NLO PDFs and $\alpha_s$ and same scale choice in order to consistently compare them with NLO  $\ttW$ results) with a minimum $\pt$ cut for the jets of 50, 100, 150 GeV, respectively. The three curves, while having a different threshold behaviour, all tend smoothly to the $\ttW$ prediction at NLO at high $\pt(\ttbar)$, clearly illustrating the fact that the dominant contributions come from kinematic configurations featuring a hard jet, such as those depicted on the right of fig.~\ref{fig:ttW_sketch}. Finally, the dashed green line is the  $\pt(\ttbar)$ as obtained from $\ttW j$ at NLO in QCD  with a minimum $\pt$ cut of the jet of 100 GeV. This prediction for $\pt(\ttbar)$ at high $\pt$  is stable and reliable, and in particular does not feature any large $K$-factor, as can be seen in the lower inset which displays the differential $K$-factor for $\ttW j$ production with $\pt$ cut of the jet of 100 GeV. For large $\pt(\ttbar)$, NLO corrections to $\ttW j$ reduce the scale dependence of LO predictions, but do not increase their central value. Consequently, as we do not expect  large effects from NNLO corrections in $\ttW$ production at large $\pt(\ttbar)$,  a simulation of NLO $\ttV$+jets merged sample à la FxFx~\cite{Frederix:2012ps} should be sufficient to provide reliable predictions over the full phase space.

\begin{table}[t]
\small
\renewcommand{\arraystretch}{1.5}
\begin{center}
\begin{tabular}{  c | c c c c }
\hline\hline
13 TeV $ \sigma$[fb] & $t \bar t Hj$ &  $t \bar t Zj$ & $t\bar t Wj$ \\
\hline
NLO & $148.3^{+3.3 \%}_{-10.1 \%}~^{+3.0 \%}_{-3.6 \%}$ & $230.7^{+6.6 \%}_{-13.4 \%}~^{+2.8 \%}_{-3.2 \%}$ & $202.9^{+11.6 \%}_{-15.6 \%}~^{+1.4 \%}_{-1.1 \%}$ \\
\hline
LO & $174.5^{+57.8 \%}_{-33.9 \%}~^{+2.8 \%}_{-2.9 \%}$ & $243.1^{+58.2 \%}_{-34.0 \%}~^{+2.7 \%}_{-2.8 \%}$ & $197.6^{+53.7 \%}_{-32.4 \%}~^{+1.5 \%}_{-1.5 \%}$\\
\hline
$K$-factor & 0.85 & 0.95 & 1.03 \\
\hline
\end{tabular}
\caption{Cross sections with $p_T(j) > 100$ GeV. The renormalisation and factorisation scales are set to $\mu_g$ of $t \bar t V$. The (N)LO cross sections are calculated with (N)LO PDFs, the relative statistical integration error is equal or smaller than one permil.
\label{tab:ttVj} }
\end{center}
\end{table}

\begin{figure}[t]
\centering
\includegraphics[width=0.475\textwidth]{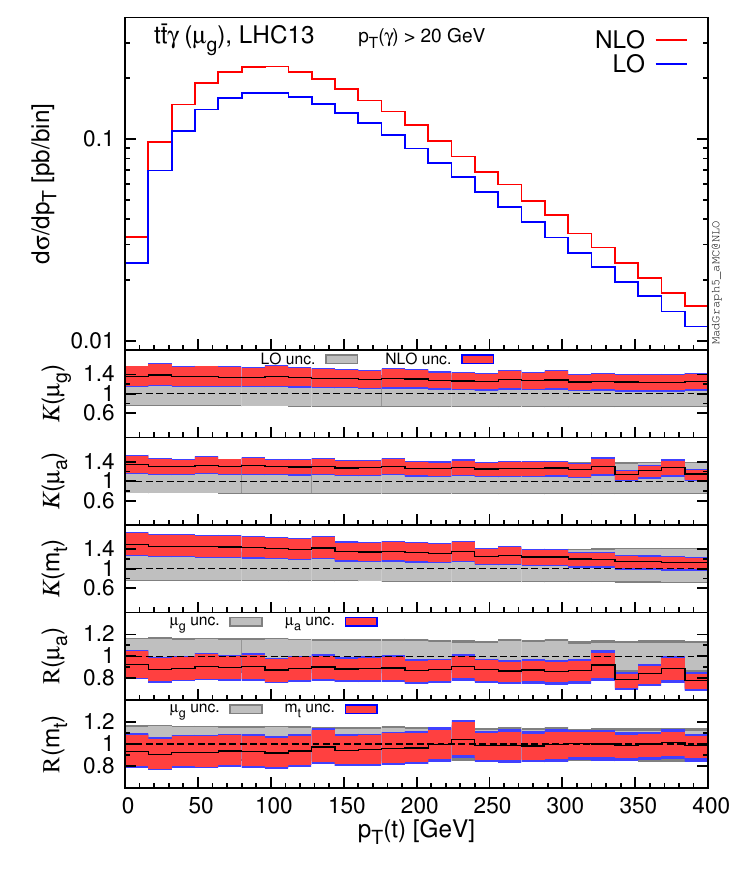}
\includegraphics[width=0.475\textwidth]{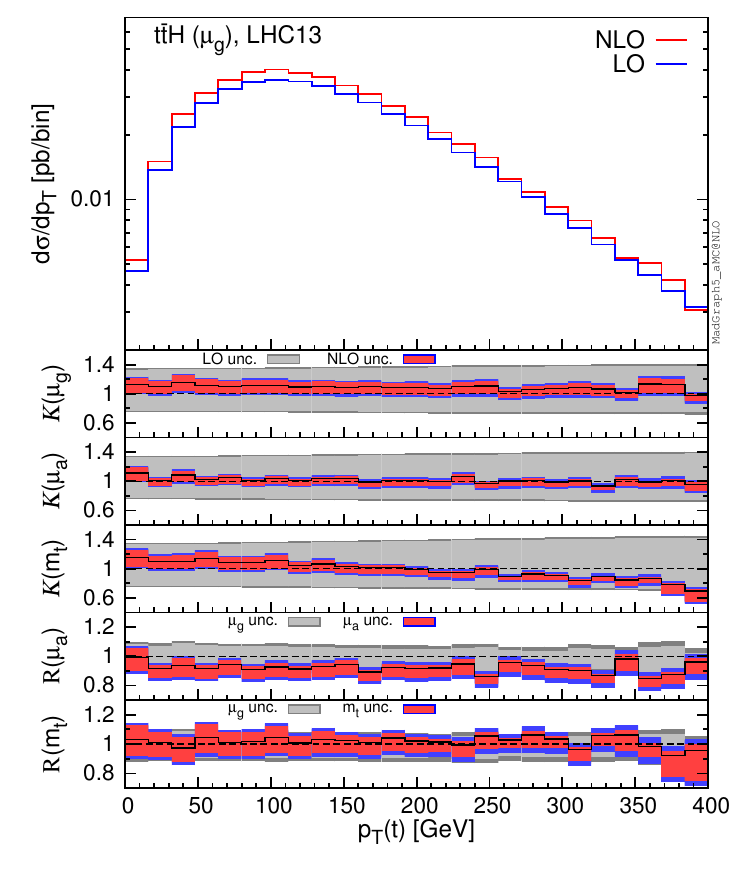}
\includegraphics[width=0.475\textwidth]{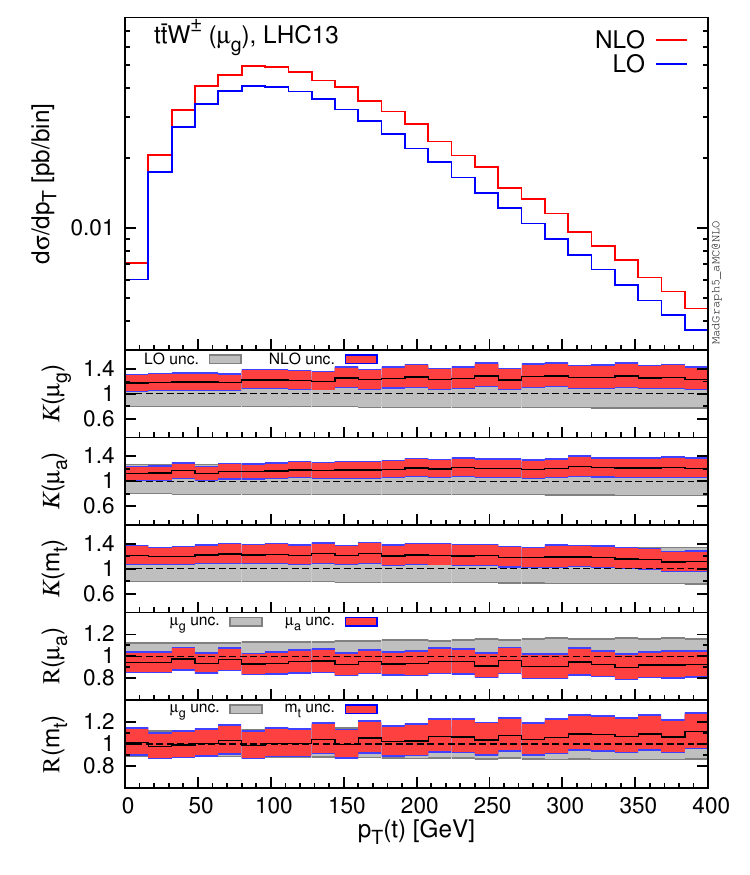}
\includegraphics[width=0.475\textwidth]{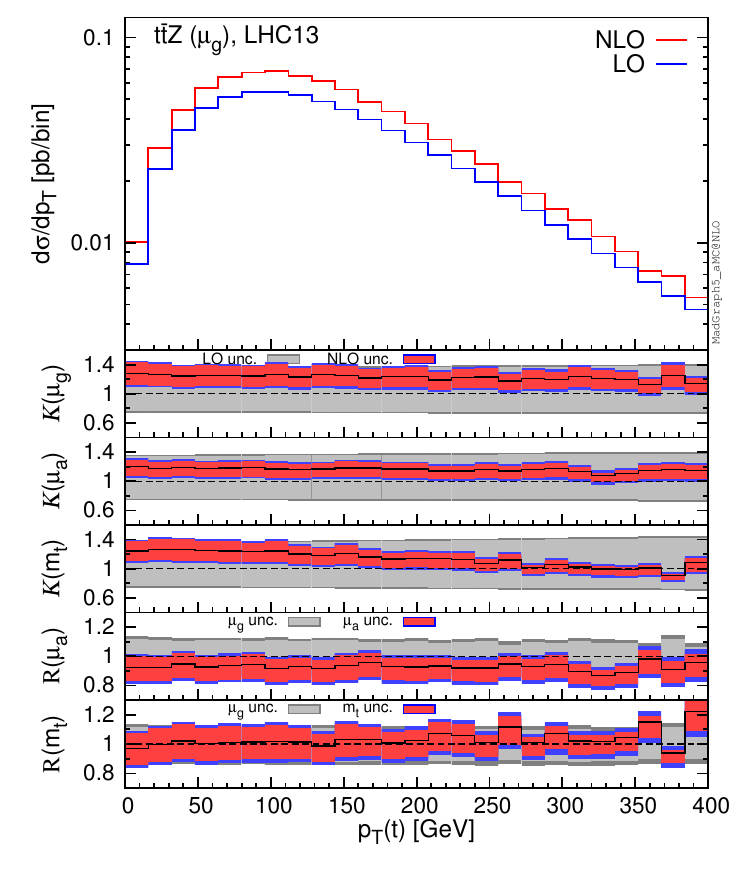}
\caption{Differential distributions for the $\pt$ of top-quark, $\pt(t)$. The format of the plots is described in detail in the text.}
\label{fig:ttV_ptt}
\end{figure}

For completeness, we provide in table \ref{tab:ttVj} the total cross sections at LO and NLO accuracy for $\ttW j$, as well as $\ttZ j$ and $\ttbar H j$ production, with a cut $p_T(j) > 100$ GeV. At variance with what has been done in fig.~\ref{fig:ttVj}, LO cross sections are calculated with LO PDFs and the corresponding $\alpha_s$, as done in the rest of the article.

The mechanism discussed in detail in previous paragraphs is also the source of the giant $K$-factors for large $\pt(\ttbar)$ in $\tta$ production, see fig.~\ref{fig:ttV_ptttx}. This process can originate from the gluon--gluon initial state at LO, however,  the emission of a photon involves soft and collinear singularities, which are not regulated by physical masses. When the photon is collinear to the final-state (anti)quark, the $qg\rightarrow \tta q$ process can be approximated as the $qg\rightarrow \ttbar q$ process times a $q\rightarrow q \gamma$ splitting. Here, soft and collinear divergencies are regulated by  both the cut on the $\pt$ of the photon ($\pt^{\rm cut}$) and the Frixione isolation parameter $R_0$. We checked that,  increasing the values of $\pt^{\rm cut}$ and/or  $R_0$, the size of the $K$-factors is reduced.  It is interesting to note also that corrections in the tail are much larger for $\mu=\mug$ than $\mu=\mua$. This is due to the fact that the softest photons, which give the largest contributions, sizeably reduce the value of $\mug$, whereas $\mua$ is by construction larger than $2\pt(\ttbar)$. This also suggests that $\mug$ might be an appropriate scale choice for this process only when the minimum $\pt$  cut and the isolation on the photon are harder.\footnote{Assuming $m_T(t)\sim m_T(\bar t)$ and $m_T(\gamma)=\pt^{\rm cut}$, the the ratio $\mua / \mug$ increases by increasing $p_T(t)$ and, when $m_T(t)>\pt^{\rm cut}$,  decreases by increasing $\pt^{\rm cut}$. Moreover, under the same assumption, $\mua = \mug$ at $m_T(t)=\pt^{\rm cut}$. For these reasons, especially for large $\pt(t \bar t)$, $\mu_g$ may underestimate the value of the scale.}

\begin{figure}[t]
\centering
\includegraphics[width=0.475\textwidth]{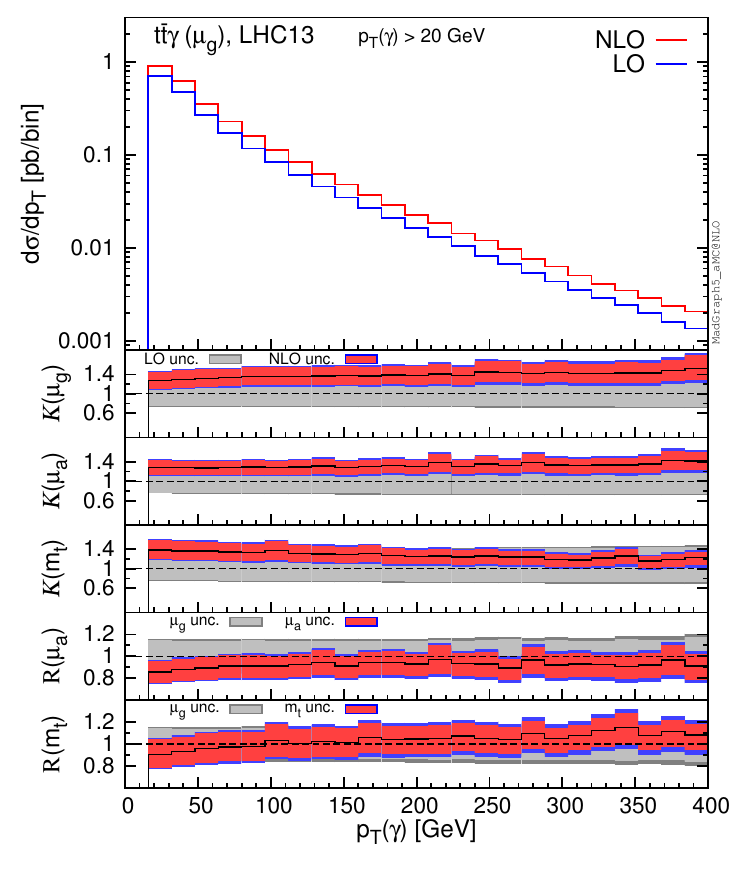}
\includegraphics[width=0.475\textwidth]{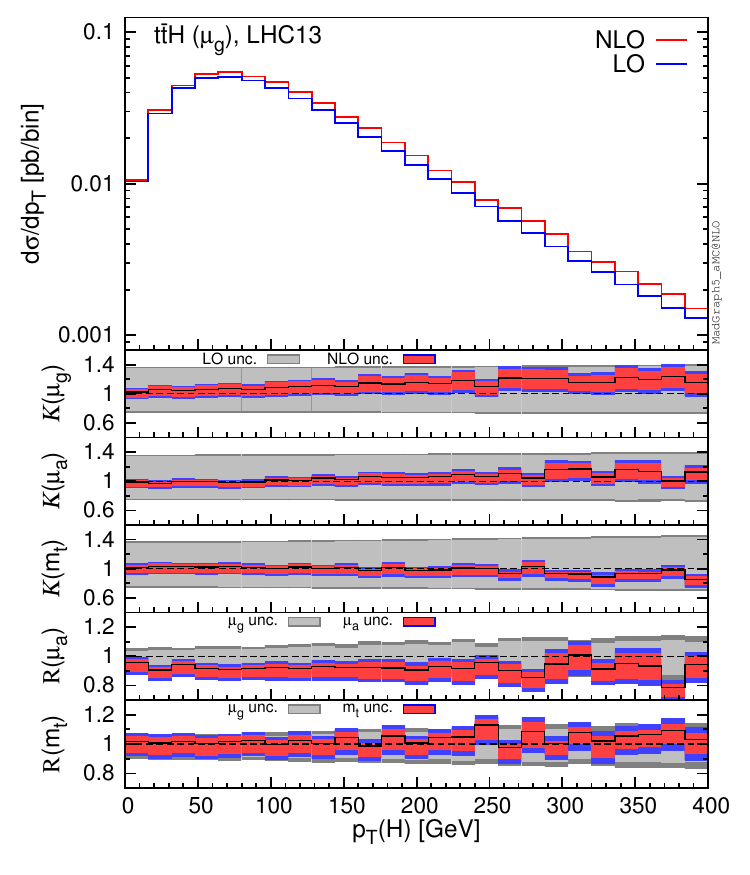}
\includegraphics[width=0.475\textwidth]{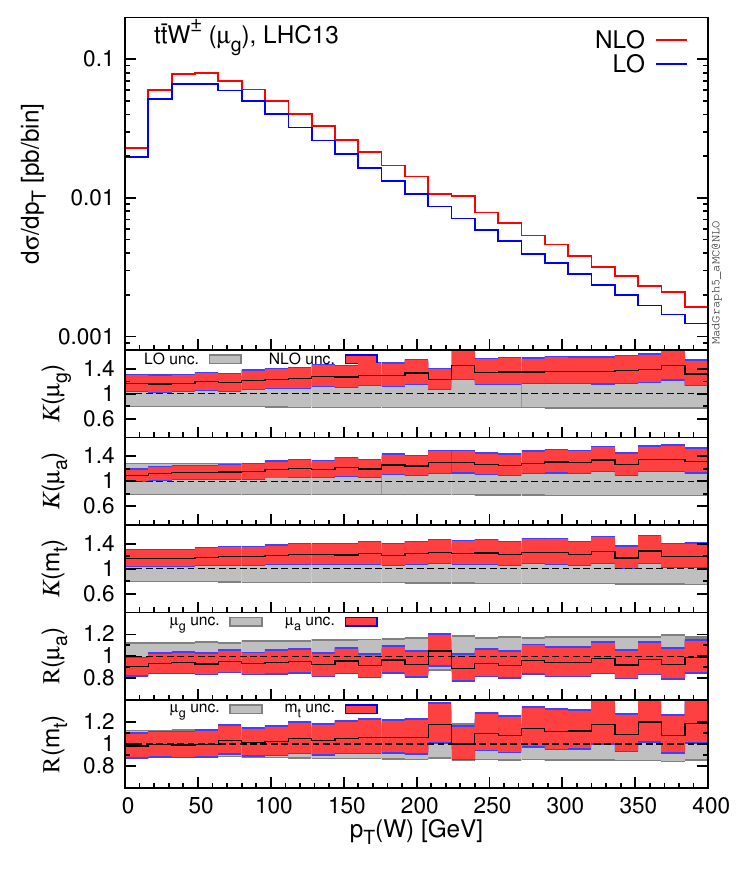}
\includegraphics[width=0.475\textwidth]{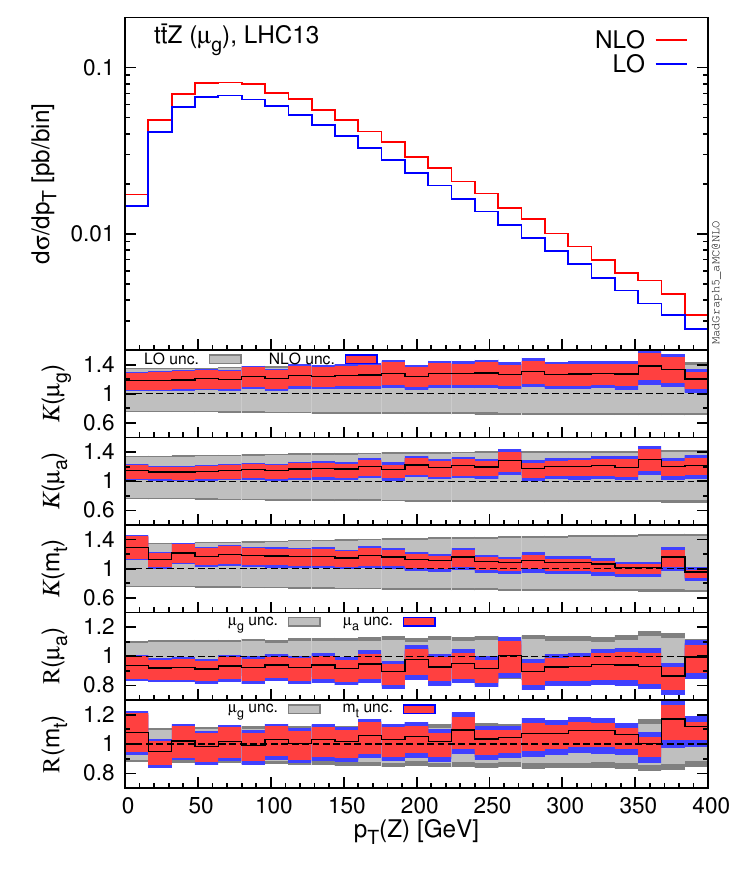}
\caption{Differential distributions for the $\pt$ of the vector or scalar boson, $\pt(V)$. The format of the plots is described in detail in the text.}
\label{fig:ttV_ptV}
\end{figure}

\begin{figure}[t]
\centering
\includegraphics[width=0.475\textwidth]{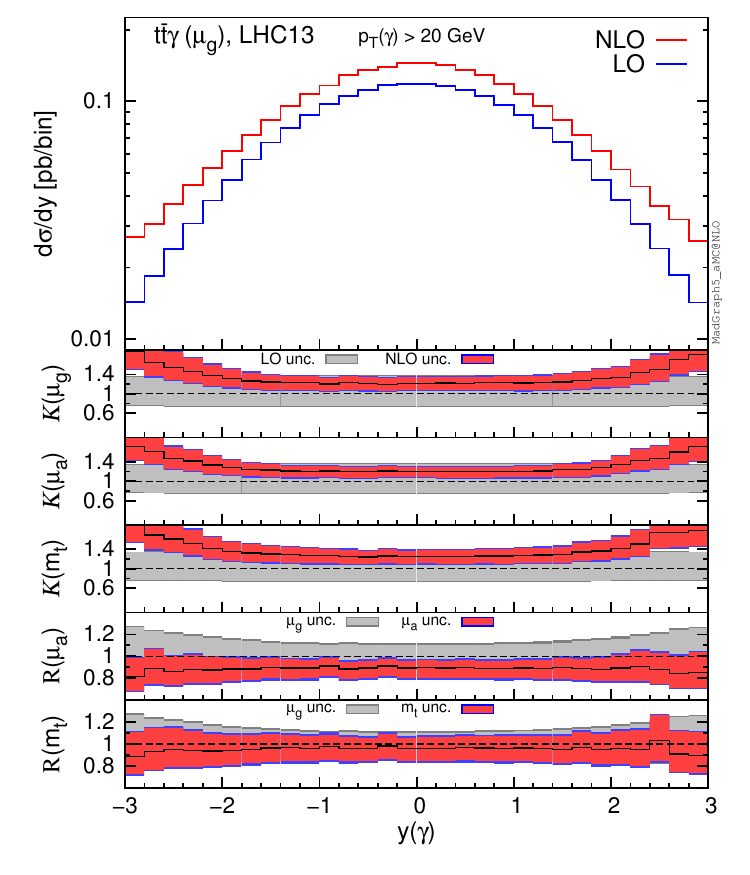}
\includegraphics[width=0.475\textwidth]{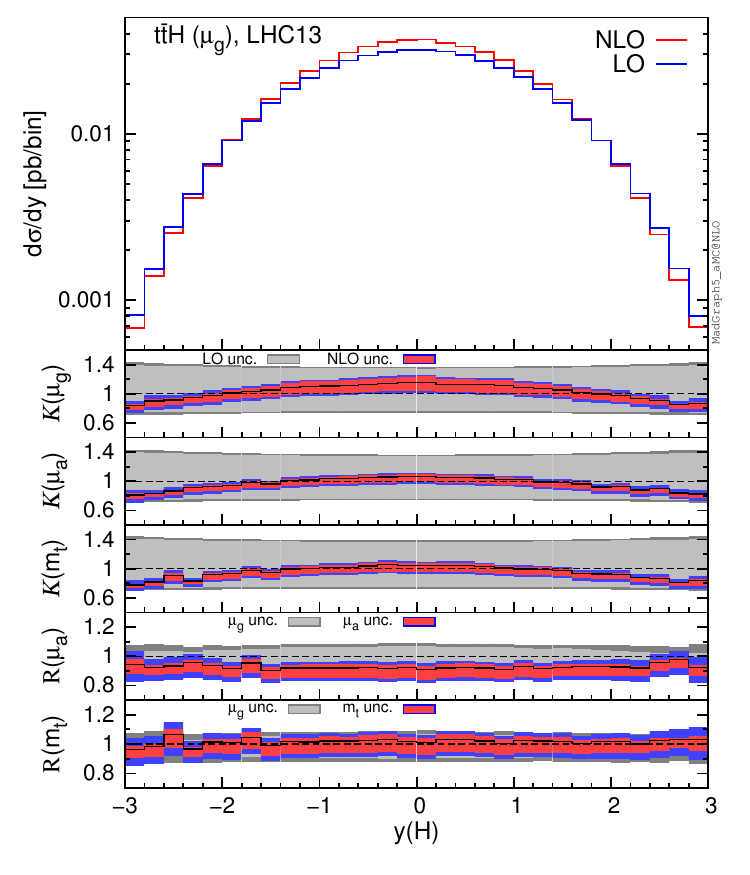}
\includegraphics[width=0.475\textwidth]{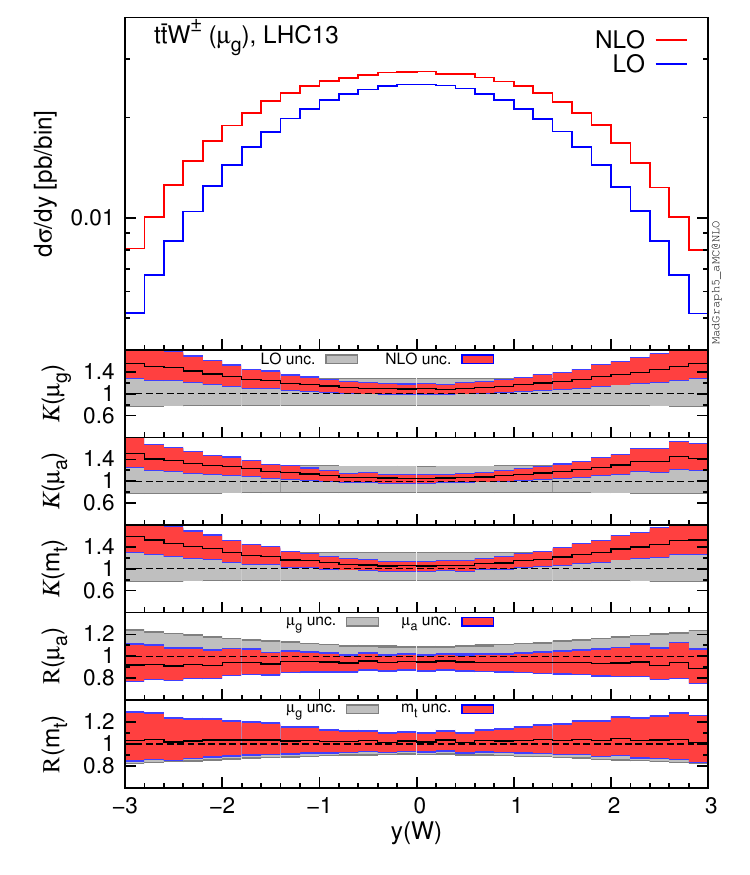}
\includegraphics[width=0.475\textwidth]{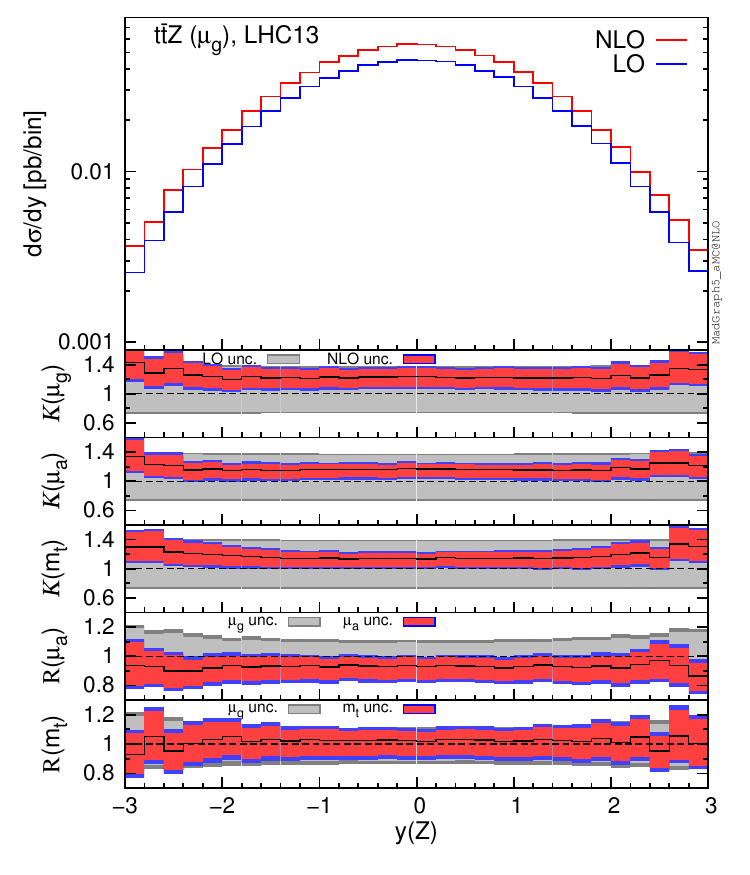}
\caption{Differential distributions for the rapidity of the vector or scalar boson, $y(V)$. The format of the plots is described in detail in the text.}
\label{fig:ttV_rapV}
\end{figure}

\begin{figure}[t]
\centering
\includegraphics[width=0.475\textwidth]{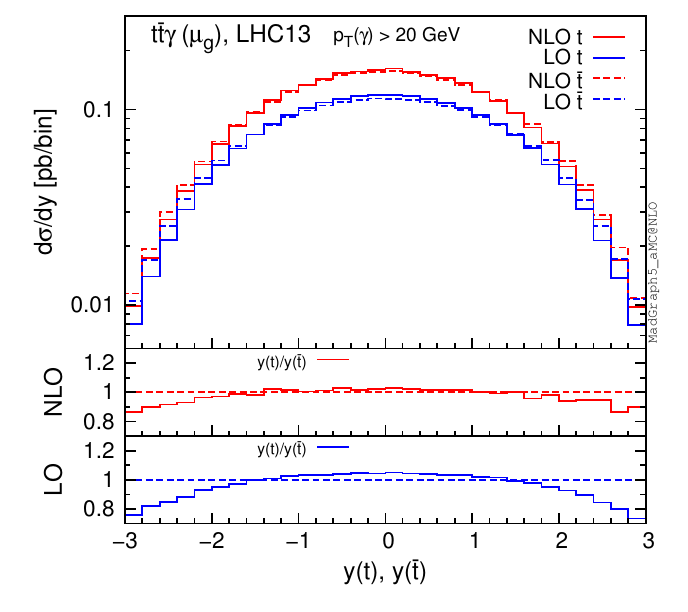}
\includegraphics[width=0.475\textwidth]{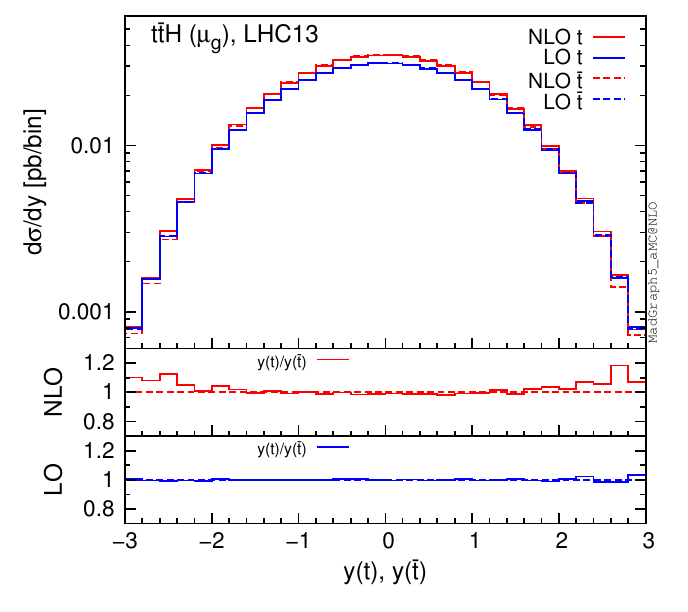}
\includegraphics[width=0.475\textwidth]{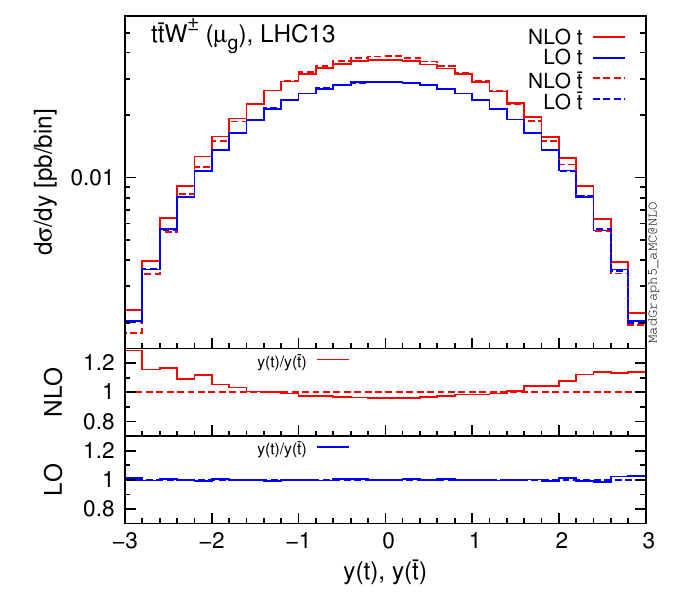}
\includegraphics[width=0.475\textwidth]{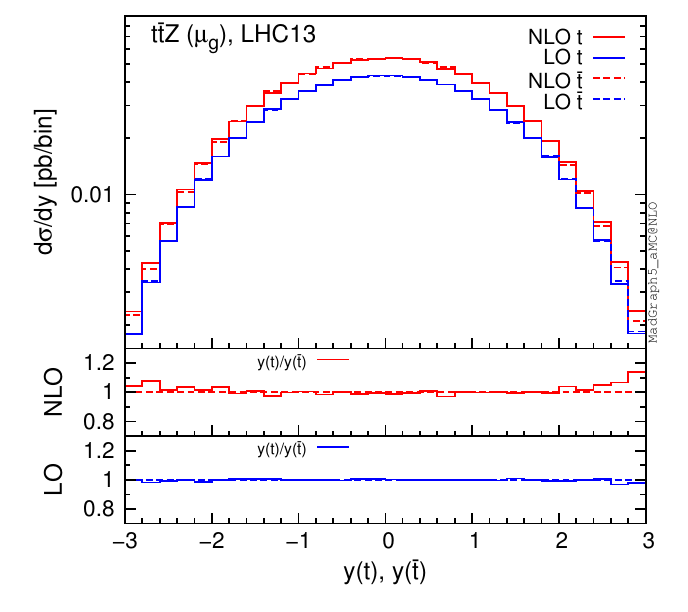}
\caption{Differential distributions for the rapidity of the top quark and antiquark, $y(t)$ and $y(\bar t)$.}
\label{fig:ttV_ptt_and_pttx}
\end{figure}

In figs.~\ref{fig:ttV_ptt} and \ref{fig:ttV_ptV} we show the $\pt$ distributions for the top quark and the vector or scalar boson, $\pt(t)$ and $\pt(V)$, respectively. For these two observables, 
we find the general features which have already been addressed for the  $m(\ttbar)$  distributions in fig.~\ref{fig:ttV_inv}.

In fig.~\ref{fig:ttV_rapV} we display the distributions for the rapidity of the vector or scalar boson, $y(V)$. In the four processes considered here, the vector or scalar boson is radiated in different ways at LO. In $\ttbar H$ production, the Higgs boson is never radiated from the initial state. In $\ttZ$ and $\tta$ production, in the quark--antiquark channel the vector boson can be emitted from the initial and final states, but in the gluon--gluon channel it can be radiated only from the final state. In $\ttW$ production, the $W$ is always emitted from the initial state.
The initial-state radiation of a vector boson is enhanced in the forward and backward direction, i.e., when it is collinear to the beam-pipe axis. Consequently, the vector boson is more peripherally distributed in $\ttW$ production, which involves only initial state radiation, than in $\tta$ and especially $\ttZ$ production. In $\ttbar H$ production, large values of $|y(V)|$ are not related to any enhancement and indeed the $y(V)$ distribution is much more central than in $\ttV$ processes. These features can be quantified by looking, e.g., at the ratio $r(V):=\frac{d\sigma}{dy}(|y|=0)/\frac{d\sigma}{dy}(|y|=3)$. At LO we find, $r(W)\sim 5$, $r(\gamma)\sim 8.5$, $r(Z)\sim 17.5$ and $r(H)\sim 40$.  As can be seen in the first three insets of the plots of  fig.~\ref{fig:ttV_rapV}, NLO QCD corrections decrease the values of $r(V)$ for $\ttW$ and $\tta$ production, i.e. the vector bosons are even more peripherally distributed ($r(W)\sim 3.5$, $r(\gamma)\sim 5.5$). A similar but milder effect is observed also in $\ttZ$ production ($r(Z)\sim 16$). On the contrary, NLO QCD corrections make the distribution of the rapidity of the Higgs boson even more central ($r(H)\sim 53$).
 In fig.~\ref{fig:ttV_rapV} one can also notice how the reduction of the scale dependence from LO to NLO results is much higher in $\ttbar H$ production than in $\ttV$ type processes. Furthermore, for this observable, $K$-factors are in general not flat also with the use of dynamical scales. From a phenomenological point of view, this is particularly important for $\ttW$ and $\tta$, since the cross section originating from  the peripheral region is not extremely suppressed, as can be seen from the aforementioned values of  $r(W)$ and $r(\gamma)$.

In fig.~\ref{fig:ttV_ptt_and_pttx} we show distributions for the rapidities of the top quark and antiquark, $y(t)$ and $y(\bar t)$. In this case we use a slightly different format for the plots. In the main panel, as in the  format of the previous plots, we show LO results in blue and NLO results in red. Solid lines correspond to $y(t)$, while dashed lines refer to $y(\bar t)$. In the first and second inset we plot the ratio of the $y(t)$ and $y(\bar t)$  distributions respectively at NLO  and LO accuracy. This ratio is helpful to easily identify which distribution is more central(peripheral) and if there is a central asymmetry for the top-quark pair. Also here, although it is not shown in the plots, $K$-factors are not in general flat. 

In the case of $\ttbar$ production the central asymmetry, or the forward-backward asymmetry in proton--antiproton collisions, originates from QCD and EW corrections. At NLO, the asymmetry arises from the interference of initial- and final-state radiation of neutral vector bosons (gluon in QCD corrections, and photons or $Z$ bosons in EW corrections) \cite{Kuhn:1998jr,Kuhn:1998kw,Bernreuther:2010ny,Hollik:2011ps,Kuhn:2011ri,Bernreuther:2012sx}. Thus, the real radiation contributions involve, at LO, the processes $pp\rightarrow \ttZ$ and  $pp\rightarrow \tta$, which are studied here both at LO and at NLO accuracy. As can be seen from  fig.~\ref{fig:ttV_ptt_and_pttx}, $\tta$ production yields an asymmetry already at LO, a feature studied in  \cite{Aguilar-Saavedra:2014vta}. The $\ttZ$ production central asymmetry is also expected to be non vanishing at LO, but the results plotted in fig.~\ref{fig:ttV_ptt_and_pttx} tell us that the actual value is very small.  
The asymmetry is instead analytically zero in $\ttW$ ($\ttbar H$) production, where the interference of initial- and final- state $W$(Higgs) bosons is not possible.\footnote{In principle, when the couplings of light-flavour quarks are considered non-vanishing, initial-state radiation of a Higgs boson is possible and also a very small asymmetry is generated. However, this possibility is ignored here.}

At NLO, all the $\ttV$ processes and the $\ttbar H$ production have an asymmetry, as can be seen in  fig.~\ref{fig:ttV_ptt_and_pttx} from the ratios of the $y(t)$ and $y(\bar t)$  distributions at NLO. In the case of $\ttW$ production the asymmetry, which is generated by NLO QCD corrections, has already been studied in detail in \cite{Maltoni:2014zpa}. In all the other cases it is analysed for the first time here.

NLO and LO results at 13 TeV for $ A_c$  defined as \begin{equation}
A_c=\frac{\sigma(|y_t|>|y_{\bar{t}}|)-\sigma(|y_t|<|y_{\bar{t}}|)}{\sigma(|y_t|>|y_{\bar{t}}|)+\sigma(|y_t|<|y_{\bar{t}}|)}
\end{equation}
are listed in table \ref{table:13tevttv_asymm}, which clearly demonstrates that NLO QCD effects  cannot be neglected, once again, in the predictions of the asymmetries. For $\ttW$ and $\ttbar H$ production, an asymmetry is actually generated only at NLO. Furthermore, NLO QCD corrections change sign and increase by a factor $\sim 7$ the asymmetry in $\ttZ$ production and they decrease it by a factor larger than two in $\tta$ production. Thus, NLO results point to the necessity of reassessing the phenomenological impact of the $\tta$ signature, which is based on a LO calculation~\cite{Aguilar-Saavedra:2014vta}. Moreover, we have also checked that for $p_T(\gamma)> 50 \gev$ both the LO and NLO  central values of the asymmetry are very similar (within 5 per cent) to the results in table \ref{table:13tevttv_asymm}, where $p_T(\gamma)> 20 \gev$.

\begin{table}[t]
\small
\renewcommand{\arraystretch}{1.5}
\begin{center}
\begin{tabular}{  c | c c c c c }
\hline\hline
13 TeV $ A_c$ [\%]  & $\ttW$ & $t \bar t \gamma$ \\
\hline
LO & - & $ -3.93^{ +0.26 }_{ -0.23 }~^{ +0.14 }_{ -0.11 } \pm 0.03$ \\
\hline
NLO   & $ 2.90^{ +0.67 }_{ -0.47 }~^{ +0.06 }_{ -0.07 } \pm 0.07$  & $ -1.79^{ +0.50 }_{ -0.39 }~^{ +0.06 }_{ -0.09 } \pm 0.06$ \\
\hline\hline
13 TeV $ A_c$ [\%] & $t \bar t H$ &  $t \bar t Z$  \\
\hline
LO & - & $ -0.12^{ +0.01 }_{ -0.01 }~^{ +0.01 }_{ -0.02 } \pm 0.03$  \\
\hline
NLO & $ 1.00^{ +0.30 }_{ -0.20 }~^{ +0.06 }_{ -0.04 } \pm 0.02$ & $ \phantom{-}0.85^{ +0.25 }_{ -0.17 }~^{ +0.06 }_{ -0.05 } \pm 0.03$  \\
\hline
\end{tabular}
\caption{NLO  and LO central asymmetries for $\ttV$-type processes and $\ttbar H$ production at 13 TeV for $\mu=\mu_g$. The first uncertainty is given by the scale variation within $\mug/2<\mu_f,\mu_r<2\mug$, the second one by PDFs. The assigned error is the absolute statistical integration error.}
\label{table:13tevttv_asymm}
\end{center}
\end{table}


\subsection{$\ttVV$ processes}
\label{sec:ttvv}

\begin{figure}[t]
\centering
\includegraphics[width=0.8\textwidth]{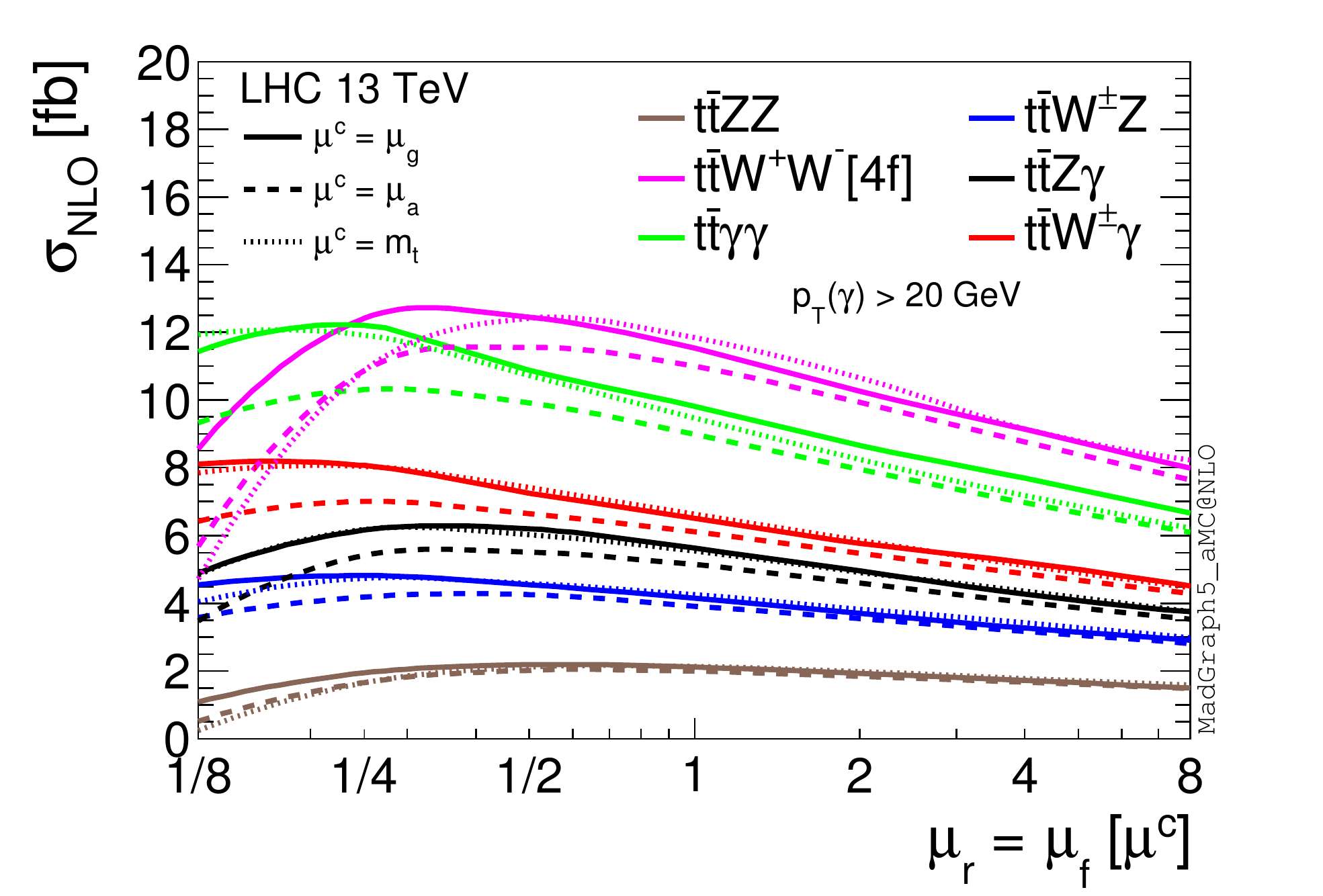}
\caption{Cross sections of $\ttVV$ processes at 13 TeV. Comparison of NLO scale dependence in the interval $\mu^c/8<\mu<8\mu^c$ for the three different choices of the central value $\mu^c$: $\mu_g$, $\mu_a$, $m_t$.}
\label{fig:scales_ttVV}
\end{figure}

We start showing for all the $\ttVV$ processes the dependence of the NLO total cross sections,  at 13 TeV, on the variation of the renormalisation and factorisation scales $\mu_r$ and $\mu_f$.
This dependence is shown in fig.~\ref{fig:scales_ttVV} and it is obtained by varying $\mu=\mu_r=\mu_f$ by a factor eight around the central value $\mu=\mug$ (solid lines), $\mu=\mua$ (dashed lines) and $\mu=m_t$ (dotted lines). 
Again, for all the processes and especially for those with a photon in the final state, we find that   $\mua$ typically leads to smaller cross sections than $\mug$ and $m_t$. For this class of processes we also investigated the effect of the independent variation of factorisation and renormalisation scales. We found that the condition $\mu_r=\mu_f$ captures the full dependence in the $(\mu_r,\mu_f)$ plane in the range $\mua/2<\mu_f,\mu_r<2\mua$. On the other hand, in the full $\mua/8<\mu_f,\mu_r<8\mua$ region off-diagonal values might differ from the values spanned at $\mu_f=\mu_r$.

\begin{figure}[t]
\centering
\includegraphics[width=0.8\textwidth]{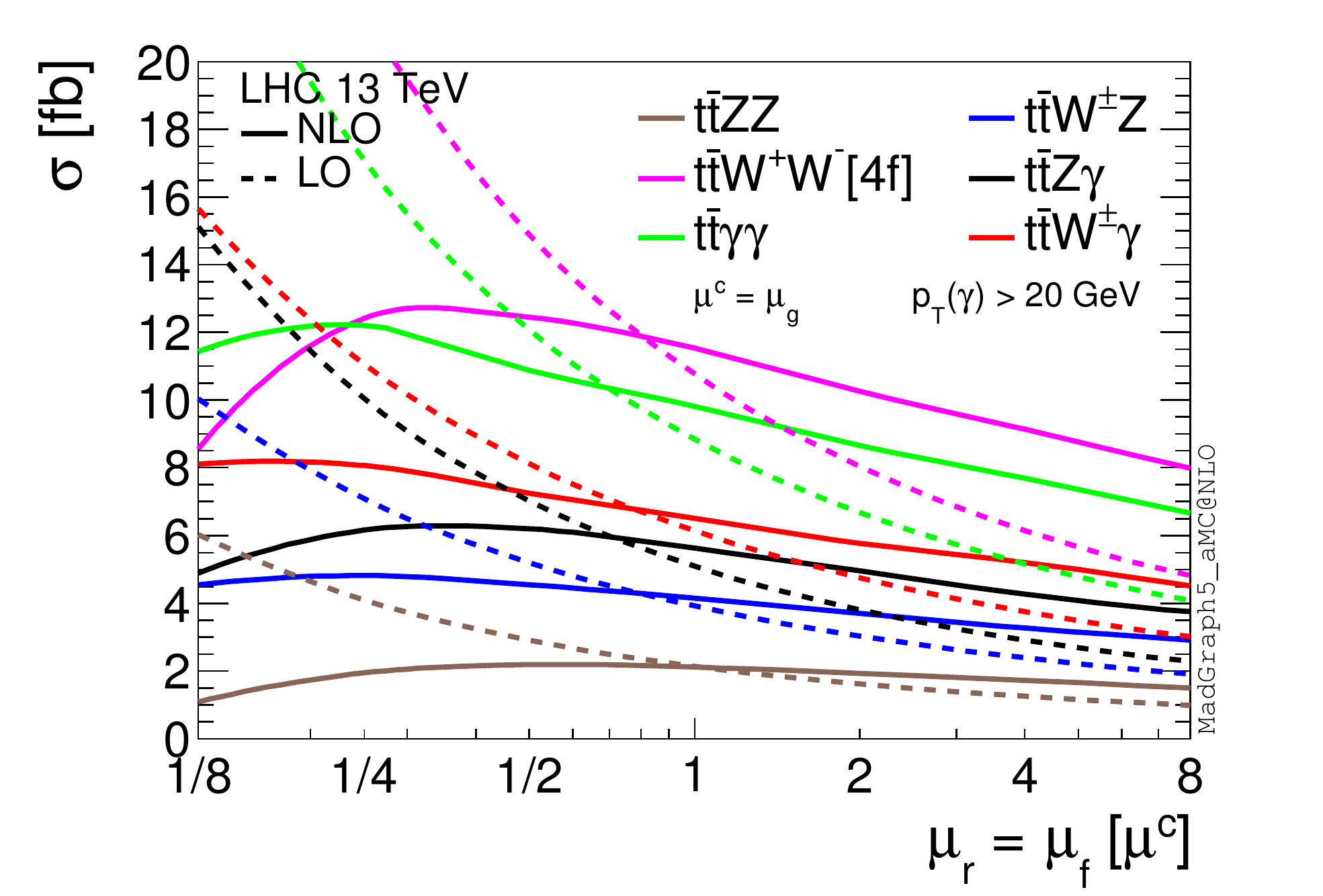}
\caption{NLO and LO cross sections at 13 TeV. Scale dependence in the interval $\mu^c/8<\mu<8\mu^c$ with $\mu^c=\mu_g$ for the $\ttVV$ processes.}
\label{fig:diagttVV}
\end{figure}

In table \ref{table:13tev} we list, for all the processes, LO and NLO cross sections together with PDF  and scale uncertainties, and $K$-factors for the central values. Again scale uncertainties are evaluated by varying independently the factorisation and the renormalisation scales in the interval $\mug/2<\mu_f, \mu_r<2\mug$.
The dependence of the LO and NLO cross sections on $\mu=\mu_r=\mu_f$ is  shown in  fig.~\ref{fig:diagttVV}  in the range $\mug/8<\mu<8\mug$. As expected,  for all the processes, the scale dependence is strongly reduced from LO to NLO predictions both in the standard interval $\mug/2<\mu<2\mug$ as well as in the full range $\mug/8<\mu<8\mug$.
 For the central scale $\mu=\mug$, $K$-factors are very close to unity.
It is interesting to note that NLO curves display  a plateau around $\mug/2$ or $\mug/4$,  corresponding to  $H_T/8$ and $H_T/16$, respectively.

\begin{table}[t]
\small
\renewcommand{\arraystretch}{1.5}
\begin{center}
\begin{tabular}{  c | c c c c c c }
\hline\hline
13 TeV $ \sigma$[fb] & $t \bar t ZZ$ &  $t \bar t W^+ W^-$[4f] & $t \bar t \gamma \gamma$ \\
\hline
NLO & $2.117^{+3.8 \%}_{-8.6 \%}~^{+1.9 \%}_{-1.8 \%}$ & $11.84^{+8.3 \%}_{-11.2 \%}~^{+2.3 \%}_{-2.4 \%}$ & $10.26^{+13.9 \%}_{-13.3 \%}~^{+1.3 \%}_{-1.3 \%}$ \\
\hline
LO & $2.137^{+36.1 \%}_{-24.4 \%}~^{+1.9 \%}_{-1.9 \%}$ & $10.78^{+38.3 \%}_{-25.4 \%}~^{+2.2 \%}_{-2.2 \%}$ & $8.838^{+36.5 \%}_{-24.5 \%}~^{+1.5 \%}_{-1.6 \%}$ \\
\hline
$K$-factor & 0.99 & 1.10 & 1.16 \\
\hline\hline
13 TeV $ \sigma$[fb] & $t \bar t W^{\pm} Z$ & $t \bar t Z \gamma$  & $t \bar t W^{\pm} \gamma$ \\[5pt]
\hline
NLO & $4.157^{+9.8 \%}_{-10.7 \%}~^{+2.2 \%}_{-1.6 \%}$ & $5.771^{+10.5 \%}_{-12.1 \%}~^{+1.8 \%}_{-1.9 \%}$ & $6.734^{+12.0 \%}_{-11.6 \%}~^{+1.8 \%}_{-1.4 \%}$  \\[5pt]
\hline
LO & $3.921^{+32.6 \%}_{-22.8 \%}~^{+2.3 \%}_{-2.2 \%}$ & $5.080^{+38.0 \%}_{-25.3 \%}~^{+1.9 \%}_{-1.9 \%}$ & $6.145^{+32.4 \%}_{-22.6 \%}~^{+2.1 \%}_{-2.0 \%}$  \\[5pt]
\hline
$K$-factor & 1.06 & 1.14 & 1.10 \\[5pt]
\hline
\end{tabular}
\caption{NLO  and LO cross sections for $\ttVV$ processes at 13 TeV for $\mu=\mu_g$. The first uncertainty is given by the scale variation within $\mug/2<\mu_f,\mu_r<2\mug$, the second one by PDFs. The relative statistical integration error is equal or smaller than one permille.}
\label{table:13tev}
\end{center}
\end{table}

\medskip

We show now the impact of NLO QCD corrections for relevant distributions and we discuss their dependence on  scale choice and its variation. 
For all the processes we have considered the distribution of the invariant mass of the top-quark pair and the  $\pt$ and the rapidity of the (anti)top quark, of the top-quark pair and of the vector bosons. Again, given the large amount of distributions that is possible to consider for such a final state, we show only representative results. We remind the interested reader that additional distributions can be easily produced via the public code \aNLO. 

\begin{figure}[h]
\centering
\includegraphics[width=0.46\textwidth]{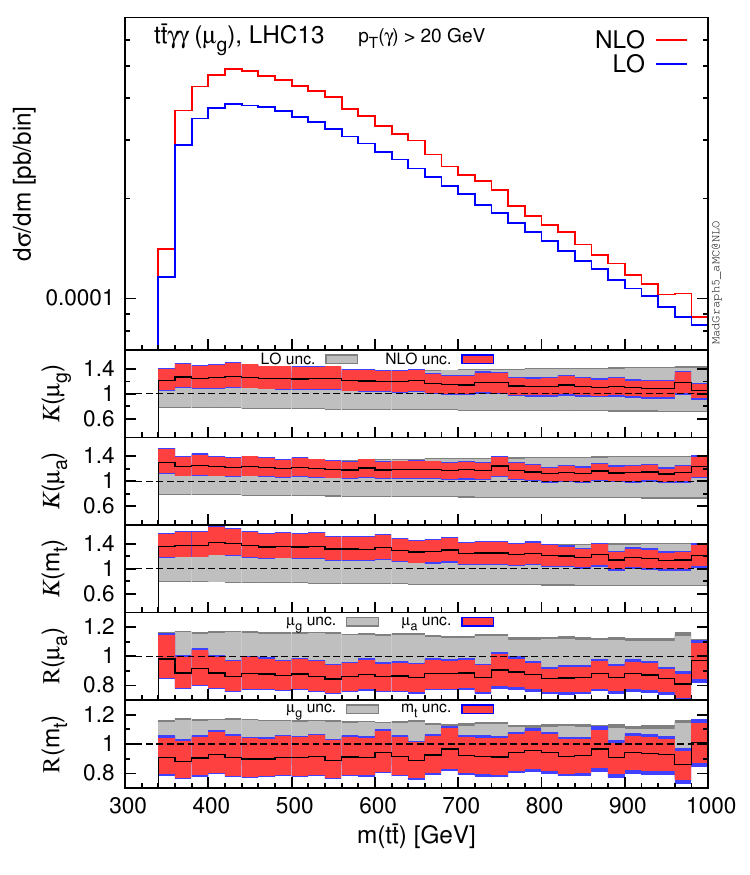}
\includegraphics[width=0.46\textwidth]{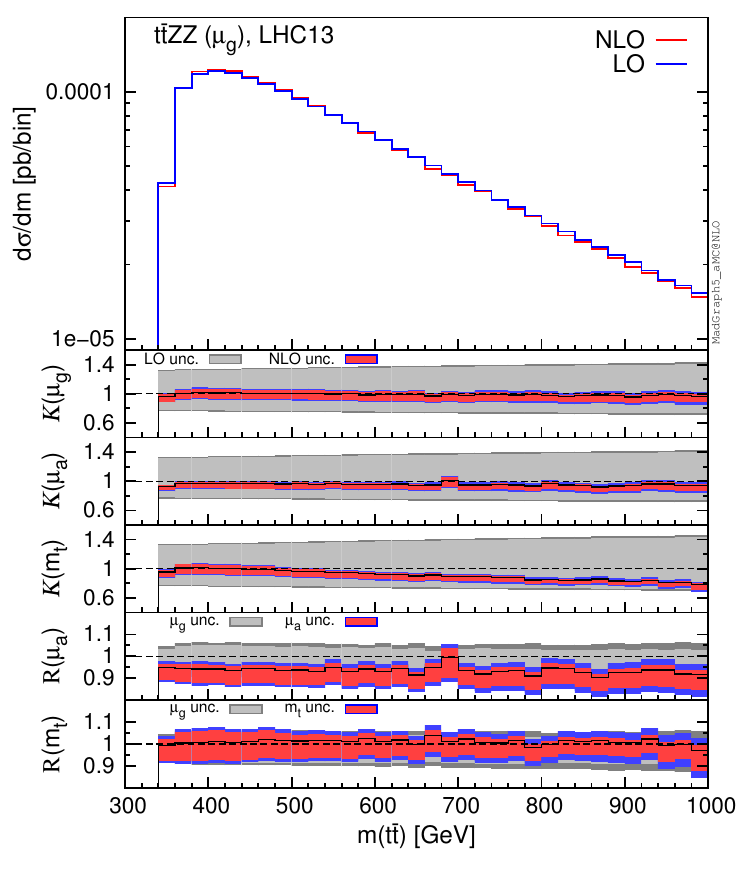}
\includegraphics[width=0.46\textwidth]{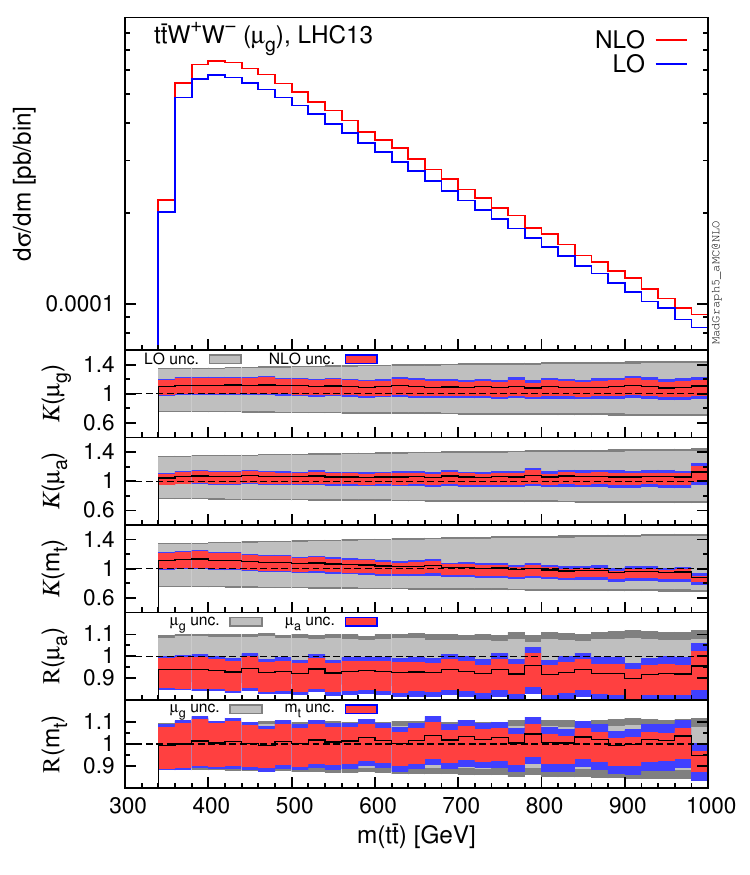}
\includegraphics[width=0.46\textwidth]{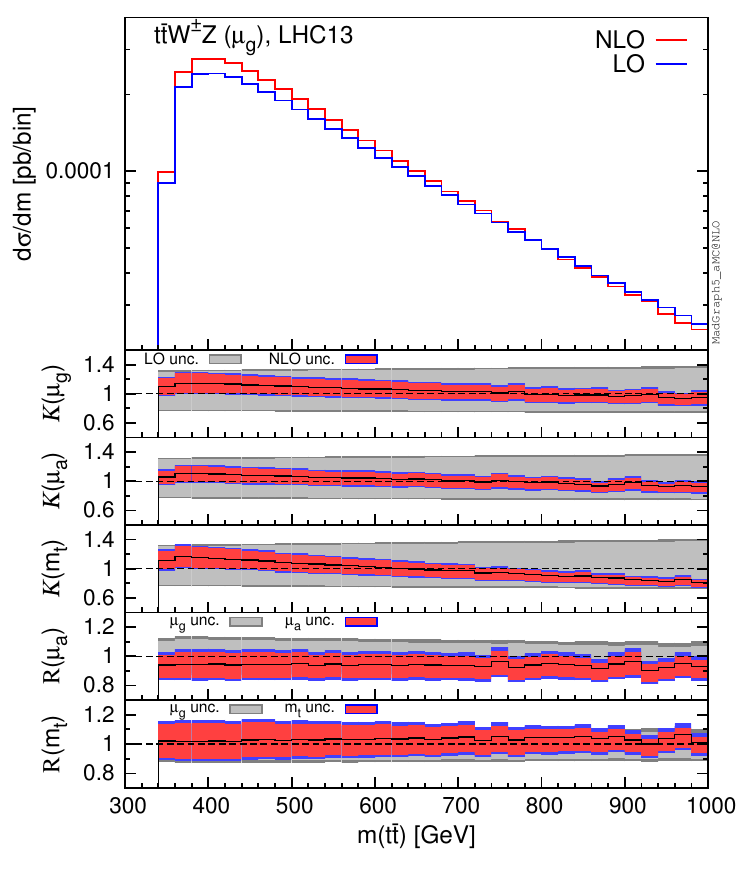}
\includegraphics[width=0.46\textwidth]{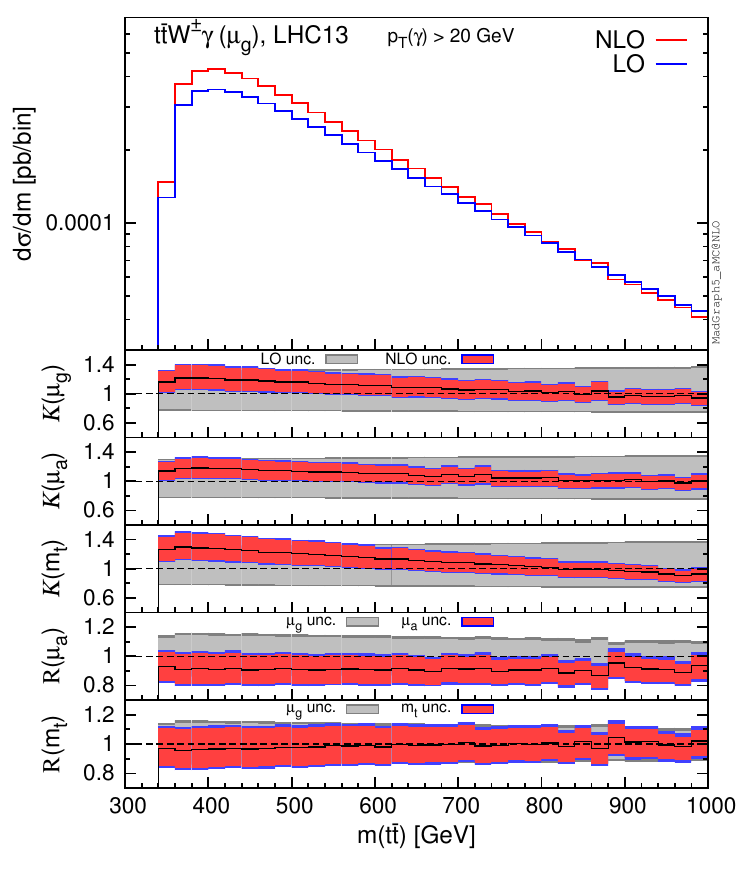}
\includegraphics[width=0.46\textwidth]{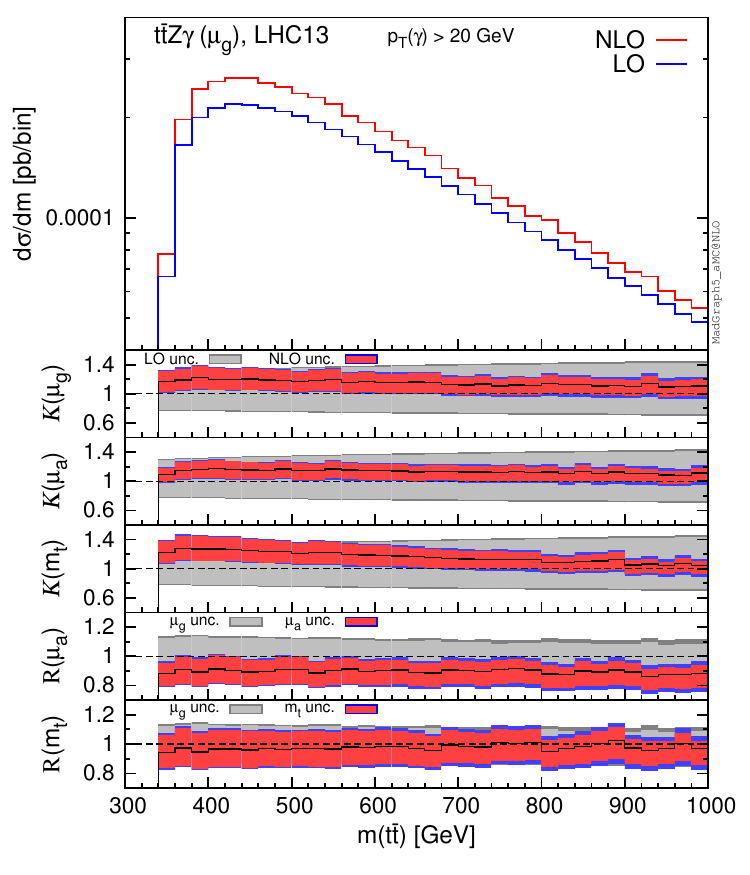}

\caption{Differential distributions for the invariant mass of top-quark pair, $m(\ttbar)$. The format of the plots is described in detail in subsection \ref{sec:ttvh}.}
\label{fig:ttVV_inv}
\end{figure}

For each figure, we display together the same type of distributions for the six different processes: $\ttaa$, $\ttZZ$, $\ttWW$, $\ttWZ$, $\ttWa$ and $\ttZa$.
We start with fig.~\ref{fig:ttVV_inv}, which shows the $m(\ttbar)$ distributions. The format of the plot is the same used for most of the distribution plots in subsection \ref{sec:ttvh}, where it is also described in detail.
 For $m(\ttbar)$ distributions, we notice features that are in general common to all the distributions and have already been addressed for $\ttV$ processes in subsection \ref{sec:ttvh}.  For instance, the use of $\mu=\mua$ leads to NLO values compatible with, but systematically smaller than, those obtained with $\mu=\mu_g$. Conversely, the choice $\mu=m_t$ leads  to scale uncertainties bands that overlaps with those obtained with $\mu=\mu_g$. 
The NLO corrections in $\ttZZ$ production are very close to zero, for $\mu=\mug$, and very stable under scale variation (see also table \ref{table:13tev}). For all the processes, the two dynamical scales $\mug$ and $\mua$ yield flatter $K$-factors than those from the fixed scale $m_t$. 

\begin{figure}[h]
\centering
\includegraphics[width=0.46\textwidth]{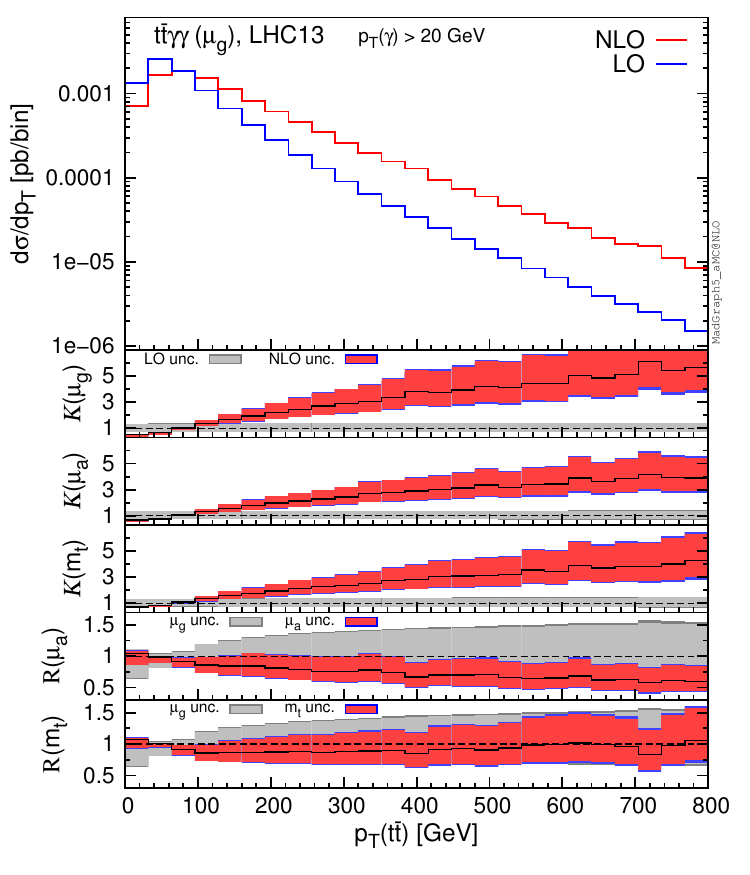}
\includegraphics[width=0.46\textwidth]{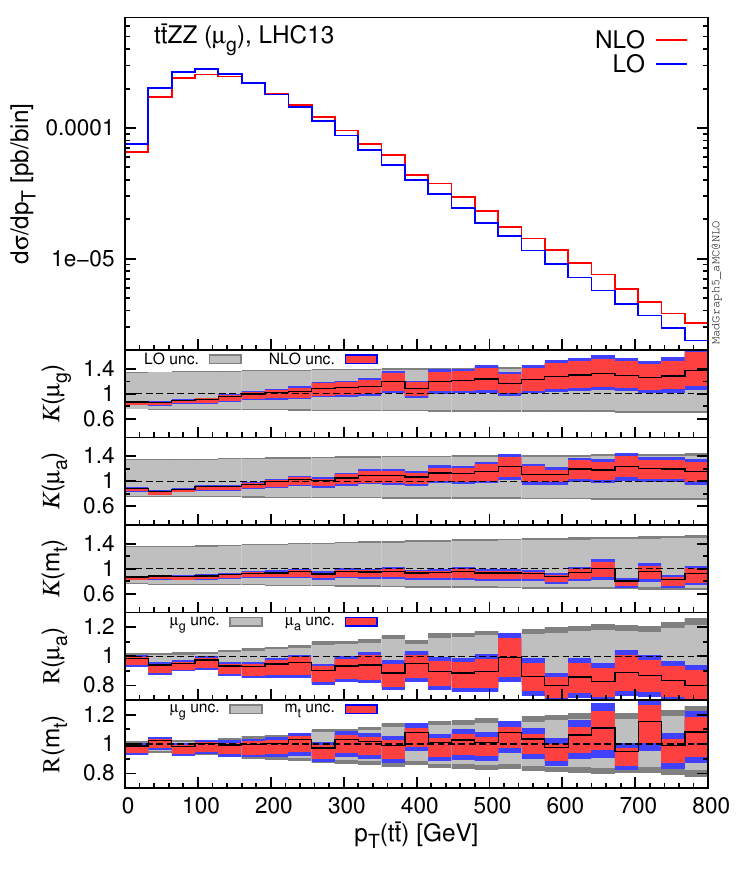}
\includegraphics[width=0.46\textwidth]{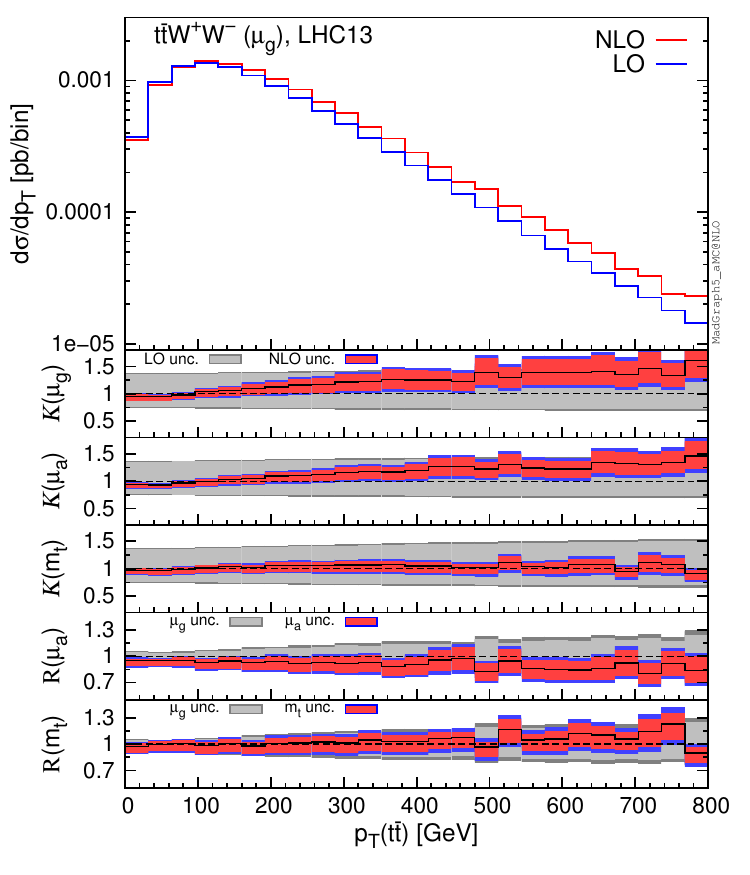}
\includegraphics[width=0.46\textwidth]{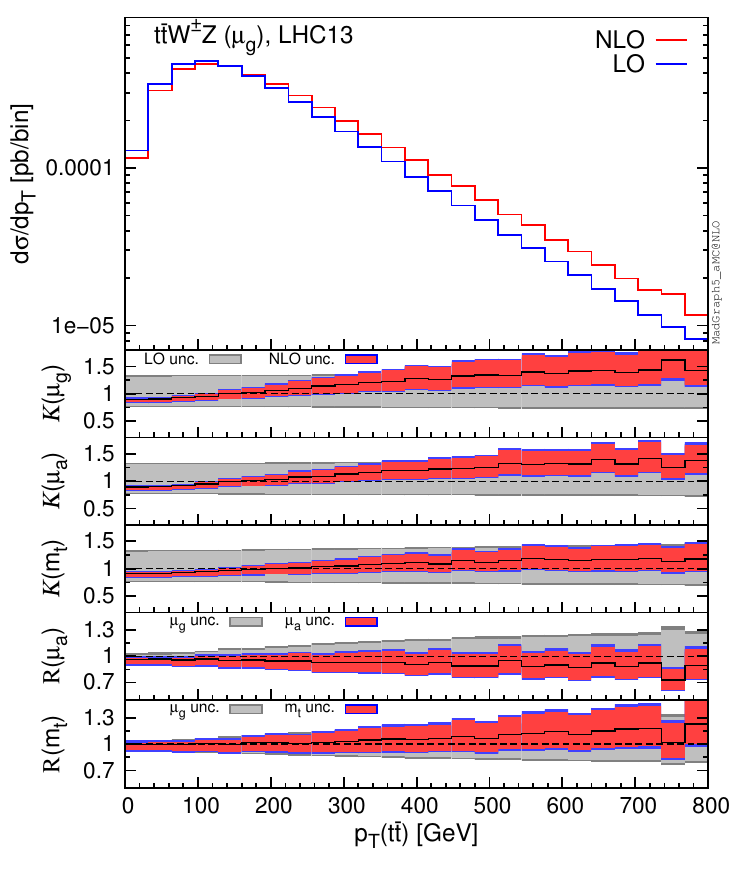}
\includegraphics[width=0.46\textwidth]{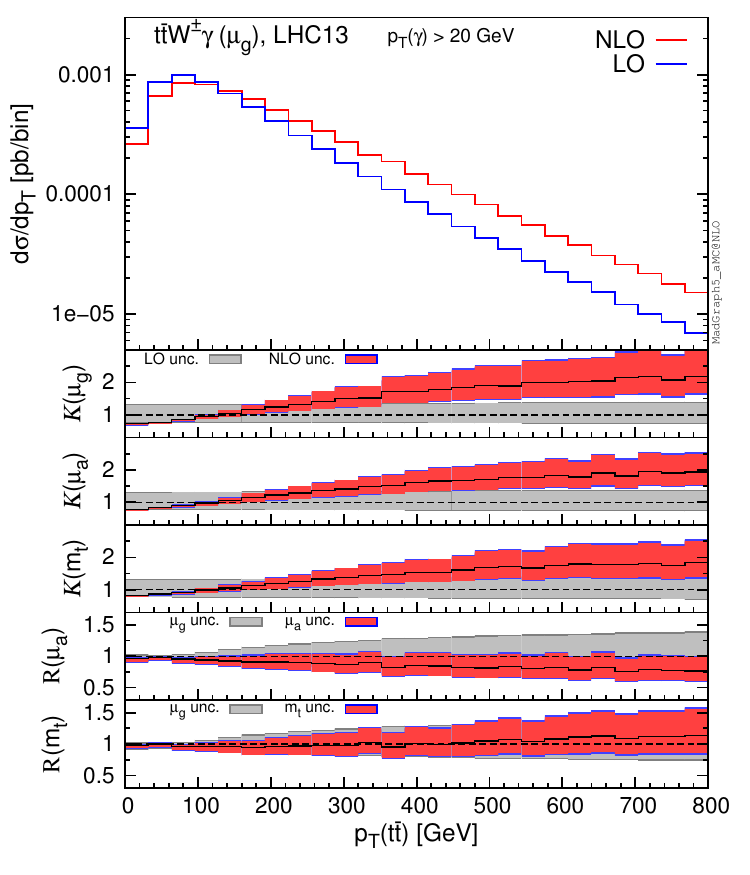}
\includegraphics[width=0.46\textwidth]{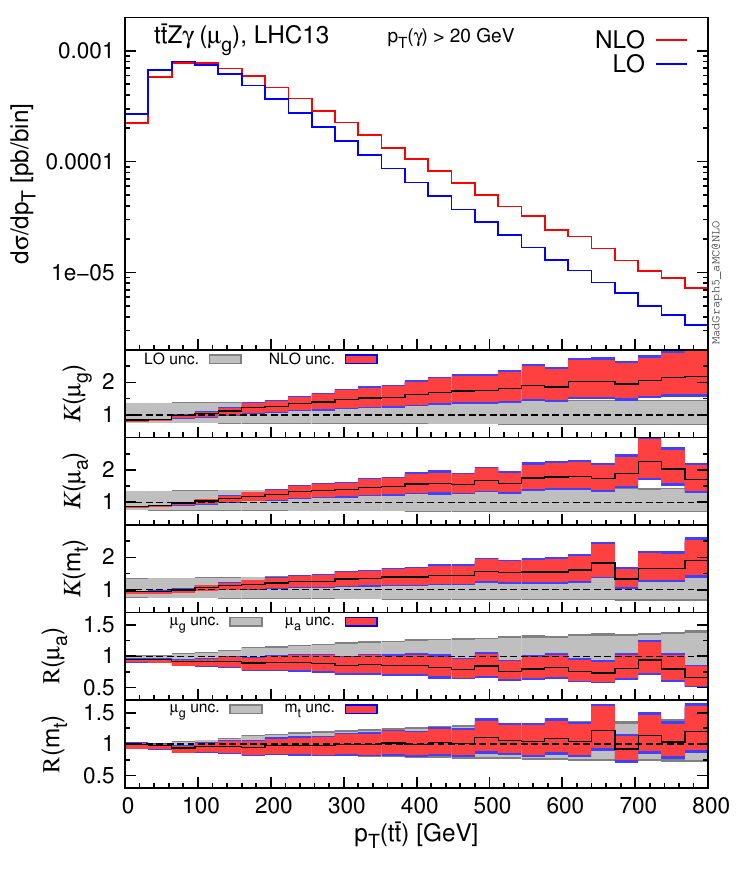}

\caption{Differential distributions for the $\pt$ of top-quark pair, $\pt(\ttbar)$. The format of the plots is described in detail in subsection \ref{sec:ttvh}.}
\label{fig:ttVV_ptttx}
\end{figure}

In fig.~\ref{fig:ttVV_ptttx} we show the distributions for $\pt(\ttbar)$. As for $\ttV$ processes (see fig.~\ref{fig:ttV_ptttx}), these distributions receive large corrections in the tails. This effect is especially strong for the processes involving a photon in the final state, namely, $\ttaa$, $\ttZa$ and $\ttWa$. Also, for all the three choices of $\mu$ employed here, $K$-factors are not flat. Surprisingly, the $K$-factors for $\ttZZ$, $\ttWZ$ and $\ttWW$ production show a larger dependence on the value of $\pt(\ttbar)$ when $\mu$ is a dynamical quantity, as can be seen from a comparison of the first ($\mu=\mug$) and second ($\mu=\mua$) insets with the third insets ($\mu=m_t$). From the fourth insets of all the six plots, it is possible to notice how the scale dependence at NLO for $\mu=\mug$ it is much larger than for $\mu=\mua$. Exactly as we argued for  $\ttV$ processes, NLO $\ttVV$+jets merged sample à la FxFx  should be used for an accurate prediction of these tails. 

\begin{figure}[h]
\centering
\includegraphics[width=0.46\textwidth]{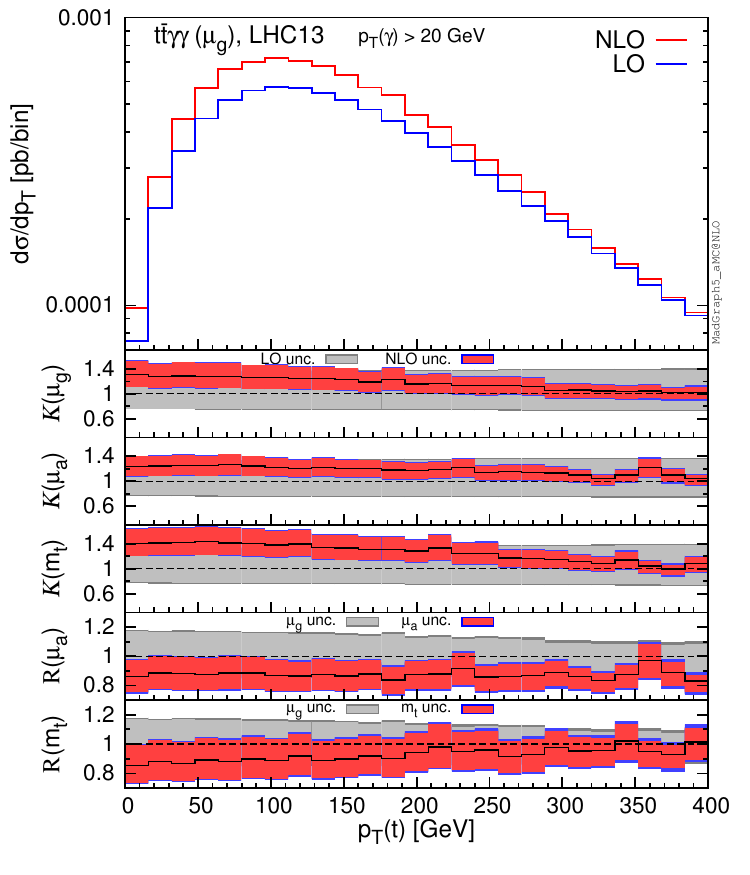}
\includegraphics[width=0.46\textwidth]{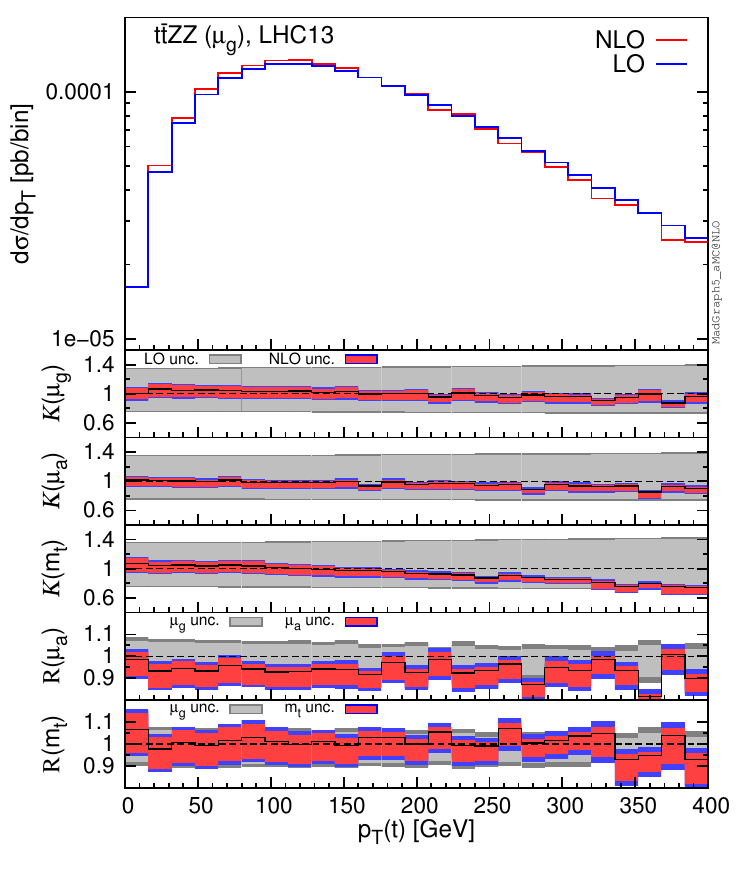}
\includegraphics[width=0.46\textwidth]{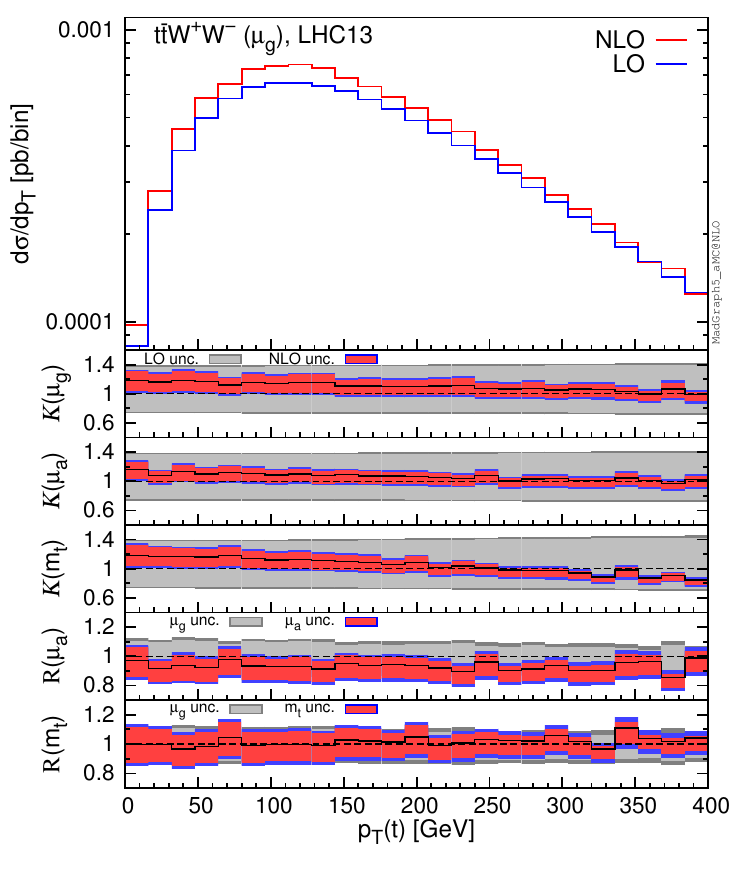}
\includegraphics[width=0.46\textwidth]{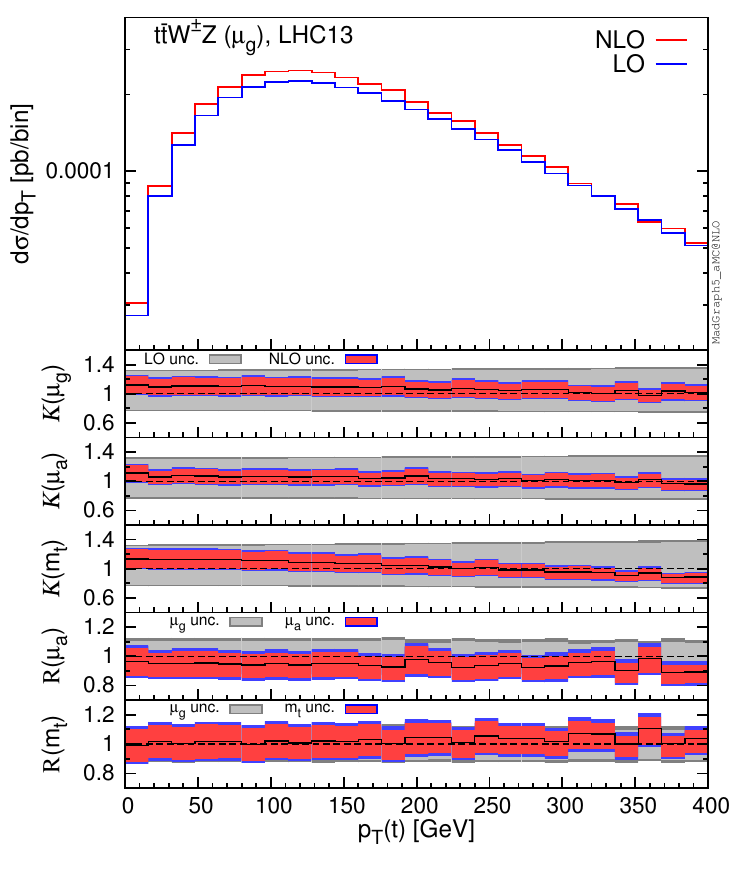}
\includegraphics[width=0.46\textwidth]{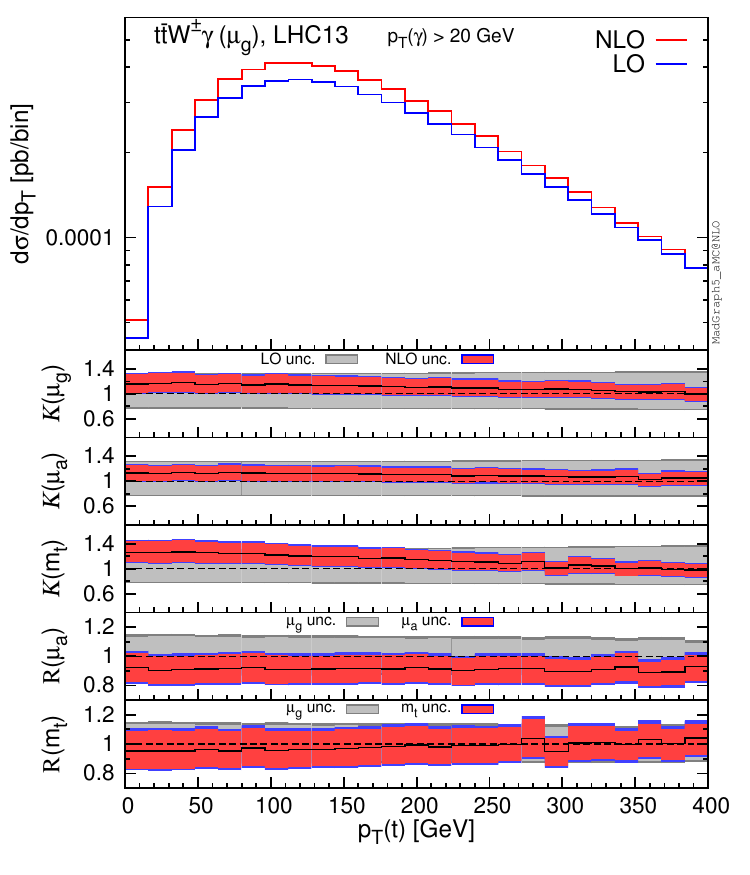}
\includegraphics[width=0.46\textwidth]{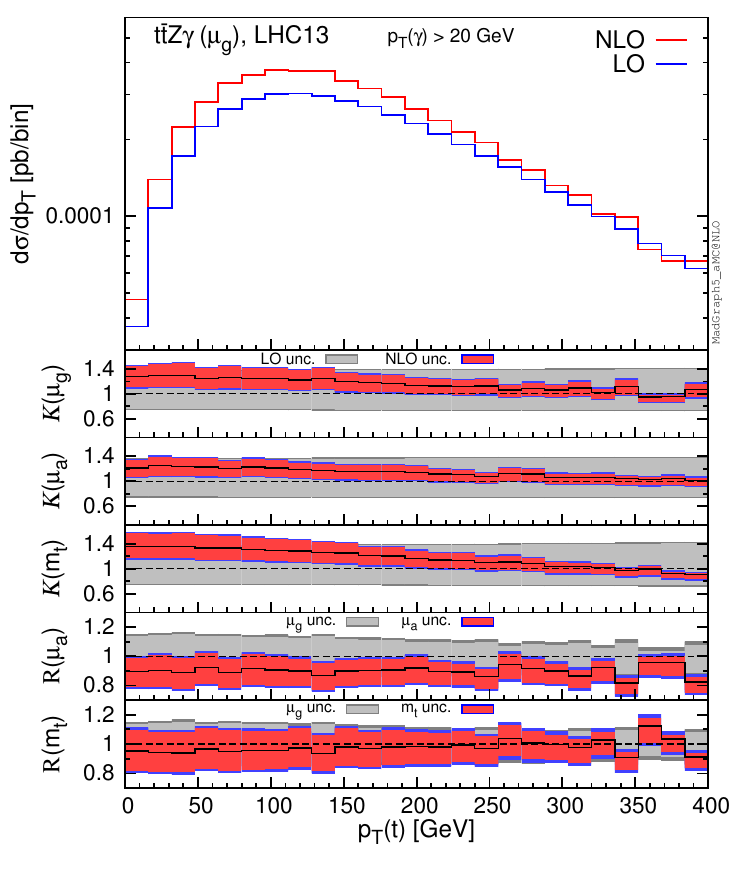}

\caption{Differential distributions for the $\pt$ of top-quark, $\pt(t)$. The format of the plots is described in detail in subsection \ref{sec:ttvh}.}
\label{fig:ttVV_ptt}
\end{figure}

\begin{figure}[h]
\centering
\includegraphics[width=0.46\textwidth]{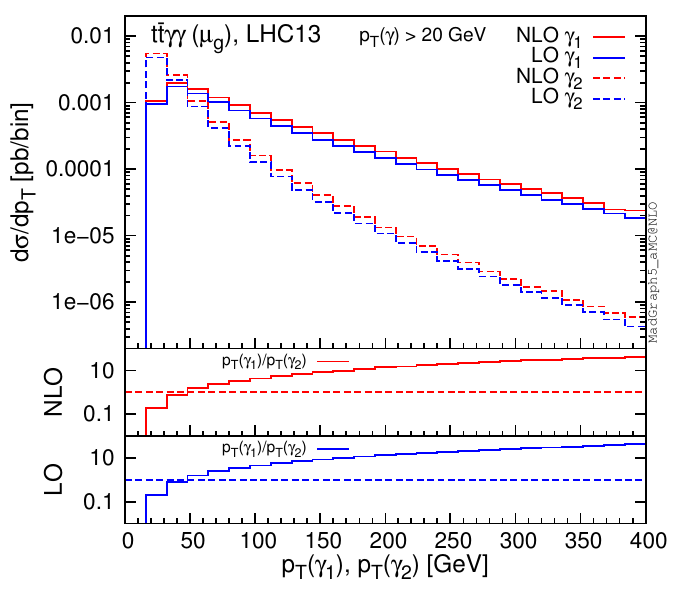}
\includegraphics[width=0.46\textwidth]{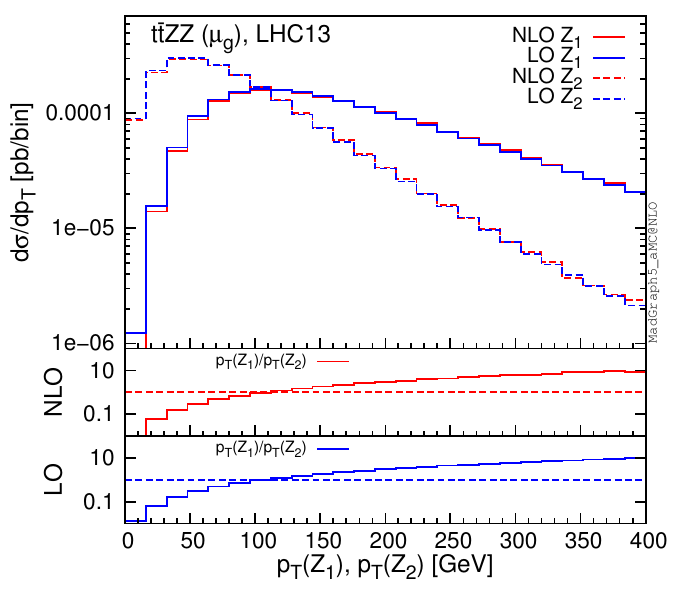}
\includegraphics[width=0.46\textwidth]{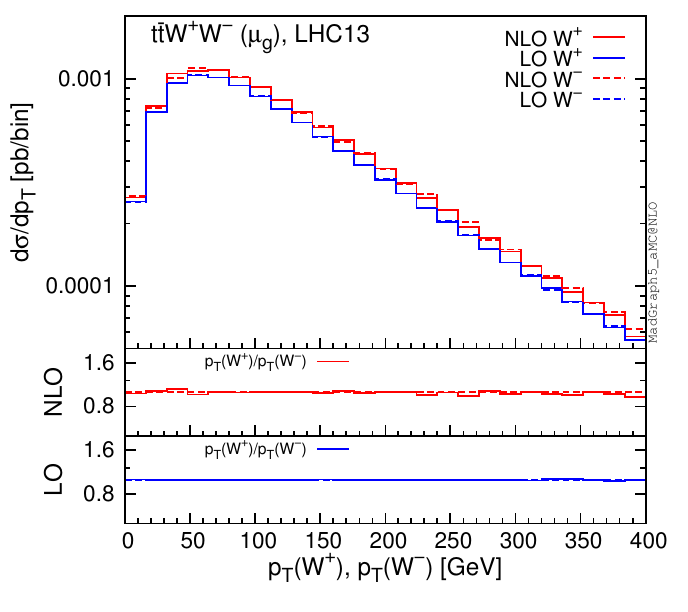}
\includegraphics[width=0.46\textwidth]{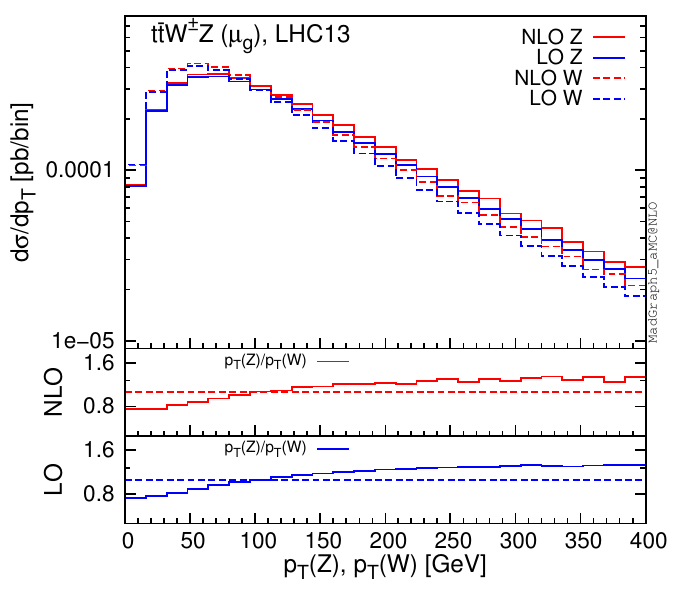}
\includegraphics[width=0.46\textwidth]{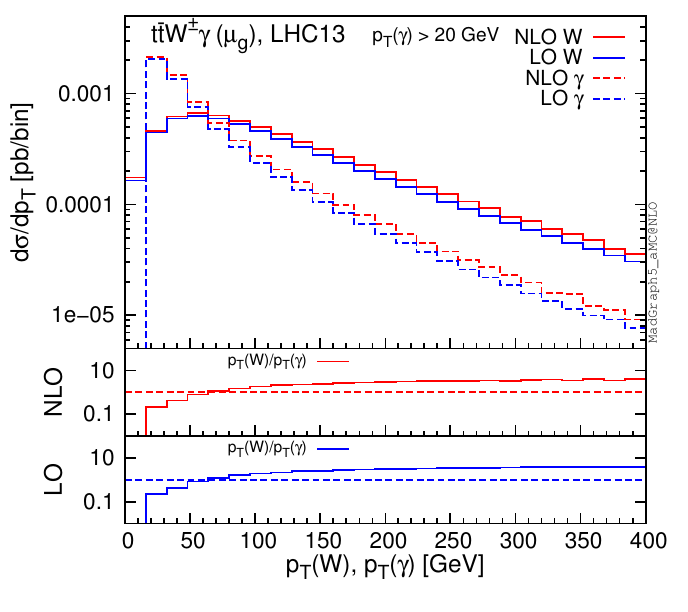}
\includegraphics[width=0.46\textwidth]{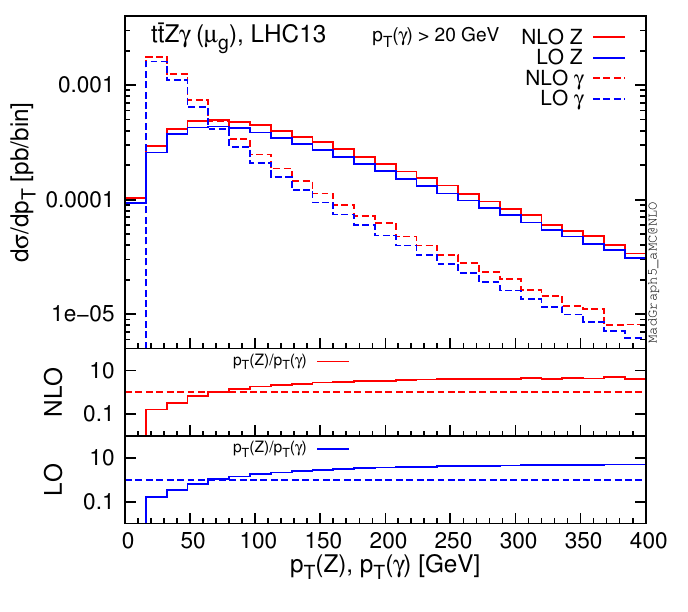}

\caption{Differential distributions for the $\pt$ of the first and second vector boson, $\pt(V_1)$ and $\pt(V_2)$.}
\label{fig:ttVV_ptV1_and_V2}
\end{figure}

In fig.~\ref{fig:ttVV_ptt} we show the distributions for $\pt(t)$. Most of the features discussed for $m(\ttbar)$ in fig.~\ref{fig:ttVV_inv} appear also for these distributions. The same applies to the distributions of the $\pt$ of the two vector bosons, which are displayed in  fig.~\ref{fig:ttVV_ptV1_and_V2}. In the plots of fig.~\ref{fig:ttVV_ptV1_and_V2} and in all the remaining figures of this section we use the same format used in subsection  \ref{sec:ttvh} for fig.~\ref{fig:ttV_ptt_and_pttx}. Thus, differential $K$-factors will not be explicitly shown. 
In the first and second inset we show the ratio of the distributions of the $\pt$ of the two vector bosons, respectively at NLO and LO accuracies. 
In the case of $\ttaa$ production, $\gamma_1$ is the hardest photon, while   $\gamma_2$ is the softest one. Similarly, in $\ttZZ$ production, $Z_1$ is the hardest $Z$ boson, while  $Z_2$ is the softest one. As can be noticed, for each process this ratio is the same at LO and NLO accuracy and thus it is not sensitive to NLO QCD corrections. 

\begin{figure}[h]
\centering
\includegraphics[width=0.46\textwidth]{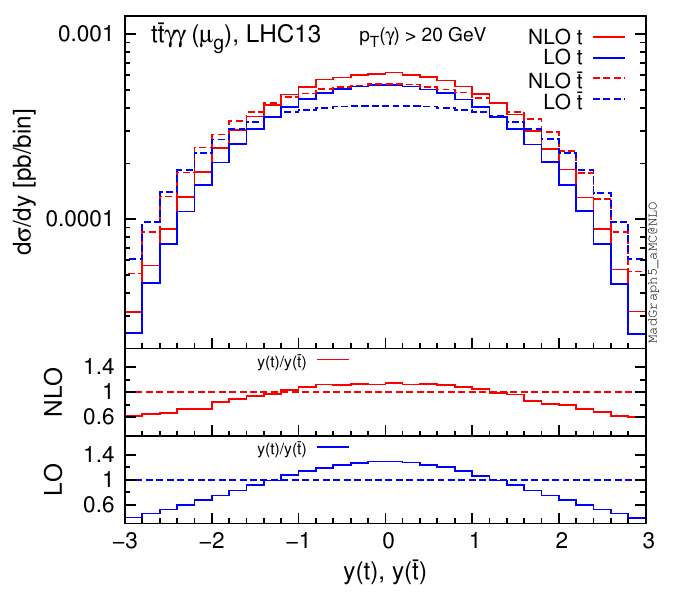}
\includegraphics[width=0.46\textwidth]{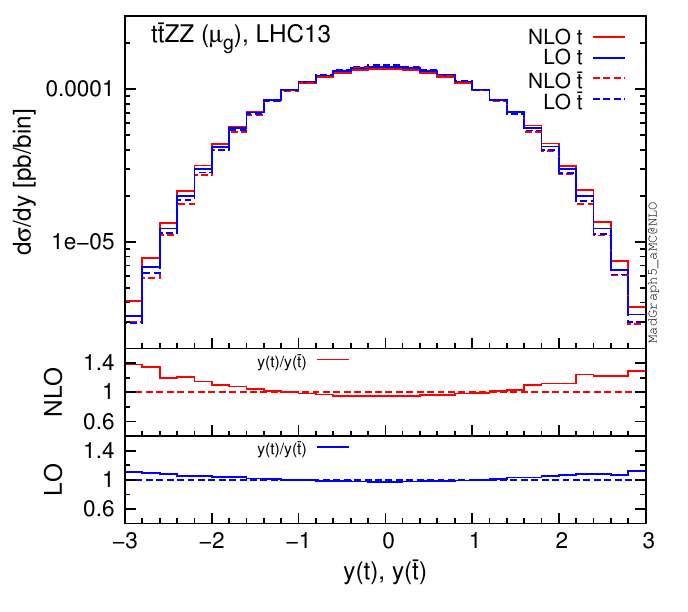}
\includegraphics[width=0.46\textwidth]{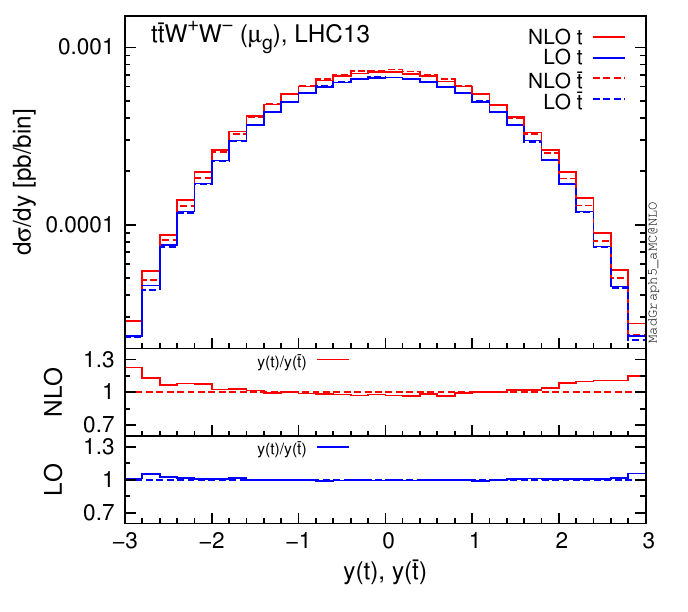}
\includegraphics[width=0.46\textwidth]{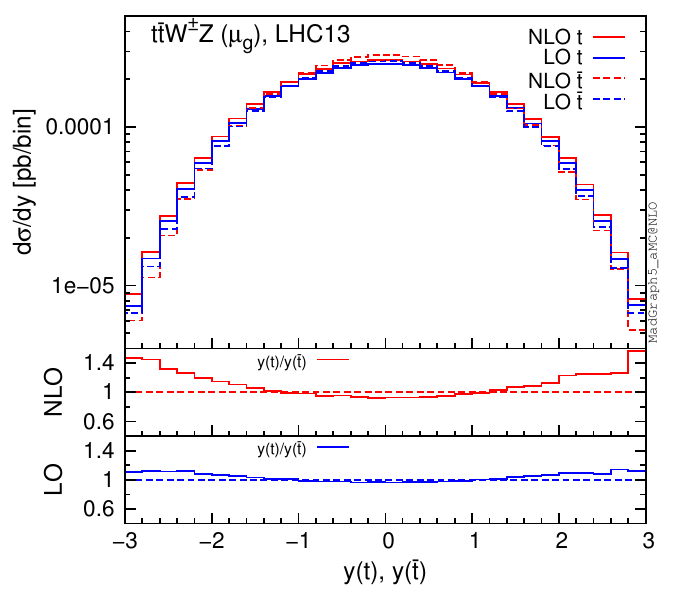}
\includegraphics[width=0.46\textwidth]{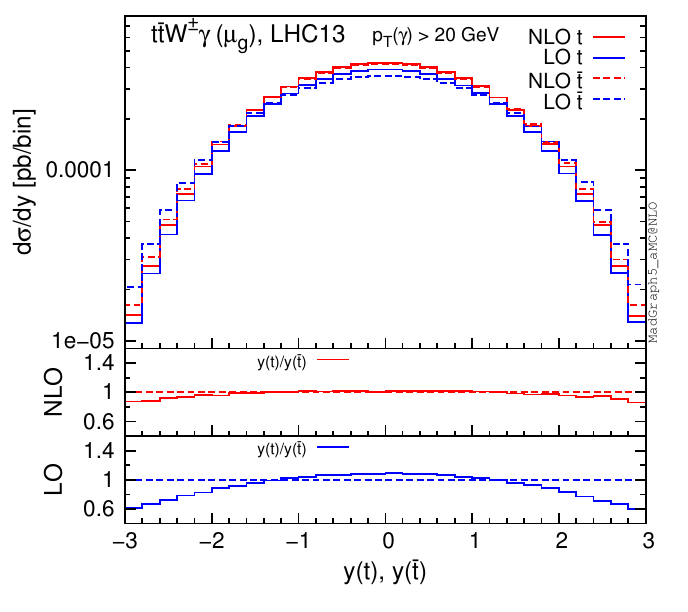}
\includegraphics[width=0.46\textwidth]{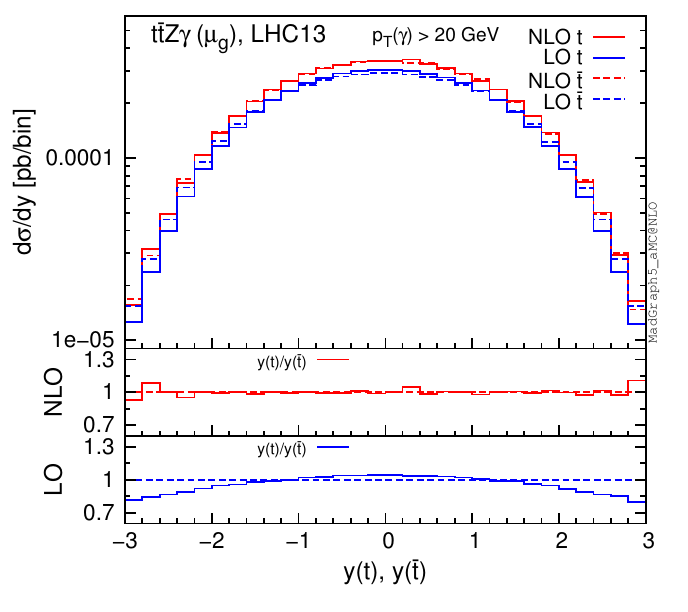}

\caption{Differential distributions for the rapidity of the top quark and antiquark, $y(t)$ and $y(\bar t)$.}
\label{fig:ttVV_rapt_and_tx}
\end{figure}

In fig.~\ref{fig:ttVV_rapt_and_tx} we show the distributions for $y(t)$ and $y(\bar t)$. The $\ttVV$ processes, with the exception of $\ttWW$~\footnote{Analytically, this process is supposed to give an asymmetry. Numerically, it turns out that it can be safely considered as zero.},  at LO  exhibit a central asymmetry for top (anti-)quarks. Top quarks are more centrally distributed than top antiquarks in  $\ttaa$, $\ttWa$ and $\ttZa$ productions, while they are more peripherally distributed in $\ttZZ$ and $\ttWZ$ production. In all the $\ttVV$ processes, NLO QCD corrections lead  to a relatively more peripheral distribution of top quarks than antiquarks. This effects yield to a non-vanishing central asymmetry for $\ttWW$ production and almost cancel the LO central asymmetry of $\ttZa$ production. Here, we refrain to present results for the central asymmetries of $\ttVV$ processes, since it is extremely unlikely that at the LHC it will be possible to accumulate enough statistics to perform these measurements.
 
\begin{figure}[h]
\centering
\includegraphics[width=0.46\textwidth]{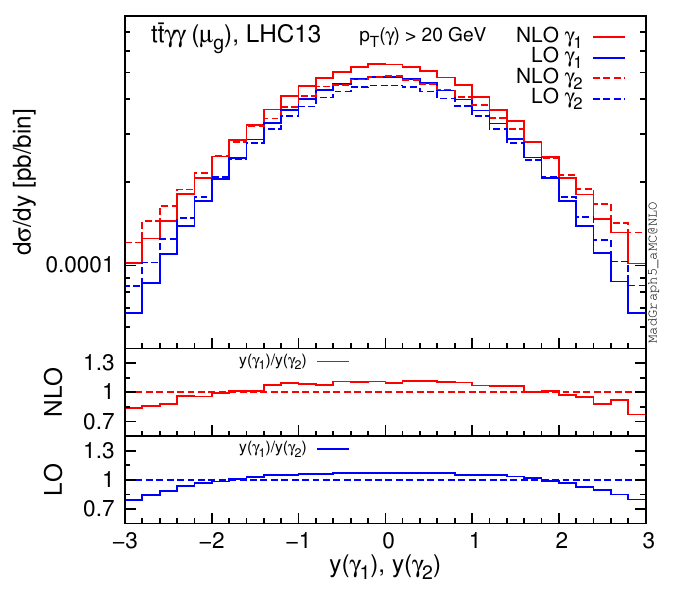}
\includegraphics[width=0.46\textwidth]{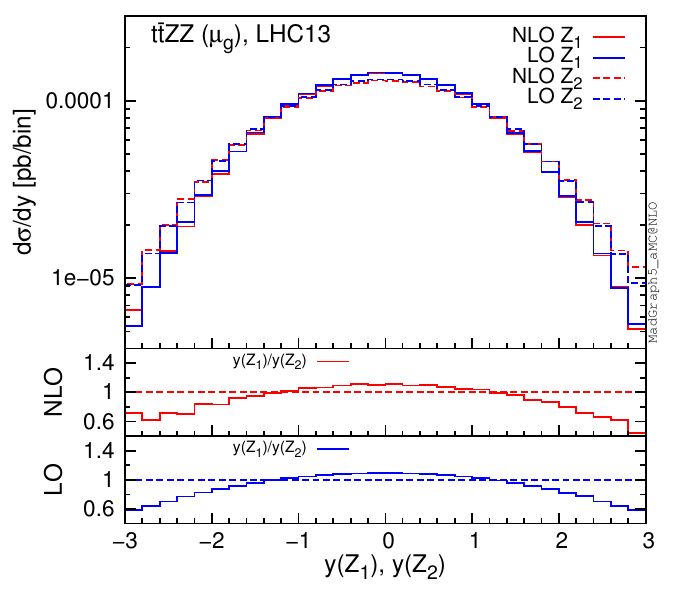}
\includegraphics[width=0.46\textwidth]{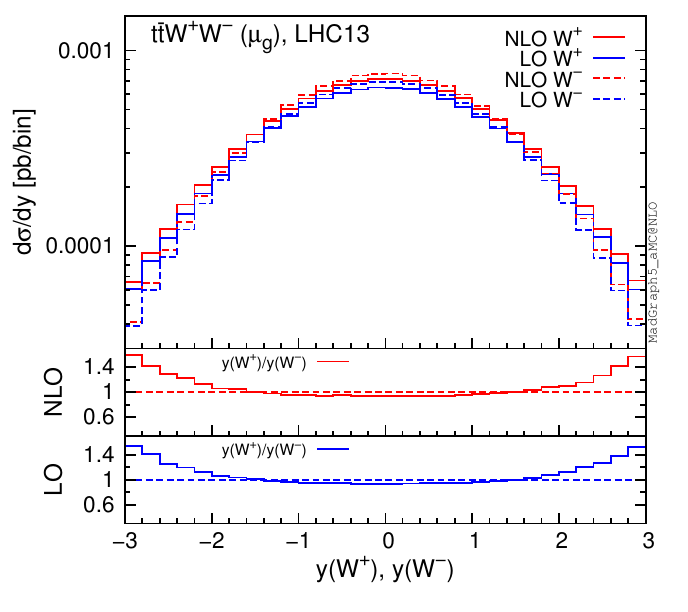}
\includegraphics[width=0.46\textwidth]{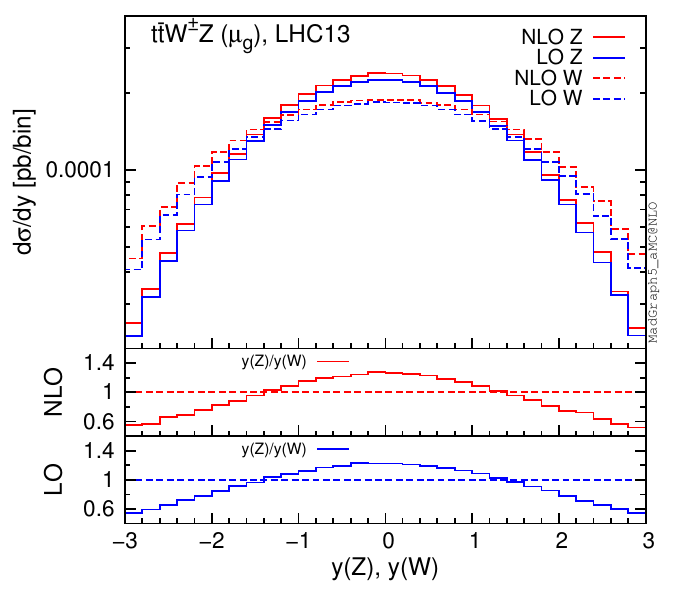}
\includegraphics[width=0.46\textwidth]{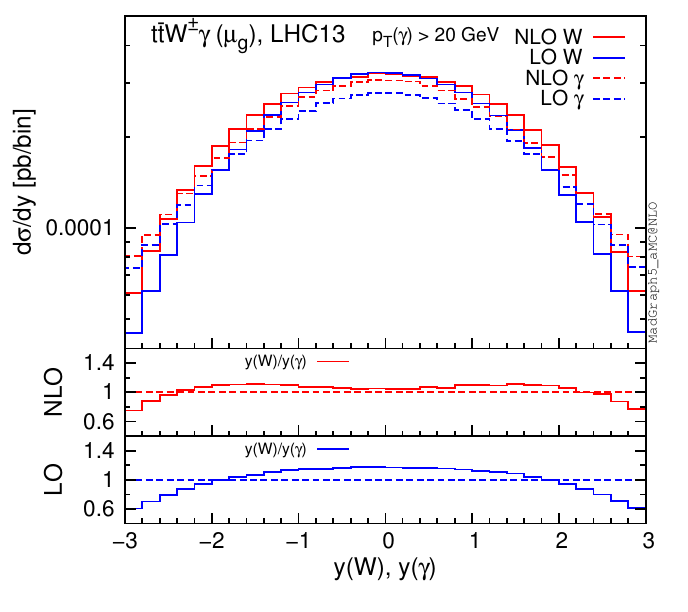}
\includegraphics[width=0.46\textwidth]{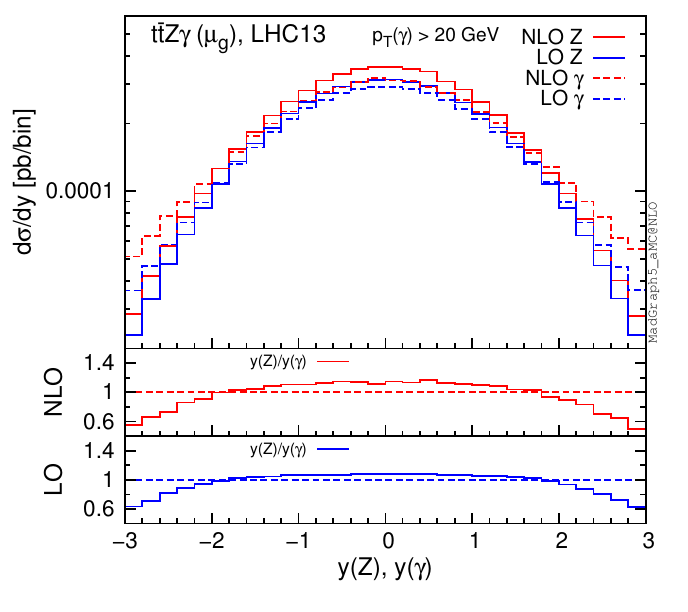}

\caption{Differential distributions for the rapidity of the first and second vector boson, $y(V_1)$ and $y(V_2)$.}
\label{fig:ttVV_rapV1_and_V2}
\end{figure}

In fig.~\ref{fig:ttVV_rapV1_and_V2} we show the distributions for $y(V_1)$ and $y(V_2)$. Comparing the first and second insets, only small differences can be seen for the ratios of the distributions at LO and NLO. Thus, unlike for the top quark and antiquark, the rapidity of the first and the second vector boson receive  NLO relative differential  corrections that are very similar in size. Both in the distributions of the rapidities of the top (anti)quark and of the vector bosons, NLO QCD corrections in general induce non-flat $K$-factors, also with the use of dynamical  scales.\footnote{We explicitly verified it and it can be easily reproduced via the public version of \aNLO, which has also been used for the phenomenological study presented here.}

\clearpage 

\subsection{$\tttt$ production}
\label{sec:tttt}

\begin{figure}[t]
\centering
\includegraphics[width=0.8\textwidth]{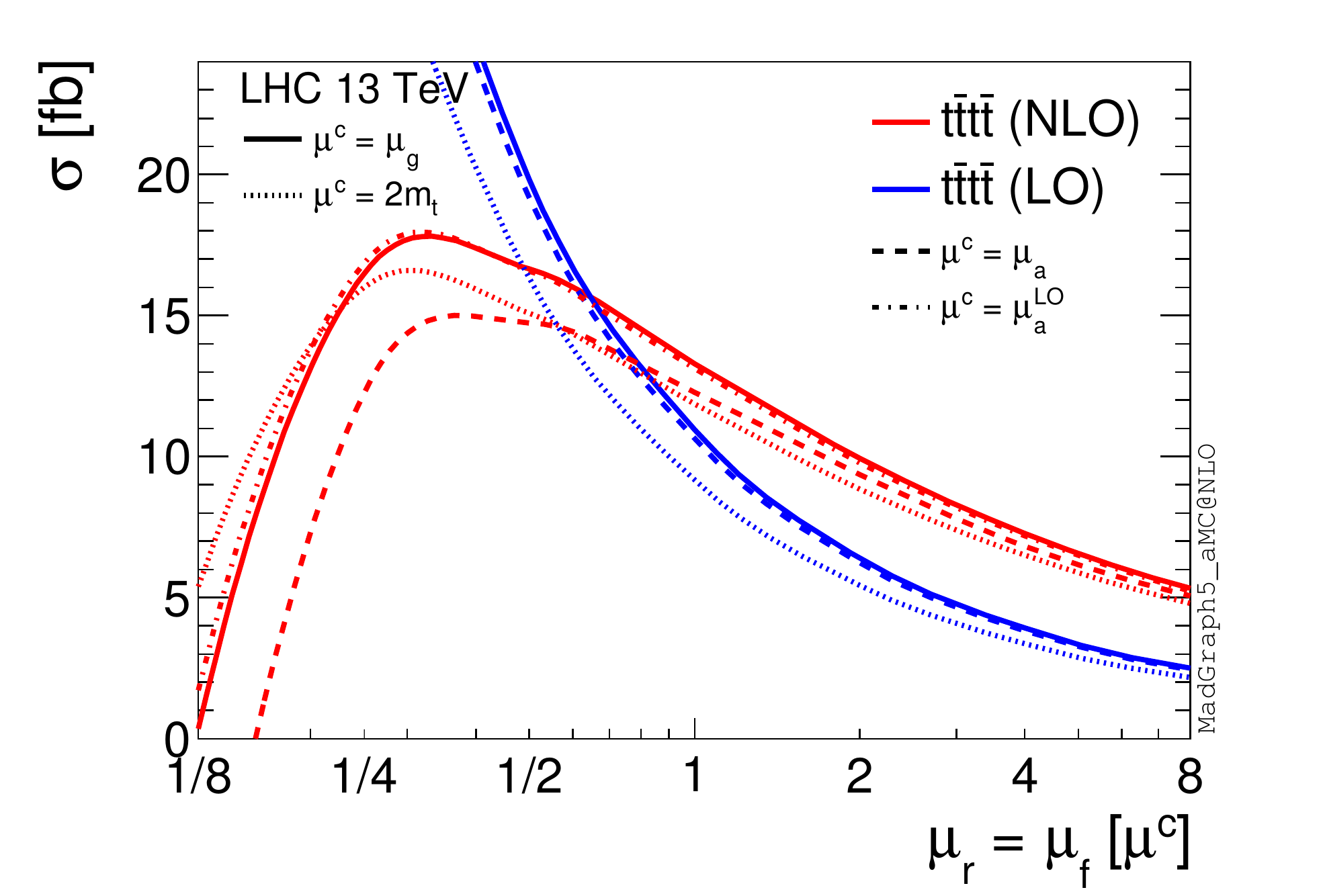}
\caption{NLO and LO cross sections for $\tttt$ production at 13 TeV. Comparison of the scale dependence in the interval $\mu^c/8<\mu<8\mu^c$ for the four different choices of the central value $\mu^c$: $\mu_g$, $\mu_a$, $\mua^{\rm LO}$, $2m_t$.}
\label{fig:scales_tttt}
\end{figure}

In this section we present results for $\tttt$ production.
We start by showing in fig.~\ref{fig:scales_tttt}  the scale dependence of the LO (blue lines) and NLO (red lines) total cross section at 13 TeV. As for the previous cases, we vary $\mu=\mu_r=\mu_f$ by a factor eight around the central value $\mu=\mug$ (solid lines), $\mu=\mua$ (dashes lines) and, due to the much heavier final state, $\mu=2m_t$ (dotted lines). In this case we also show with a dot-dashed line the dependence of the NLO cross section on an alternative definition of average scale  $\mua^{\rm LO}=\frac{1}{N}\sum_{   i=1,N  }m_{T,i}$, where possible additional partons appearing in the final state do not contribute.  

As expected, predictions relative to $\mug$ and $\mua^{\rm LO}$  are very close. Conversely, $\mua$ and $\mua^{\rm LO}$ show a non-negligible difference. Note that the value of $\mua$ and $\mua^{\rm LO}$ is the same for Born and and virtual contributions for any kinematic configuration. Thus, the difference between dashed and dot-dashed lines is formally an NNLO effect that arise from differences in the scale renormalisation for real radiation events only. To investigate the origin of this effect, we have explicitly checked that the difference is mainly induced by the corresponding change in the renormalisation scale and not of the factorisation scale. Similar behaviour is also found in $\ttV$ and $\ttVV$ processes, yet since the masses of the final-state particles are different and the $\alpha_s$ coupling order lower, $\mug$ and $\mua^{\rm LO}$ lines are more distant than in $\tttt$ production.

Since the LO cross section is of $\ord(\alpha_s^4)$, it strongly depends on the value of the renormalisation scale, as can be seen in fig.~\ref{fig:scales_tttt}. This dependence is considerably reduced at NLO QCD accuracy in the standard interval $\mug/2<\mu<2\mug$. Conversely, for $\mu<\mug/4$ the value of the cross section falls down rapidly, reaching zero for $\mu\sim \mug/8$. This is a signal that in this region the dependence of the cross section on $\mu$ is not under control. Qualitatively similar considerations apply also for the different choices of scales, as can be seen in fig.~\ref{fig:scales_tttt}.

In eqs. \eqref{ttttNLO} and \eqref{ttttLO}, we list the NLO and LO cross sections evaluated at the scale $\mu=\mug$ together with scale and PDF uncertainties. As done in previous subsections, scale uncertainties are evaluated by varying the factorisation and renormalisation scales  in the standard interval $\mug/2<\mu_f,\mu_r<2\mug$. As a result the total cross section at LHC 13 TeV for  the $\mu=\mug$ central scale
choice reads
\begin{eqnarray}
\sigma_{\rm NLO}&=&13.31^{+25.8 \%}_{-25.3 \%}~^{+5.8 \%}_{-6.6 \%} \; \;{\rm fb} \label{ttttNLO}\, ,\\ 
\sigma_{\rm LO}&=&10.94^{+81.1 \%}_{-41.6 \%}~^{+4.8 \%}_{-4.7 \%} \; \; {\rm fb}\label{ttttLO}\, ,\\
K{\rm -factor}&=&1.22 \label{ttttK}\,.
\end{eqnarray}
Different choices for the central value and functional form of the scales, as well as the interval of variation,
lead to predictions that are compatible with the result above, see also e.g.~\cite{Bevilacqua:2012em}.

\begin{figure}[h]
\centering
\includegraphics[width=0.457\textwidth]{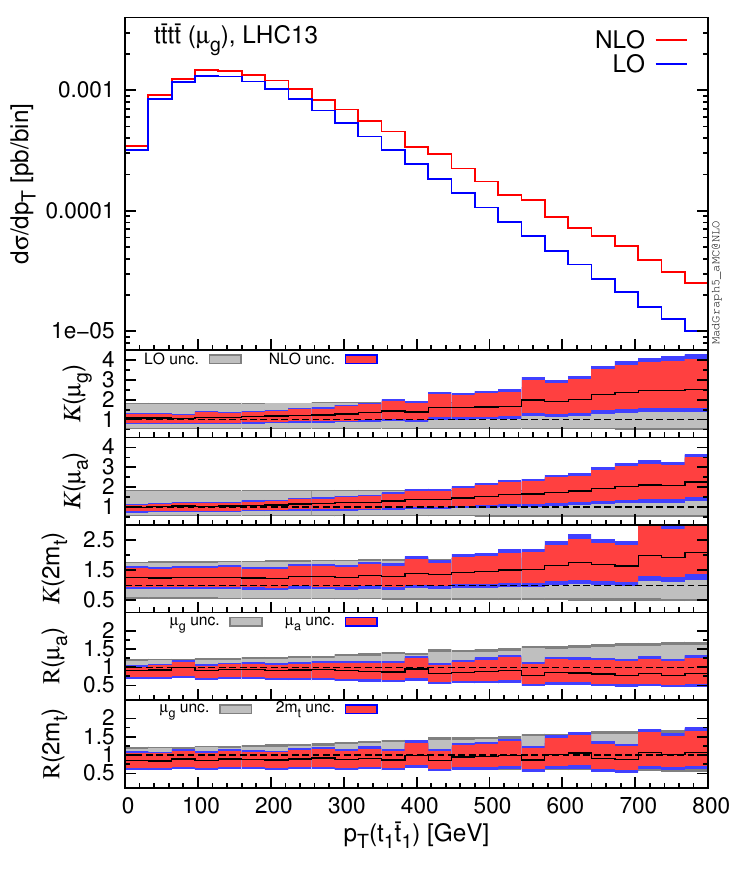}
\includegraphics[width=0.457\textwidth]{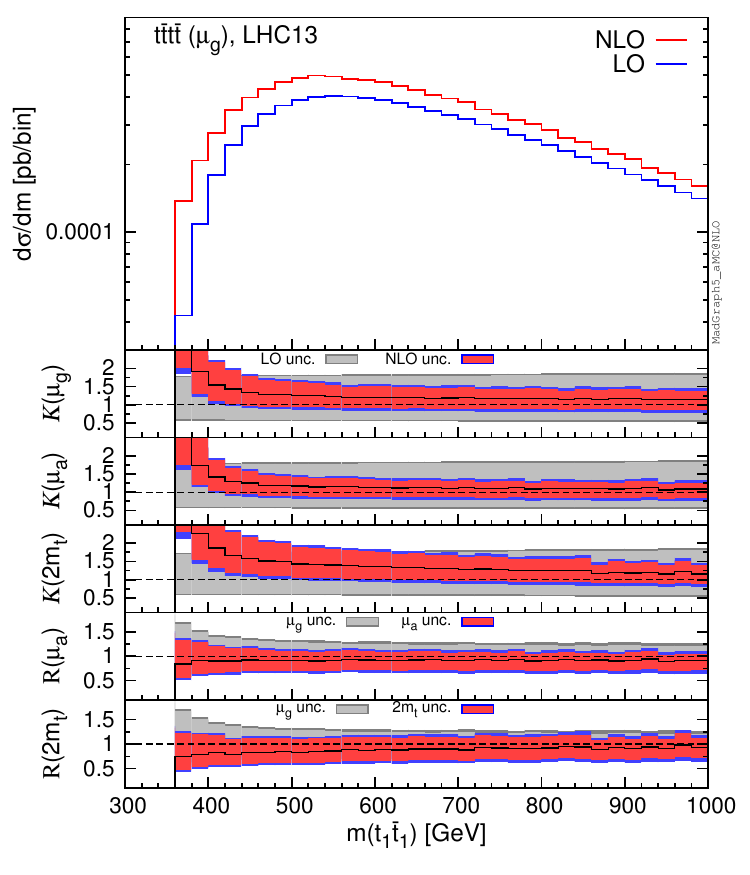}
\includegraphics[width=0.457\textwidth]{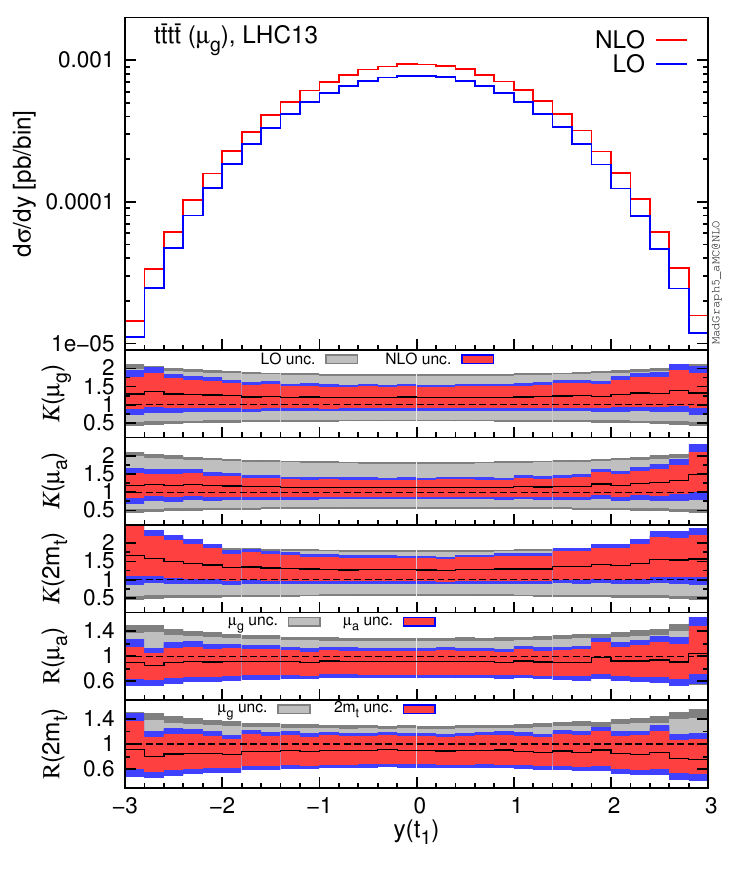}
\includegraphics[width=0.457\textwidth]{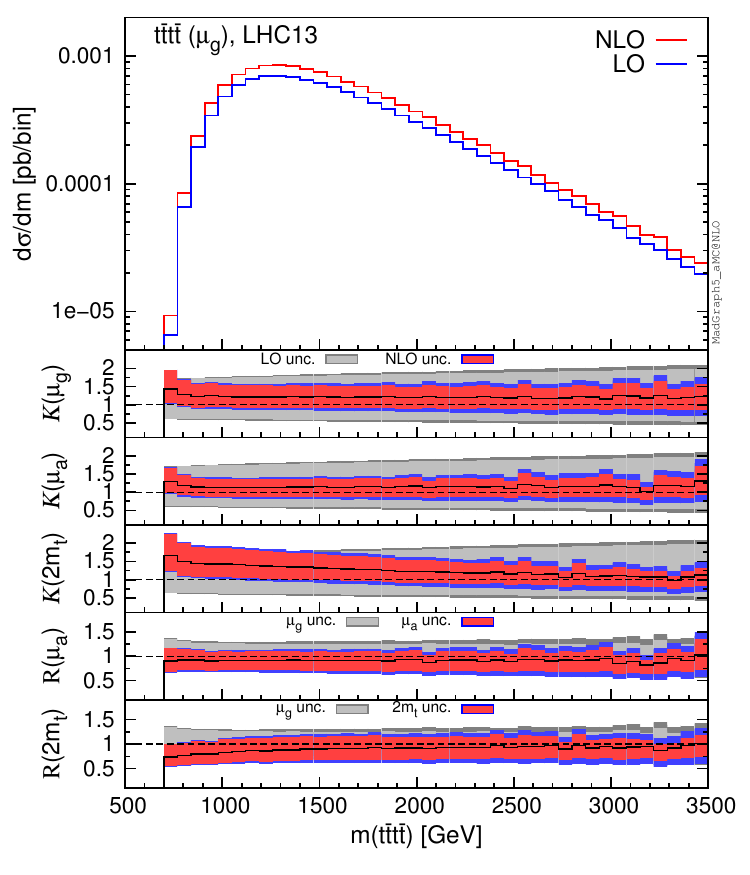}
\includegraphics[width=0.457\textwidth]{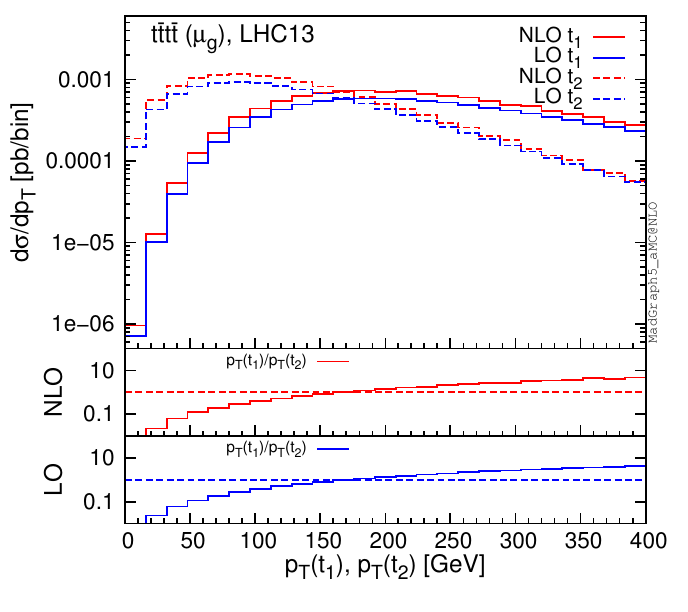}
\caption{Differential distributions for $\tttt$ production.}
\label{fig:tttt_distributions}
\end{figure}

We now discuss the effect of NLO QCD corrections on differential distributions. We analysed the distribution of the invariant mass, the $\pt$ and the rapidity of top (anti-)quark and the possible top-quark pairs.  Again, given the large amount of distributions, we show only representative results. All the distributions considered and additional ones can be produced via the public code \aNLO. For this process the scale dependence of many distributions has been studied also in \cite{Bevilacqua:2012em} and our results are in agreement with those therein. In fig.~\ref{fig:tttt_distributions} we show plots with the same formats as those used and described in the previous sections. Specifically, we display  the distributions for the total $\pt$ of the two hardest top quark and antiquark ($\pt(t_1 \bar{t}_1)$), their invariant mass ($m(t_1 \bar{t}_1)$), the rapidity of the hardest top quark $y(t_1)$ and the invariant mass of the $\tttt$ system ($m(\tttt)$). Also, in the last plot of fig.~\ref{fig:tttt_distributions}, we show the $\pt$ distributions of the hardest together with the softest top quarks, $\pt(t_1)$ and $\pt(t_2)$, and their ratios at NLO and LO.

We avoid  repeating once again the general features that have already been pointed out several times in the previous two sections; they are still valid for $\tttt$ production. Here, we have found, interestingly, that NLO corrections give a sizeable enhancement in the threshold region for $m(t_1 \bar{t}_1)$.  It is  worth to notice that also for this process NLO QCD corrections are very large in the tail of the $\pt(t_1 \bar{t}_1)$ distribution, especially with the use of dynamical scales. We have verified that in these regions of phase space the $qg \to \ttbar \ttbar q$ contributions are important. Finally, as can be seen in the last plot, we find that the ratios of $\pt(t_1)$ and $\pt(t_2)$ distributions are not sensitive to NLO QCD corrections.

\clearpage

\subsection{Total cross sections from 8 to 100 TeV}
\label{sec:energy}
\begin{figure}[t]
\centering
\includegraphics[width=0.77\textwidth]{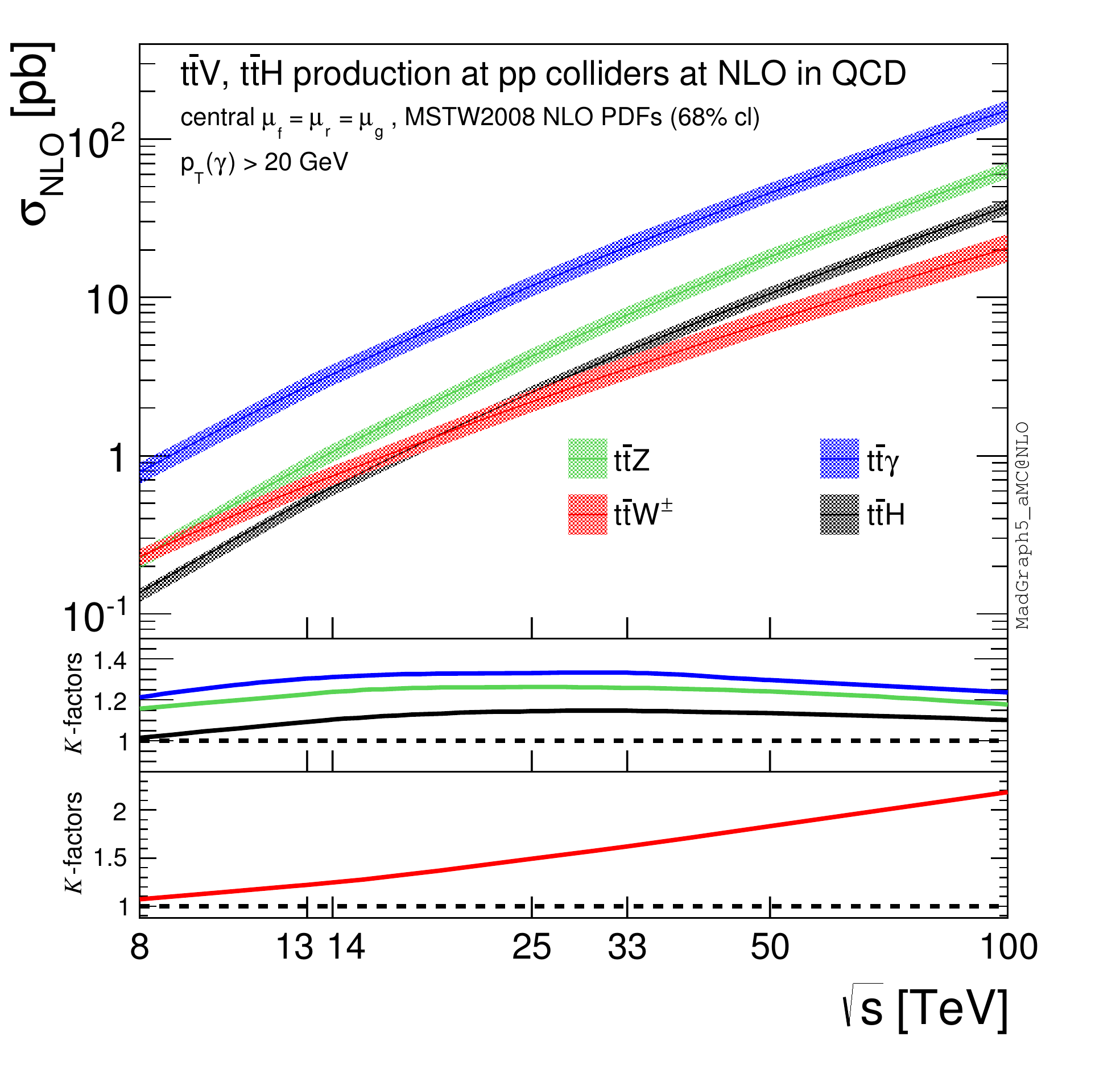}
\includegraphics[width=0.77\textwidth]{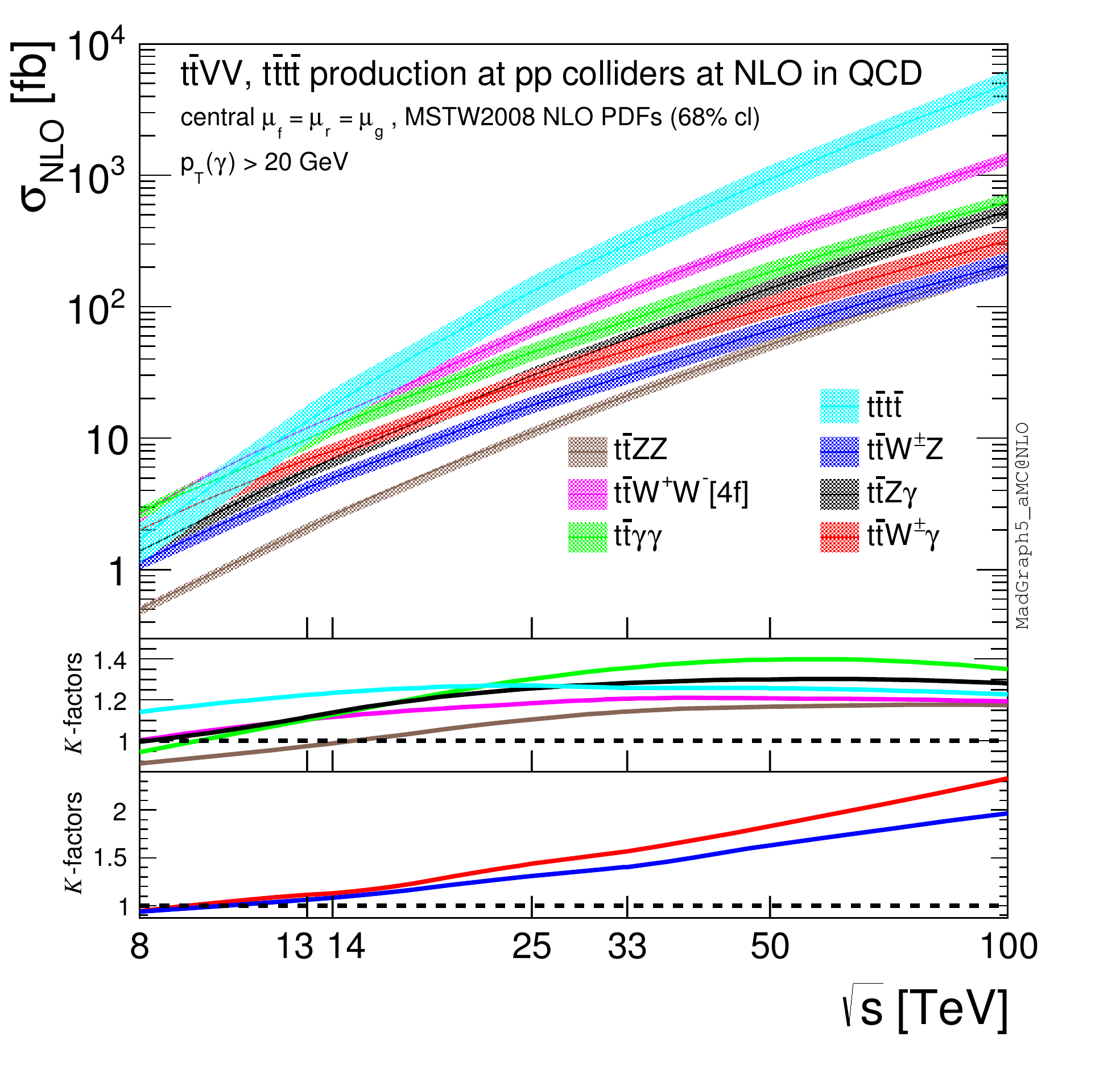}
\caption{NLO total cross sections from 8 to 100 TeV. The error bands include scale and PDF uncertainties (added linearly). The upper plot refers to $\ttV$ processes and $\ttbar H$ production, the lower plot to $\ttVV$ processes and $\tttt$ production. For final states with photons the $\pt(\gamma)> 20~\gev$ cut is applied.}
\label{fig:xsec_8to100}
\end{figure}

In addition to the studies performed for the LHC at 13 TeV, in this subsection we discuss and show results for the dependence of the total cross section on the energy of the  proton--proton collision. In fig.~\ref{fig:xsec_8to100} NLO QCD total cross sections are plotted from 8 to 100 TeV, as bands  including scale and PDF uncertainties. The corresponding numerical values are listed in table \ref{table:xsec_8_100}. As usual, central values refers to $\mu=\mug$, and scale uncertainties are obtained by varying independently $\mu_r$ and $\mu_f$ in the standard interval $\mug/2<\mu_f,\mu_r<2\mug$.

In the upper plot of fig.~\ref{fig:xsec_8to100} we show the results for $\ttbar H$ production and $\ttV$ processes, whereas $\tttt$ production and $\ttVV$ processes results are displayed in the lower plot.
In both plots we show the dependence of the $K$-factors at $\mu=\mug$ on the energy (the first and  the second inset). The first insets refer to processes with zero-total-charge final states, whereas the second insets refer to processes with charged final states.  
The very different qualitative behaviours between the two classes of processes is due to the fact that the former include already at LO an initial state with gluons, whereas the latter do not. The gluon appears in the partonic initial states of charged processes only at NLO via the (anti)quark--gluon channel. At small Bjorken-$x$'s, the gluon PDF grows much faster than the (anti)quark PDF. Thus, increasing the energy of the collider, the relative corrections induced by the (anti)quark--gluon initial states leads to the growth of the $K$-factors and dominates in their energy dependence.
Also, as can be seen in fig.~\ref{fig:xsec_8to100} and table \ref{table:xsec_8_100}, these processes  present a larger dependence on the scale variation than the uncharged processes. 

The differences in the slopes of the curves in the main panels of the plots are also mostly due to the gluon PDF. Charged processes do not originate from the gluon--gluon initial state neither at LO nor at NLO. For this reason, their growth with the increasing of the energy is smaller than for the uncharged processes. All these arguments point to the fact that, at 100 TeV collider, it will be crucial to have NNLO QCD corrections for $\ttW$, $\ttWa$ and $\ttWZ$ processes, if precise measurements to be compared with theory will be available.

\begin{figure}[t]
\centering
\includegraphics[width=0.77\textwidth]{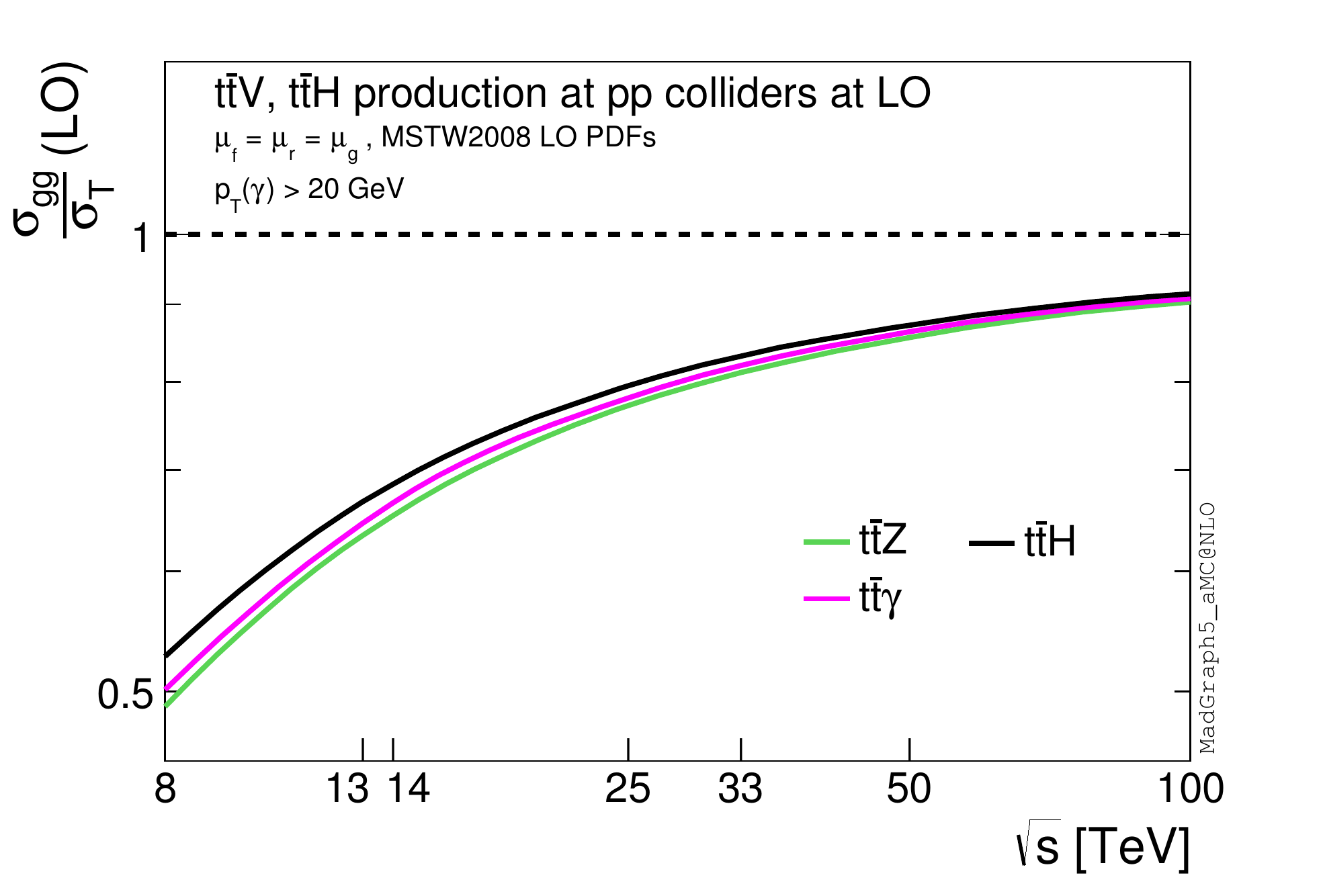}
\includegraphics[width=0.77\textwidth]{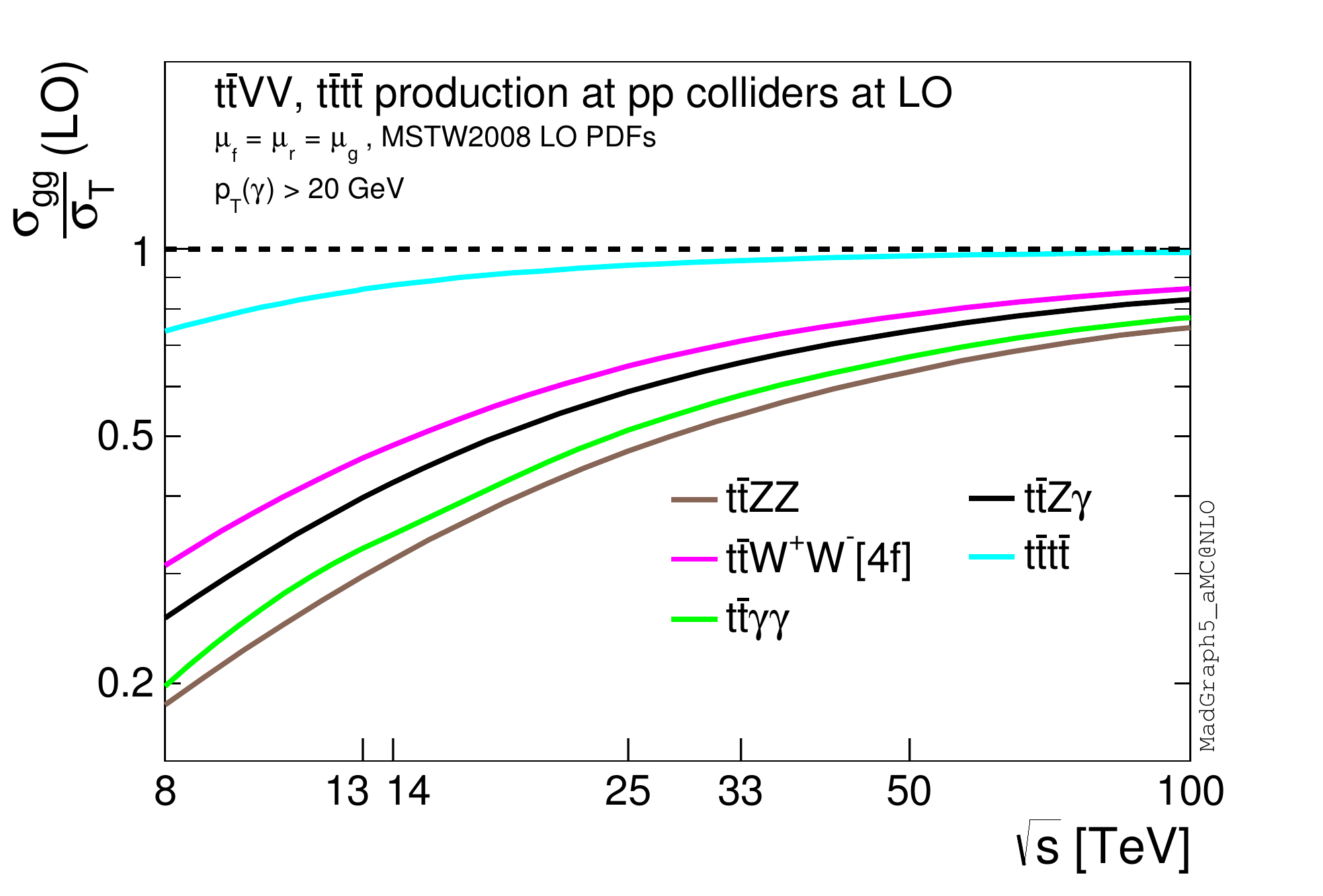}
\caption{Relative contribution of the $gg$ channel  to the total cross section at LO for
$\ttbar V,\ttbar H, \ttbar VV$ and $\ttbar\ttbar$ processes for $pp$ collisions from 8 to 100 TeV centre-of-mass energy. For final states with photons the $\pt(\gamma)> 20~\gev$ cut is applied. }
\label{fig:gg8to100}
\end{figure}

The fact that $\tttt$ production is the process with the rapidest growth is again due to percentage content of gluon--gluon-initiated channels, which is higher than for all the other processes, see fig.~\ref{fig:gg8to100}.  From the left plot  of fig.~\ref{fig:xsec_8to100}, it is easy also to note that the scale  uncertainty of $\tttt$ production is larger than for the $\ttVV$ processes. In this case, the difference originates from the different powers of
$\alpha_s$ at LO; $\tttt$ production is of $\ord(\alpha_s^4)$ whereas $\ttVV$
processes are of $\ord(\alpha_s^2 \alpha^2)$.

\noindent

\noindent

\begin{landscape}

\begin{table}[h]
\renewcommand{\arraystretch}{1.7}
\medskip
\resizebox{\linewidth}{!}{%
\tabcolsep=2pt
\begin{tabular}{  c | c c c c c c c }
\hline\hline
$\sigma$ [fb] & 8 TeV &  13 TeV & 14 TeV & 25 TeV & 33 TeV & 50 TeV & 100 TeV \\[5pt]
\hline
$t \bar t ZZ$ & $0.502^{+2.9 \%}_{-8.6 \%}~^{+2.7 \%}_{-2.2 \%}$ & $2.12^{+3.8 \%}_{-8.6 \%}~^{+1.9 \%}_{-1.8 \%}$ & $2.59^{+4.3 \%}_{-8.7 \%}~^{+1.8 \%}_{-1.8 \%}$ & $11.1^{+6.9 \%}_{-9.1 \%}~^{+1.2 \%}_{-1.4 \%}$ & $21.1^{+8.1 \%}_{-9.4 \%}~^{+1.1 \%}_{-1.3 \%}$ & $51.6^{+9.9 \%}_{-9.8 \%}~^{+0.9 \%}_{-1.1 \%}$ & $204^{+11.3 \%}_{-9.9 \%}~^{+0.8 \%}_{-1.0 \%}$ \\[5pt]
\hline
$t \bar t W^+ W^-$[4f] & $2.67^{+6.2 \%}_{-11.1 \%}~^{+2.9 \%}_{-2.7 \%}$ & $11.8^{+8.3 \%}_{-11.2\%}~^{+2.3 \%}_{-2.4 \%}$ & $14.4^{+12.2 \%}_{-12.8 \%}~^{+2.6 \%}_{-2.9 \%}$ & $66.6^{+9.5 \%}_{-10.8 \%}~^{+1.6 \%}_{-2.0 \%}$ & $130^{+10.2 \%}_{-10.8 \%}~^{+1.5 \%}_{-1.8 \%}$ & $327^{+10.9 \%}_{-10.6 \%}~^{+1.3 \%}_{-1.6 \%}$ & $1336^{+10.3 \%}_{-9.9 \%}~^{+1.0 \%}_{-1.3 \%}$ \\[5pt]
\hline
$t \bar t \gamma \gamma$ & $2.77^{+6.4 \%}_{-10.5 \%}~^{+1.9 \%}_{-1.5 \%}$ & $10.3^{+13.9 \%}_{-13.3 \%}~^{+1.3 \%}_{-1.3 \%}$ & $12^{+12.5 \%}_{-12.6 \%}~^{+1.2 \%}_{-1.2 \%}$ & $ 44.8^{+15.7 \%}_{-13.5 \%}~^{+0.9 \%}_{-0.9 \%}$ & $78.2^{+16.4 \%}_{-13.6 \%}~^{+0.8 \%}_{-0.9 \%}$ & $184^{+19.2 \%}_{-14.7 \%}~^{+0.8 \%}_{-0.9 \%}$ & $624^{+15.5 \%}_{-13.4 \%}~^{+0.7 \%}_{-1.0 \%}$ \\[5pt]
\hline
$t \bar t W^{\pm} Z$ & $1.13^{+5.8 \%}_{-9.8 \%}~^{+3.1 \%}_{-2.1 \%}$ & $4.16^{+9.8 \%}_{-10.7 \%}~^{+2.2 \%}_{-1.6 \%}$ & $4.96^{+10.4 \%}_{-10.8 \%}~^{+2.1 \%}_{-1.6 \%}$ & $17.8^{+15.1 \%}_{-12.6 \%}~^{+1.5 \%}_{-1.1 \%}$ & $30.2^{+18.3 \%}_{-14.1 \%}~^{+1.2 \%}_{-0.9 \%}$ & $66^{+18.9 \%}_{-14.3 \%}~^{+1.1 \%}_{-0.8 \%}$ & $210^{+21.6 \%}_{-15.8 \%}~^{+1.0 \%}_{-0.8 \%}$ \\[5pt]
\hline
$t \bar t Z \gamma$ & $1.39^{+6.9 \%}_{-11.2 \%}~^{+2.5 \%}_{-2.2 \%}$ & $5.77^{+10.5 \%}_{-12.1 \%}~^{+1.8 \%}_{-1.9 \%}$ & $6.95^{+10.7 \%}_{-12.1 \%}~^{+1.8 \%}_{-1.9 \%}$ & $29.9^{+12.9 \%}_{-12.4 \%}~^{+1.3 \%}_{-1.5 \%}$ & $56.5^{+13.2 \%}_{-12.2 \%}~^{+1.2 \%}_{-1.4 \%}$ & $138^{+13.7 \%}_{-12.0 \%}~^{+1.0 \%}_{-1.1 \%}$ &  $533^{+13.3 \%}_{-11.1 \%}~^{+0.8 \%}_{-1.0 \%}$ \\[5pt]
\hline
$t \bar t W^{\pm} \gamma$ & $2.01^{+7.9 \%}_{-10.5 \%}~^{+2.6 \%}_{-1.8 \%}$ & $6.73^{+12.0 \%}_{-11.6 \%}~^{+1.8 \%}_{-1.4 \%}$ & $7.99^{+12.8 \%}_{-11.9 \%}~^{+1.7 \%}_{-1.3 \%}$ & $27.6^{+18.7 \%}_{-14.4 \%}~^{+1.2 \%}_{-0.9 \%}$ & $46.3^{+20.2 \%}_{-15.1 \%}~^{+1.1 \%}_{-0.8 \%}$ & $98.4^{+21.9 \%}_{-15.9 \%}~^{+1.0 \%}_{-0.7 \%}$ & $318^{+22.5 \%}_{-17.7 \%}~^{+1.0 \%}_{-0.7 \%}$ \\[5pt]
\hline
$t \bar t t \bar t $ & $1.71^{+24.9 \%}_{-26.2 \%}~^{+7.9 \%}_{-8.4 \%}$ & $13.3^{+25.8 \%}_{-25.3\%}~^{+5.8 \%}_{-6.6 \%}$ & $17.8^{+26.6 \%}_{-25.4 \%}~^{+5.5 \%}_{-6.4 \%}$ & $130^{+26.7 \%}_{-24.3 \%}~^{+3.8 \%}_{-4.6 \%}$ & $297^{+25.5 \%}_{-23.3 \%}~^{+3.1 \%}_{-3.9 \%}$ & $929^{+24.9 \%}_{-22.4 \%}~^{+2.4 \%}_{-3.0 \%}$ & $4934^{+25.0 \%}_{-21.3 \%}~^{+1.7 \%}_{-2.1 \%}$ \\[5pt]
\hline
\hline
$\sigma$ [pb] & 8 TeV &  13 TeV & 14 TeV & 25 TeV & 33 TeV & 50 TeV & 100 TeV \\[5pt]
\hline
$t \bar t Z $ & $ 0.226^{+9.0 \%}_{-11.9 \%}~^{+2.6 \%}_{-3.0 \%}$ & $ 0.874^{+10.3 \%}_{-11.7 \%}~^{+2.0 \%}_{-2.5 \%}$ & $ 1.057^{+10.4 \%}_{-11.7 \%}~^{+1.9 \%}_{-2.4 \%}$ & $ 4.224^{+11.0 \%}_{-11.0 \%}~^{+1.5 \%}_{-1.8 \%}$ & $ 7.735^{+11.2 \%}_{-10.8 \%}~^{+1.3 \%}_{-1.5 \%}$ & $ 18^{+11.1 \%}_{-10.2 \%}~^{+1.1 \%}_{-1.3 \%}$ & $ 64.17^{+11.1 \%}_{-11.0 \%}~^{+0.9 \%}_{-1.2 \%}$ \\[5pt]
\hline
$t \bar t W^{\pm} $ & $ 0.23^{+9.6 \%}_{-10.6 \%}~^{+2.3 \%}_{-1.7 \%}$ & $ 0.645^{+13.0 \%}_{-11.6 \%}~^{+1.7 \%}_{-1.3 \%}$ & $ 0.745^{+13.5 \%}_{-11.8 \%}~^{+1.6 \%}_{-1.3 \%}$ & $ 2.188^{+17.0 \%}_{-13.2 \%}~^{+1.3 \%}_{-0.9 \%}$ & $ 3.534^{+18.1 \%}_{-13.7 \%}~^{+1.2 \%}_{-0.8 \%}$ & $ 7.03^{+19.2 \%}_{-14.3 \%}~^{+1.1 \%}_{-0.8 \%}$ & $ 20.55^{+21.5 \%}_{-18.1 \%}~^{+1.1 \%}_{-0.8 \%}$ \\[5pt]
\hline
$t \bar t \gamma $ & $ 0.788^{+12.7 \%}_{-13.5 \%}~^{+2.1 \%}_{-2.4 \%}$ & $ 2.746^{+14.2 \%}_{-13.5 \%}~^{+1.6 \%}_{-1.9 \%}$ & $ 3.26^{+14.2 \%}_{-13.4 \%}~^{+1.6 \%}_{-1.9 \%}$ & $ 11.77^{+14.5 \%}_{-12.7 \%}~^{+1.2 \%}_{-1.4 \%}$ & $ 20.84^{+14.9 \%}_{-12.5 \%}~^{+1.1 \%}_{-1.3 \%}$ & $ 45.68^{+14.2 \%}_{-11.7 \%}~^{+1.0 \%}_{-1.2 \%}$ & $ 152.6^{+14.3 \%}_{-13.7 \%}~^{+0.9 \%}_{-1.2 \%}$ \\[5pt]
\hline
$t \bar t H $ & $ 0.136^{+3.3 \%}_{-9.1 \%}~^{+2.8 \%}_{-3.2 \%}$ & $ 0.522^{+6.0 \%}_{-9.4 \%}~^{+2.1 \%}_{-2.6 \%}$ & $ 0.631^{+6.3 \%}_{-9.4 \%}~^{+2.0 \%}_{-2.5 \%}$ & $ 2.505^{+8.3 \%}_{-9.4 \%}~^{+1.6 \%}_{-1.9 \%}$ & $ 4.567^{+8.8 \%}_{-9.2 \%}~^{+1.4 \%}_{-1.7 \%}$ & $ 10.55^{+9.5 \%}_{-9.0 \%}~^{+1.2 \%}_{-1.4 \%}$ & $ 37.65^{+10.0 \%}_{-9.8 \%}~^{+1.0 \%}_{-1.3 \%}$ \\[5pt]
\hline
\end{tabular}}
\caption{NLO cross sections for $t \bar t V V , t \bar t t \bar t , t \bar t V , t \bar t H $ processes using the geometrical average scale. The first uncertainty is given by scale variation, the second by PDFs. For final states with photons the $\pt(\gamma)> 20~\gev$ cut is applied. The cross sections for the four final-state particle processes are calculated with percent accuracy, whereas for the processes with three final-state particles with per mill.}
\label{table:xsec_8_100}
\end{table}

\end{landscape}
\clearpage

\section{Analyses of $\ttbar H$ signatures}
\label{sec:analysis}

In this section we provide numerical results for the contributions of signal and irreducible background processes to two different classes of $\ttbar H$ signatures at the LHC. In subsection \ref{sec:photon} we consider a signature involving two isolated photons emerging from the decay of the Higgs boson into photons, $H \rightarrow \gamma \gamma$. In subsection \ref{sec:leptons} we analyse three different signatures involving two or more leptons, where $\ttbar H$ production can contribute via the $H\TO Z Z^*$, $H\TO W W^*$ and $H\TO \tau^{+}\tau^{-}$ decays.  We perform both the analyses at 13 TeV and we adopt the cuts of \cite{Khachatryan:2014qaa}.\footnote{In our simulation we do not take into account particle identification efficiencies and possible misidentification effects.} The preselection cuts, which are common for both the analyses, are: 
\begin{eqnarray}
\pt(e)>7 \gev \,, & \quad  |\eta(e)<2.5|\, , \quad &  \pt(\mu)>5 \gev \,,  \quad |\eta(\mu)|<2.4\, , \nonumber \\
|\eta(\gamma)| < 2.5\, , & \quad  \pt(j)>25 \gev \,, \quad  & |\eta(j)|<2.4 \, ,
\label{eq:preselection}
\end{eqnarray}
where jets are clustered via anti-$k_T$ algorithm \cite{Cacciari:2008gp} with the distance parameter $R=0.5$. Event by event, only particles satisfying the preselection cuts in eq.\eqref{eq:preselection} are considered and, for each jet $j$ and lepton $\ell$, if $\Delta R(j,\ell) < 0.5$ the lepton  $\ell$  is clustered into the jet $j$. With the symbol $\ell$, unless otherwise specified, we always refer  to electrons(positrons) and (anti)muons, not to $\tau$ (anti)leptons.

All the simulations for the signal and the background processes have been performed at NLO QCD accuracy matched with parton shower effects ($\rm NLO+PS$). Events are generated via {\aNLO}, parton shower  and hadronization effects are realised in {\Pythiae} \cite{Sjostrand:2007gs}, and   jets are clustered via  {\FastJet} \cite{Cacciari:2011ma}.\footnote{In our simulation, $b$-tagging is performed by looking directly at $B$ hadrons, which we keep stable.}  Unless differently specified, decays of the heavy states, including $\tau$ leptons, are performed in {\Pythiae}. In the showering, only QCD effects have been included; QED and  purely weak  effects are not included. Furthermore, multi-parton interaction and underlying event effects are not taken into account.  

In order to discuss NLO effects at the analysis level,  in the following  we will also report results for events generated at LO accuracy including shower and hadronization effects  ($\rm LO+PS$). As done for the fixed-order studies in section \ref{sec:prod}, $\rm LO+PS$ and $\rm NLO+PS$ central values are evaluated at $\mu_f=\mu_r=\mug$ and scale uncertainties are obtained by varying independently  the factorisation and the renormalisation scale in the interval $\mug/2<\mu_f , \mu_r<2\mug$.

\subsection{Signature with two photons}
\label{sec:photon} 

The present analysis focuses on the Higgs boson decaying into two photons in $\ttbar H$ production, which presents as irreducible background the  $\ttaa$ production. In our simulation, top quark pairs are decayed via {\Madspin} for both the signal and the background, whereas the loop-induced $H \rightarrow \gamma \gamma$ decay is forced in {\Pythiae} and event weights are rescaled by the branching ratio ${\rm BR}(H \rightarrow \gamma \gamma)=2.28 \times 10^{-3}$, which is taken from \cite{BR-HXSWG}.

In this analysis, at least two jets are required and one of them has to be $b$-tagged. In addition, the following cuts are applied:
\begin{eqnarray}
100\gev < m(\gamma_1 \gamma_2) < 180\gev\,, & p_T(\gamma_1) > \dfrac{m(\gamma_1 \gamma_2)}{2} \, , \quad &  p_T(\gamma_2) > 25 \gev \, , \nonumber \\
\Delta R(\gamma_1 , \gamma_2) ,\, \Delta R(\gamma_{1,2} , j)  > 0.4\, , & \quad  \Delta R(\gamma_{1,2} , \ell)  > 0.4\ \,, \quad  & \quad p_T(\ell_1)  > 20 \gev \, , 
\label{eq:cuts-diphoton}
\end{eqnarray}
and an additional cut 
\begin{equation}
 \Delta R(\ell_i, \ell_j) > 0.4   
\label{eq:cuts-diphoton-2}
\end{equation}
is applied if leptons are more than one.
With $\gamma_1$ and $\gamma_2$ we respectively denote the hard and the soft photon, analogously $\ell_1$ indicates the hardest lepton. Cuts on lepton(s) imply that the fully and semileptonic decays of the top-quark pair are selected.

Results at $\rm LO+PS$ and $\rm NLO+PS$ accuracy are listed in table \ref{table:results_photon} for the signal and the $\ttaa$ background. Also, we display fixed order results (LO, NLO) at production level only, without including top decays, shower and hadronization effects. In order to be as close as possible to the analyses level, we apply the cuts in eqs. \eqref{eq:preselection} and \eqref{eq:cuts-diphoton} that involve only photons. Thus, the difference between LO and NLO results of $\ttaa$ in tables \ref{table:13tev} and  \ref{table:results_photon} are solely due to these cuts. 

\begin{figure}[!h]
\centering
\includegraphics[width=0.475\textwidth]{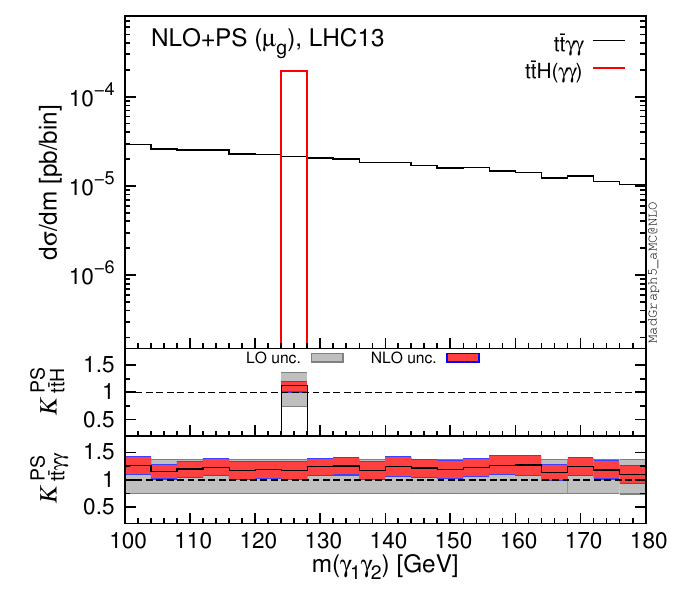}
\includegraphics[width=0.475\textwidth]{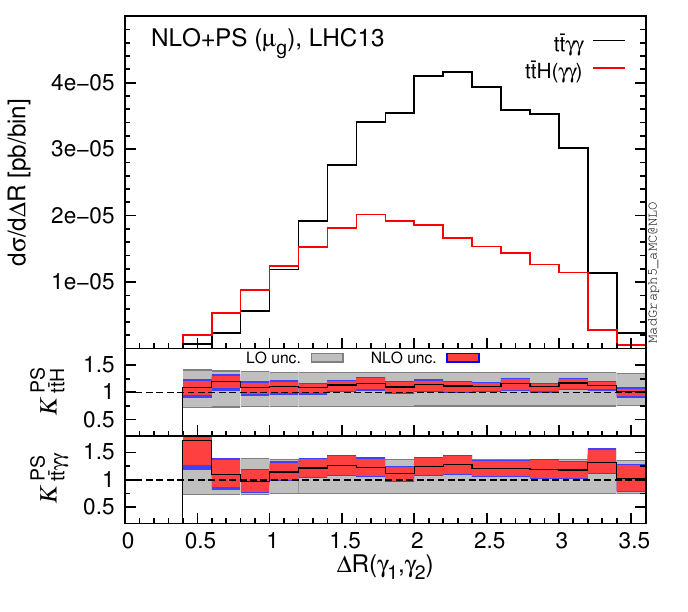}
\includegraphics[width=0.475\textwidth]{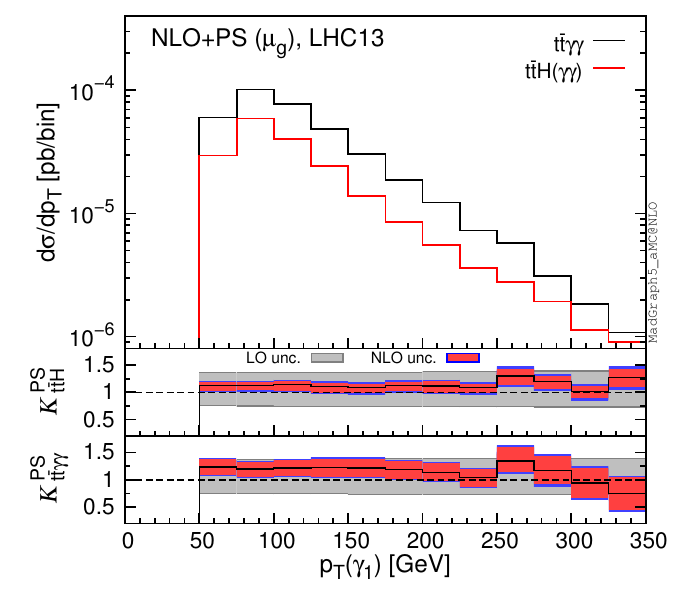}
\includegraphics[width=0.475\textwidth]{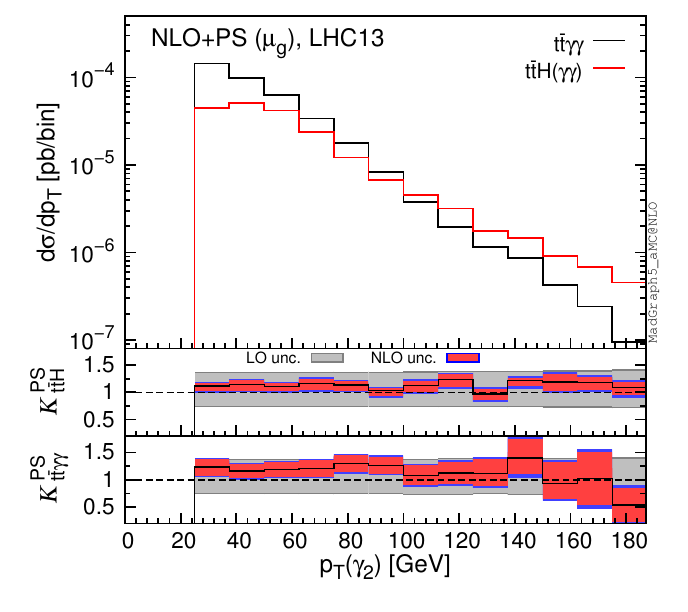}
\caption{Differential distributions for signal and background processes for the diphoton analysis.}
\label{fig:diphot_analys}
\end{figure}

In table  \ref{table:results_photon}, we show global $K$-factors both at fixed order ($K:={\rm NLO}/{\rm LO}$) and including decays, shower and hadronization effects, and all the cuts employed in the analysis ($K^{\rm PS}:={\rm NLO+ PS}/{\rm LO+PS}$). Comparing $K^{\rm PS}$ and $K$ it is possible to directly quantify the difference between a complete NLO simulation ($K^{PS}$) and the simulation typically performed at experimental level, i.e., a ${\rm LO+PS}$ simulation rescaled by a $K$-factor from production only ($K$).
As shown in table  \ref{table:results_photon}, e.g., the second approach would underestimate the prediction for $\ttaa$ production wrt. a complete ${\rm NLO+PS}$ simulation. This difference is not of particular relevance at the level of discovery, which mostly relies on an identification of a peak in the $m(\gamma_1 \gamma_2)$ (see also fig.~\ref{fig:diphot_analys}), but could be important in the determination of signal rates and in the extraction of Higgs couplings. Conversely, the difference between $K$ and $K^{\rm PS}$ is much smaller for the signal.

In fig.~\ref{fig:diphot_analys} we show representative differential distributions at ${\rm NLO+PS}$ accuracy for the signal (red) and background (black) processes. In the two insets we display the differential $K$-factor for the signal (${K^{\rm PS}_{\ttbar H}}$) and the background (${K^{\rm PS}_{\ttaa}}$) using the same layout and conventions adopted in the plots of section \ref{sec:prod}. In particular, we plot the invariant mass of the two photons ($m(\gamma_1 \gamma_2)$) their distance ($\Delta R (\gamma_1, \gamma_2)$) and the transverse momentum of the hard ($\pt(\gamma_1)$) and the soft ($\pt(\gamma_2)$) photon. We note that predictions for key discriminating observables, such as the $\Delta R (\gamma_1, \gamma_2)$ and $\pt(\gamma_2)$ are in good theoretical control.

\begin{table}[t]
\renewcommand{\arraystretch}{1.5}
\begin{center}
\tiny
\begin{tabular}{  c | c c | c | c c }
\hline\hline
13 TeV $ \sigma$[fb] & $t \bar t H \times {\rm BR}(H \rightarrow \gamma \gamma) $ &  $t \bar t \gamma \gamma $ &  & $t \bar t H (H \rightarrow \gamma \gamma)$ & $t \bar t \gamma \gamma $ \\[5pt]
\hline
NLO & $1.191^{+6.0 \%}_{-9.4 \%}~^{+2.1 \%}_{-2.6 \%}$ & $1.466^{+8.7 \%}_{-11.0 \%}~^{+1.6 \%}_{-1.8 \%}$ & NLO+PS & $0.194^{+5.9 \%}_{-9.3 \%}~^{+2.0 \%}_{-2.6 \%} \pm 0.002$ & $ 0.374^{ +11.4 \%}_{ -12.2 \%}~^{ +1.5 \%}_{ -1.7 \%} \pm 0.004 $ \\[5pt]
\hline
LO & $1.087^{+35.5 \%}_{-24.2 \%}~^{+2.0 \%}_{-2.1 \%}$ & $1.340^{+37.0 \%}_{-24.8 \%}~^{+1.7 \%}_{-1.8 \%}$ & LO+PS & $0.172^{+35.2 \%}_{-24.1 \%}~^{+2.0 \%}_{-2.2\%} \pm 0.001$ & $ 0.310^{ +36.4 \%}_{ -24.5 \%}~^{ +1.7 \%}_{ -1.8 \%} \pm 0.002 $ \\[5pt]
\hline
$K$ & 1.10 & 1.09 & $K^{\rm PS}$ & $1.13 \pm 0.01$ & $ 1.21 \pm 0.01$ \\[5pt]
\hline
\end{tabular}
\caption{NLO  and LO cross sections for $t \bar t H(H \rightarrow \gamma \gamma) , t \bar t \gamma \gamma $ processes at 13 TeV. The first uncertainty is given by scale variation, the second by PDFs. The assigned error is the statistical Monte Carlo uncertainty.}
\label{table:results_photon}
\end{center}
\end{table}

\subsection{Signatures with leptons}
\label{sec:leptons} 
 This  analysis involves three different signatures and signal regions that includes two or more leptons and it is specifically  designed for $\ttbar H$ production with subsequent $H\TO Z Z^*$, $H\TO W W^*$ and $H\TO \tau^{+}\tau^{-}$ decays.  In the simulation,  all the decays of the massive particles are performed in {\Pythiae}. In the case of the signal processes, the Higgs boson is forced to decay to the specific final state  ($H\TO Z Z^*$, $H\TO W W^*$ or $H\TO \tau^{+}\tau^{-}$) and event weights are rescaled by the corresponding branching ratios, which are taken from \cite{BR-HXSWG}: ${\rm BR}(H \rightarrow W W^*) = 2.15 \times 10^{-1}$, ${\rm BR}(H \rightarrow Z Z^*) = 2.64 \times 10^{-2}$, ${\rm BR}(H \rightarrow \tau^+ \tau^-) = 6.32 \times 10^{-2} $. The isolation of leptons from the hadronic activity is performed by directly selecting only prompt leptons in the analyses, i.e., only leptons emerging from $Z$, $W$ or  from $\tau$ leptons which emerge from $Z$, $W$ or Higgs bosons.\footnote{We observed that applying hadronic isolation cuts as done in \cite{Khachatryan:2014qaa} we obtain results with at most $10\%$ difference with those presented here by selecting prompt leptons.  $K$-factors are independent of the application of hadronic isolation cuts.} 
 
We consider as irreducible background the contribution from $t \bar t W^{\pm}$, $t \bar t Z/\gamma^*$, $t \bar t W^+W^-$, $t \bar t ZZ$, $t \bar t W^{\pm}Z$ and $t \bar t t \bar t $ production.\footnote{In principle also $t \bar t W \gamma$ and $t\bar t Z \gamma$ production can contribute to the signatures specified in the following. However, they are a small fraction of $t \bar t W$ and $t\bar t Z$ production and indeed are not taken into account in the analyses of \cite{Khachatryan:2014qaa}.}  Precisely, with the notation $t \bar t Z/\gamma^*$ we mean the full process $t \bar t \ell^+ \ell^- (\ell = e , \mu , \tau)$, where $Z$ and photon propagators, from which the $\ell^+ \ell^- $ pair emerges, can both go off-shell and interfere.\footnote{To this purpose, we excluded Higgs boson propagators in order to avoid a double count of the $\ttbar H (H\to \tau^+ \tau ^-)$ contributions.} All the processes, with the exception of $t \bar t Z/\gamma^*$, have been  also studied at fixed-order accuracy in section \ref{sec:prod}.

In the analyses the following common cuts are applied in order to select at least two leptons
\begin{equation}
m(\ell_1 \ell_2) > 12\, , \qquad \Delta R(\ell_i , \ell_j)  > 0.4\, .
\label{common:lep}
\end{equation} 
Then, the three signatures and the corresponding signal regions are defined as described in the following:

\noindent
\textbullet \; \textbf{Signal region one (SR1): two same-sign leptons} \\
Exactly two same-sign leptons with with $p_T(\ell) > 20$ GeV  are requested. The event is selected if it includes at least four jets with one or more of them that are $b$-tagged. Furthermore it is required that $p_T(\ell_1)+p_T(\ell_2)+E_T^{\text{miss}}>100\gev$ and, for the dielectron events,  $|m(e^{\pm}e^{\pm}) - m_Z| > 10$ GeV and $E_T^{\text{miss}} > 30$ GeV, in order to suppress background from electron sign misidentification in $Z$ boson decays.\\
\noindent
\textbullet \; \textbf{Signal region two (SR2):   three leptons}\\
Exactly three leptons with $p_T(\ell_1) > 20$ GeV, $p_T(\ell_2) > 10$ GeV, $p_T(\ell_3=e(\mu)) > 7 (5)$ GeV are requested. The event is selected if it includes at least two jets with one or more of them that are $b$-tagged. For a $Z$ boson background suppression, events with an opposite-sign same-flavour lepton pair are required  to have $|m(\ell^+ \ell^-) - m_Z| > 10$ GeV. Also, for this kind of events if the number of jets is equal or less than three, the cut  $E_T^{\text{miss}} > 80$ GeV is applied.\\
\noindent
\textbullet \; \textbf{Signal region three (SR3):  four leptons}\\
Exactly four leptons with $p_T(\ell_1) > 20$ GeV, $p_T(\ell_2) > 10$ GeV, $p_T(\ell_{3,4}=e(\mu)) > 7 (5)$ GeV are requested. The event is selected if it includes at least two jets with one or more of them that are $b$-tagged. Also here, for a $Z$ boson background suppression, events with an opposite-sign same-flavor lepton pair are required  to have $|m(\ell^+ \ell^-) - m_Z| > 10$ GeV.\\

For both signal and background processes, results at ${\rm LO+PS}$ and ${\rm NLO+PS}$ accuracy as well as ${K^{\rm PS}}$-factors are listed in table \ref{table:results_leptons} for the three signal regions. Also,  for each process we display the value of the global $K$-factor (listed also in section \ref{sec:prod}), which does not take into account shower effects, cuts and decays.  A posteriori, we observe that in these analyses the $K$-factors are almost insensitive of shower effects and the applied cuts. This is evident from a comparison  of the values of $K$ and  ${K^{\rm PS}}$ in table  \ref{table:results_leptons}, where the largest discrepancy stems from the  $t \bar t Z/\gamma^*$ process in SR1.  We also verified, with the help of {\Madspin}, that results in the SR3 (SR2 for $\ttW$) do not change when  spin-correlation effects are taken into account in the decays.\footnote{SR2 and especially SR1 involves a rich combinatoric of leptonic and hadronic $Z$, $W$ and $\tau$ decays, which render the simulation with spin-correlation non-trivial. However, we checked also here for representative cases that spin-correlation effects do not sensitively alter the results.} 
It is important to note that, a priori, with different cuts and/or at different energies, $K$ and  ${K^{\rm PS}}$ could be in principle different and spin correlation effects may be not negligible. Thus, a genuine NLO+PS simulation is always preferable.
\begin{table}[t]
\renewcommand{\arraystretch}{1.5}
\begin{center}
\tiny
\begin{tabular}{ c | c | c c c }
\hline\hline
13 TeV $ \sigma$[fb] &  &  SR1 & SR2 & SR3 \\[5pt]
\hline
 & NLO+PS & $ 1.54^{ +5.1 \%}_{ -9.0 \%}~^{ +2.2 \%}_{ -2.6 \%} \pm 0.02 $ & $ 1.47^{ +5.2 \%}_{ -9.0 \%}~^{ +2.0 \%}_{ -2.4 \%} \pm 0.02 $  & $ 0.095^{ +7.4 \%}_{ -9.7 \%}~^{ +2.0 \%}_{ -2.4 \%} \pm 0.002 $  \\[5pt]
$t \bar t H (H \rightarrow W W^*)$ & LO+PS & $ 1.401^{ +35.6 \%}_{ -24.4 \%}~^{ +2.1 \%}_{ -2.2 \%} \pm 0.008 $ & $ 1.355^{ +35.2 \%}_{ -24.1 \%}~^{ +2.0 \%}_{ -2.2 \%} \pm 0.008 $  & $ 0.0855^{ +34.9 \%}_{ -24.0 \%}~^{ +2.0 \%}_{ -2.2 \%} \pm 0.0007 $   \\[5pt]
$K = 1.10$ & $K^{\rm PS}$ & $ 1.10 \pm 0.02$  & $ 1.09 \pm 0.02$ & $ 1.11 \pm 0.02$   \\[5pt]
\hline
 & NLO+PS & $ 0.0437^{ +5.5 \%}_{ -9.2 \%}~^{ +2.3 \%}_{ -2.8 \%} \pm 0.0004 $  & $ 0.119^{ +6.3 \%}_{ -9.6 \%}~^{ +2.1 \%}_{ -2.5 \%} \pm 0.002 $  & $ 0.0170^{ +5.0 \%}_{ -8.5 \%}~^{ +2.0 \%}_{ -2.4 \%} \pm 0.0003 $   \\[5pt]
$t \bar t H (H \rightarrow Z Z^* )$ & LO+PS & $ 0.0404^{ +36.1 \%}_{ -24.6 \%}~^{ +2.2 \%}_{ -2.3 \%} \pm 0.0002 $ & $ 0.1092^{ +35.3 \%}_{ -24.2 \%}~^{ +2.0 \%}_{ -2.2 \%} \pm 0.0008 $  & $ 0.0152^{ +34.7 \%}_{ -23.9 \%}~^{ +1.9 \%}_{ -2.1 \%} \pm 0.0001 $   \\[5pt]
$K = 1.10$ & $K^{\rm PS}$ & $ 1.08 \pm 0.01$  & $ 1.09 \pm 0.02$ & $ 1.12 \pm 0.02$  \\[5pt]
\hline
 & NLO+PS & $ 0.563^{ +4.6 \%}_{ -8.8 \%}~^{ +2.2 \%}_{ -2.7 \%} \pm 0.007 $ & $ 0.669^{ +6.0 \%}_{ -9.4 \%}~^{ +2.1 \%}_{ -2.6 \%} \pm 0.008 $ & $ 0.0494^{ +7.1 \%}_{ -9.9 \%}~^{ +2.1 \%}_{ -2.5 \%} \pm 0.0007 $ \\[5pt]
$t \bar t H (H \rightarrow \tau^+ \tau^- )$ & LO+PS & $ 0.513^{ +35.9 \%}_{ -24.5 \%}~^{ +2.2 \%}_{ -2.3 \%} \pm 0.003 $ & $ 0.611^{ +35.4 \%}_{ -24.2 \%}~^{ +2.1 \%}_{ -2.2 \%} \pm 0.003 $ & $ 0.0438^{ +35.1 \%}_{ -24.1 \%}~^{ +2.0 \%}_{ -2.2 \%} \pm 0.0003 $    \\[5pt]
$K = 1.10$ & $K^{\rm PS}$ & $ 1.10 \pm 0.02$  & $1.10\pm0.01$ & $ 1.13 \pm 0.02$ \\[5pt]
\hline
\hline
 & NLO+PS & $ 5.77^{ +15.1 \%}_{ -12.7 \%}~^{ +1.6 \%}_{ -1.2 \%} \pm 0.07 $  & $ 2.44^{ +13.1 \%}_{ -11.6 \%}~^{ +1.7 \%}_{ -1.4 \%} \pm 0.01 $  & -  \\[5pt]
$t \bar t W^{\pm}$ & LO+PS & $ 4.57^{ +27.7 \%}_{ -20.2 \%}~^{ +1.8 \%}_{ -1.9 \%} \pm 0.03 $  & $ 1.989^{ +27.5 \%}_{ -20.0 \%}~^{ +1.8 \%}_{ -1.9 \%} \pm 0.007 $ & -  \\[5pt]
$K = 1.22$ & $K^{\rm PS}$ & $ 1.26 \pm 0.02$ & $ 1.23 \pm 0.01$ & -  \\[5pt]
\hline
 & NLO+PS & $ 1.61^{ +7.7 \%}_{ -10.5 \%}~^{ +2.0 \%}_{ -2.5 \%} \pm 0.02 $  & $ 2.70^{ +9.0 \%}_{ -11.2 \%}~^{ +2.0 \%}_{ -2.5 \%} \pm 0.03 $  & $ 0.280^{ +9.8 \%}_{ -11.0 \%}~^{ +1.9 \%}_{ -2.3 \%} \pm 0.003 $   \\[5pt]
$t \bar t Z/\gamma^*$ & LO+PS & $ 1.422^{ +36.8 \%}_{ -24.9 \%}~^{ +2.2 \%}_{ -2.3 \%} \pm 0.008$  & $ 2.21^{ +36.4 \%}_{ -24.7 \%}~^{ +2.1 \%}_{ -2.2 \%} \pm 0.01 $ & $ 0.221^{ +35.8 \%}_{ -24.4 \%}~^{ +2.0 \%}_{ -2.2 \%} \pm 0.001 $    \\[5pt]
$K = 1.23$ & $K^{\rm PS}$ & $ 1.13 \pm 0.02$ & $ 1.23 \pm 0.01$ & $ 1.27 \pm 0.01$  \\[5pt]
\hline
 & NLO+PS &$ 0.288^{ +8.0 \%}_{ -11.1 \%}~^{ +2.3 \%}_{ -2.6 \%} \pm 0.003 $  & $ 0.201^{ +7.4 \%}_{ -10.7 \%}~^{ +2.1 \%}_{ -2.3 \%} \pm 0.003 $ & $ 0.0116^{ +6.9 \%}_{ -10.2 \%}~^{ +2.2 \%}_{ -2.3 \%} \pm 0.0002 $   \\[5pt]
$t \bar t W^+W^-$ & LO+PS & $ 0.260^{ +38.4 \%}_{ -25.5 \%}~^{ +2.3 \%}_{ -2.3 \%} \pm 0.001$  & $ 0.181^{ +38.0 \%}_{ -25.3 \%}~^{ +2.2 \%}_{ -2.2 \%} \pm 0.001$  & $ 0.01073^{ +37.7 \%}_{ -25.1 \%}~^{ +2.2 \%}_{ -2.2 \%} \pm 0.00008 $  \\[5pt]
$K = 1.10$ & $K^{\rm PS}$ & $ 1.11 \pm 0.01$  & $ 1.11 \pm 0.01$ & $ 1.08 \pm 0.02$  \\[5pt]
\hline
 & NLO+PS & $ 0.340^{ +27.5 \%}_{ -25.8 \%}~^{ +5.5 \%}_{ -6.4 \%} \pm 0.004$ & $ 0.211^{ +27.4 \%}_{ -25.6 \%}~^{ +5.2 \%}_{ -6.1 \%} \pm 0.003$ & $ 0.0110^{ +27.0 \%}_{ -25.5 \%}~^{ +5.0 \%}_{ -5.9 \%} \pm 0.0002 $ \\[5pt]
$t \bar t t \bar t$ & LO+PS & $ 0.271^{ +80.9 \%}_{ -41.5 \%}~^{ +4.6 \%}_{ -4.6 \%} \pm 0.001$ & $ 0.166^{ +80.3 \%}_{ -41.4 \%}~^{ +4.4 \%}_{ -4.4 \%} \pm 0.001$  & $ 0.00871^{ +79.8 \%}_{ -41.2 \%}~^{ +4.2 \%}_{ -4.2 \%} \pm 0.00007 $ \\[5pt]
$K = 1.22$ & $K^{\rm PS}$  & $ 1.26 \pm 0.02$ & $ 1.27 \pm 0.02$ & $ 1.26 \pm 0.03$   \\[5pt]
\hline
13 TeV $ \sigma$[ab] &  &  SR1 & SR2 & SR3 \\[5pt]
\hline
 & NLO+PS & $ 9.60^{ +3.5 \%}_{ -8.4 \%}~^{ +1.8 \%}_{ -1.8 \%} \pm 0.06$ & $5.02^{ +3.7 \%}_{ -8.3 \%}~^{ +1.8 \%}_{ -1.7 \%} \pm 0.04 $ & $ 0.249^{ +7.2 \%}_{ -9.6 \%}~^{ +1.9 \%}_{ -1.8 \%} \pm 0.009 $ \\[5pt]
$t \bar t Z Z$ & LO+PS & $ 9.71^{ +36.3 \%}_{ -24.5 \%}~^{ +1.9 \%}_{ -1.9 \%} \pm 0.02 $ & $ 5.08^{ +35.9 \%}_{ -24.3 \%}~^{ +1.9 \%}_{ -1.9 \%} \pm 0.02 $  & $ 0.250^{ +35.5 \%}_{ -24.2 \%}~^{ +1.9 \%}_{ -1.9 \%} \pm 0.004 $ \\[5pt]
$K = 0.99$ & $K^{\rm PS}$ & $ 0.99 \pm 0.01$ & $ 0.99 \pm 0.01$  &$ 1.00 \pm 0.04$  \\[5pt]
\hline
 & NLO+PS & $ 62.0^{ +9.0 \%}_{ -10.2 \%}~^{ +2.2 \%}_{ -1.6 \%} \pm 0.7 $   & $ 27.9^{ +9.2 \%}_{ -10.3 \%}~^{ +2.3 \%}_{ -1.7 \%} \pm 0.5 $ & $ 0.91^{ +7.2 \%}_{ -9.2 \%}~^{ +2.4 \%}_{ -1.7 \%} \pm 0.02   $   \\[5pt]
$t \bar t W^{\pm} Z$ & LO+PS & $ 60.2^{ +32.2 \%}_{ -22.6 \%}~^{ +2.4 \%}_{ -2.3 \%} \pm 0.3 $ & $ 26.4^{ +32.0 \%}_{ -22.5 \%}~^{ +2.4 \%}_{ -2.2 \%} \pm 0.2 $ & $ 0.893^{ +31.9 \%}_{ -22.4 \%}~^{ +2.4 \%}_{ -2.2 \%} \pm 0.009 $ \\[5pt]
$K = 1.06$ & $K^{\rm PS}$ & $ 1.03 \pm 0.01$ & $ 1.06 \pm 0.02$ & $ 1.02 \pm 0.02$  \\[5pt]
\hline
\end{tabular}
\caption{ NLO  and LO cross sections for signal and background processes for $t \bar t H$ to multileptons at 13 TeV. The first uncertainty is given by scale variation, the second by PDFs. The assigned error is the statistical Monte Carlo uncertainty.}
\label{table:results_leptons}
\end{center}
\end{table}

\section{Conclusions}
\label{sec:conclusion} 

In this paper we have presented a thorough study at NLO QCD accuracy for $\ttV$ and $\ttVV$ processes as well as for $\ttbar H$ and $\tttt$ production within the same computational framework and using the same input parameters. In the case of $\ttVV$ processes, with the exception of $\ttaa$ production, NLO cross sections have been studied for the first time here. Moreover, we  have performed a complete analysis with realistic selection cuts on final states at NLO QCD accuracy including the matching to parton shower and decays, for both signal and background processes relevant for searches at the LHC for the $\ttbar H$ production. Specifically, we have considered the cases where the Higgs boson decays either into leptons, where  $\ttV$ and $\ttVV$ processes and $\tttt$ production provide  backgrounds, or into two photons giving  the same signature as $\ttaa$ production.

We have investigated the behaviour of fixed order NLO QCD corrections for several distributions and we have analysed their dependence on (the definition of) the renormalisation and factorisation scales.  We have found  that QCD corrections on key distributions cannot be described by overall $K$-factors. However, dynamical scales in general, even though not always, reduce the dependence of the corrections on kinematic variables and thus lead to flatter $K$-factors. In addition, our study shows that while it is not possible to identify a ``best scale'' choice for all  processes and/or differential distributions in $\ttV$ and $\ttVV$, such processes present similar features and can be studied together. 
For all the processes considered, NLO QCD corrections are in general necessary in order to provide precise and reliable predictions at the LHC. In particular cases they are  also essential for a realistic phenomenological description. Notable examples discussed in the text are, e.g., the giant corrections in the tails of $\pt(\ttbar)$  distributions for $\ttV$ processes and the large decrement of the top-quark central asymmetry for $\tta$ production. In the case of future (hadron) colliders also inclusive cross sections receive sizeable corrections, which lead, e.g., to $K$-factors larger than two at 100 TeV for $\ttV$ and $\ttVV$ processes with a charged final state. 

In the searches at the LHC for the $\ttbar H$ production with the Higgs boson decaying either into leptons or photons, NLO QCD corrections are important for precise predictions of the signal and the background. We have explicitly studied the sensitivity of NLO+PS QCD corrections on experimental cuts by comparing genuine NLO+PS QCD predictions with LO+PS predictions rescaled by global $K$-factors from the fixed order calculations without cuts.{ \it A posteriori}, we have verified that these two approximations give compatible results for analyses at the 13 TeV Run-II of the LHC with the cuts specified in the text. {\it A priori}, this feature is not guaranteed for analyses with different cuts and/or at different energies. In general, a complete NLO+PS prediction for both signal and background processes is more reliable an thus preferable for any kind of simulation. 

All the results presented in this paper have been obtained automatically in the publicly available {\aNLO} framework and they can be reproduced starting from the input parameters specified in the text.

\section*{Acknowledgements}
We thank  the $\ttbar H$ subgroup of the LHCHXSWG and in particular Stefano Pozzorini for many stimulating conversations. We thank also all the members of the {{\sc\small MadGraph5\_\-aMC@NLO}} collaboration for their help and for providing a great framework for pursuing this study. This work is done in the context of and supported in part (DP) by the ERC grant 291377 ``LHCtheory: Theoretical predictions and analyses of LHC physics: advancing the precision frontier" and under the Grant Agreement number PITN-GA-2012-315877 (MCNet).   The work of FM and IT is supported by the IISN ``MadGraph'' convention 4.4511.10, by the IISN ``Fundamental interactions''
convention 4.4517.08, and in part by the Belgian Federal Science Policy Office
through the Interuniversity Attraction Pole P7/37.

\bibliographystyle{JHEP}
\bibliography{article}

\end{document}